\UseRawInputEncoding
\documentclass[aps,rmp,reprint,amsmath,amssymb,graphicx,longbibliography]{revtex4-1}
\usepackage{graphicx} 
\usepackage{bm}
\usepackage{hyperref}
\usepackage{xcolor}

\newcommand{\Lb}{\left(}
\newcommand{\Rb}{\right)}

\begin{document}

\title{The proton charge radius}

\author{H. Gao}
\affiliation{Department of Physics, Duke University and the Triangle Universities Nuclear Laboratory, Science Drive, Durham, NC 27708, USA}
\author{M. Vanderhaeghen} 
\affiliation{Institut f\"ur Kernphysik and $\text{PRISMA}^+$ Cluster of Excellence, Johannes Gutenberg Universit\"at, D-55099 Mainz, Germany}

\date{\today{}}

\begin{abstract}

Nucleons (protons and neutrons) are the building blocks of atomic nuclei, 
and are responsible for more than 99\% of the visible matter in the universe.
 Despite decades of efforts in studying its internal structure, there are still a
 number of puzzles surrounding the proton such as its spin, and charge radius. 
Accurate knowledge about the proton charge radius is not only essential for understanding 
how quantum chromodynamics (QCD) works in the non-perturbative region, but also 
important for bound state quantum electrodynamics (QED) calculations of atomic 
energy levels. It also has an impact on the Rydberg constant, one of the most precisely measured
 fundamental constants in nature. This article reviews the latest situation concerning 
the proton charge radius in light of the new experimental results from both atomic hydrogen spectroscopy and electron scattering measurements, with particular focus on the latter. We also present the related theoretical developments and backgrounds concerning the determination 
of the proton charge radius using different experimental techniques. We discuss upcoming experiments, and briefly mention the deuteron charge radius puzzle at the end.
\end{abstract}


\maketitle

\tableofcontents{}

\section{Introduction}
\label{intro}

Nucleons (protons and neutrons) are the building blocks of atomic nuclei, and are responsible for more than 99\% of the visible matter in the universe. The force that is responsible for binding nucleons into nuclei -- and responsible for the composite nature of nucleons -- is the strong force, one of the four fundamental forces in nature. The ultimate goal of modern nuclear physics is to predict properties of nucleons, atomic nuclei and nuclear reactions from the first principles of Quantum Chromodynamics (QCD), the theory of strong interaction with quarks and gluons as the fundamental degrees of freedom. While QCD has been well tested experimentally at high energies where perturbative calculations can be carried out, in the low-energy regio, how QCD works still requires much better understanding. Nucleons therefore become important QCD laboratories through studies of their rich internal structure. 

Despite decades of efforts in studying the internal structure of the proton, there are still a number of puzzles and open questions surrounding the proton such as its spin and charge radius. Following the original deep-inelastic scattering (DIS) experiments utilizing the electron beam at the Stanford Linear Accelerator Center (SLAC)~\cite{Bloom:1969kc,Breidenbach:1969kd}, polarized DIS measurements were carried out for the first time at the European Organization for Nuclear Research (CERN) by the European Muon Collaboration (EMC)~\cite{Ashman88} in which polarized muons were scattering off polarized nucleon targets -- and quarks were found to contribute little to the proton spin. This triggered the so-called ``Proton Spin Crisis" as the EMC finding was so different from the naive expectation based on the quark model in which the quarks' spins are expected to make up the entire proton spin. After more than three decades of polarization experiments worldwide -- the emerging picture about the proton spin is that -- quark spin contributes about a third to the proton spin with comparable contribution likely by the spins of the gluons -- and the remaining comes from the orbital angular momenta of the quarks and gluons inside.  The effort to resolve the proton spin puzzle has now been focusing on more precise information about the gluon spin contribution, and quantitative information about the orbital angular momentum contributions by the quarks and gluons to the proton spin. For a review of the proton spin, we refer readers to~\cite{Kuhn09,ji2020proton}.

Another interesting aspect concerning the proton is its mass.
The discovery of the Higgs boson at the LHC -- the particle responsible for the masses of fundamental particles in nature -- marked another great triumph of the Standard Model. However, the Higgs particle is not so relevant when it comes to the mass of the proton. The proton mass decomposition has been a topic of increasing interest in recent years motivated by the experimental capability offered by the energy upgraded 12-GeV Continuous Electron Beam Accelerator Facility (CEBAF) at Jefferson Lab~\cite{12GeVWP}, and the future Electron-Ion Collider (EIC)~\cite{Accardi16} to be built at the Brookhaven National Laboratory.  Various approaches in the proton mass decomposition exist in the literature, and a few examples are given in~\cite{Ji1994,Lorce2018,Shifman1978}. 
Following Ji's decomposition, 
the quark mass contribution to the proton mass is found to be $\sim$ 11\%, trace anomaly is about 22\%, and the rest is due to the quark and gluon energy~\cite{Gao2015}. 
There have been major theoretical and experimental efforts in studying the QCD trace anomaly contribution to the proton mass. Near-threshold electro- and photo-production cross sections of $J/\Psi$ and $\Upsilon$ particles~\cite{Kharzeev:1998bz,PhysRevD.98.074003,Gryniuk:2016mpk,Gryniuk:2020mlh}, from the proton have been proposed as effective ways to access the trace anomaly contribution, and experiments ~\cite{E12-12-006,Gryniuk:2020mlh} are being planned at Jefferson Lab, and also the future EIC.

The proton root-mean-square (rms) charge radius (a.k.a. proton charge radius) is a quantity that is not only of importance to QCD, but also important for bound state QED calculations of atomic energy levels, and has a direct impact on the determination of the Rydberg constant, one of the most well-known fundamental quantities in Nature. Conventionally, the proton charge radius can be determined from electron-proton elastic scattering, a method pioneered by Hofstadter, and atomic spectroscopic measurements using ordinary hydrogen atoms. In the former case, one determines the proton electric form factor from scattering cross sections first from which one then extracts the proton charge radius. In the latter case, experimentally measured atomic transitions combined with state-of-the-art QED calculations allow for an extraction of the proton charge radius.  

The proton charge radius puzzle originated in 2010 following a discrepancy of 5-7 standard deviations between the ultrahigh precise values of the proton charge radius determined from muonic hydrogen Lamb shift measurements and the Committee on Data for Science and Technology (CODATA) values compiled from electron-proton scattering experiments and ordinary hydrogen spectroscopy measurements, and with the muonic results being significantly smaller than the CODATA recommended values. In the last ten years, major progress has been made in resolving this puzzle, which is the focus of this review paper. While we cover the latest progress in atomic spectroscopy concerning the proton charge radius, special emphasis will be given in this review to the progress from lepton scattering, and the associated challenges.
The rest of the paper is organized as the following. We set the stage and introduce the proton charge radius puzzle in Section II. In Section III we describe how the charge radius is defined, how it can be properly understood in terms of a quark charge distribution, and how it is connected to the quark structure of the proton. We subsequently describe the  experimental techniques in determining the proton charge radius from elastic electron-proton scattering in Section III and from atomic hydrogen spectroscopy in Section IV. Section V and VI review the results from the recent lepton scattering, and spectroscopy measurements, respectively. In Section VII, we review ongoing and planned lepton scattering experiments. Section VIII provides a brief introduction of another charge radius puzzle which concerns the deuteron before we conclude in Section IX.

\section{The Proton Charge Radius Puzzle}
\label{background}

\subsection{Before the millennium} 
Although the release of the proton charge radius result from a muonic hydrogen spectroscopic measurement by the CREMA collaboration in 2010~\cite{Pohl10} -- triggered a major proton charge radius puzzle -- there was a puzzle even before that, known perhaps only to a much smaller community. Prior to the muonic hydrogen measurement, all results of the proton charge radius came from electron-proton scattering experiments and also ordinary hydrogen spectroscopic measurements.  An important motivation to improve the precision in determining the proton charge radius from electron scattering experiments is for precision tests of QED through hydrogen Lamb shift measurements. The standard hydrogen Lamb shift
measurement probes the 1057 MHz fine structure transition between the
$2S_{1/2}$ and $2P_{1/2}$ states -- and can be calculated to high precision with higher-order
corrections in QED with the proton rms charge radius as an important input for finite size and other
hadronic structure contributions. 
However, the two most precise values from electron scattering experiments in the literature before 2010 -- each with a relative uncertainty of less than 1.5\% but differing by about 7\% (relative) -- are  $r_{p} = 0.805(11)$ fm \cite{Hand63} and $ r_{p} =
0.862(12)$ fm \cite{Simon80}. The result from \cite{Hand63} includes data from several experiments. In late 1990s, several groups published high precision results from hydrogen spectroscopic measurements~\cite{Weitz94, Hagley94, Berkeland95, Bourzeix96, Wijingaarden98}, and these results support a larger value of the proton charge radius (0.862 fm) when compared with QED predictions including the two-loop binding effects. Melnikov and van Ritbergen~\cite{Melnikov00} calculated the three-loop slope of the Dirac form factor -- the last known contribution to the hydrogen energy levels at order $m\alpha^7$ -- and extracted a proton charge radius value of $0.883(14)$ fm combining the QED calculation of the 1$S$ Lamb shift and the experimental measurement~\cite{Schwob99}.  

\subsection{After the millennium}
The situation surrounding the proton charge radius became more intriguing in 2010 when the CREMA collaboration~\cite{Pohl10} reported the first determination of the proton charge radius from a muonic hydrogen spectroscopic method ever -- giving a value of 0.84184(67) fm by measuring the transition between the $2S_{1/2} (F=1)$ and the $2P_{3/2}(F=2)$ energy levels  -- that was most precise at the time, but 7 $\sigma$ smaller than the 2010 CODATA recommended value of 0.8775(51) fm ~\cite{Mohr10}. This discrepancy created the much more widely known proton charge radius puzzle compared with the one before 2000, see~\cite{Pohl:2013yb,Carlson:2015jba} 
for some early reviews. 
In 2013, the CREMA collaboration reported~\cite{Antognini13} a value of 0.84087(39) fm from combined analyses of the original transition they reported in 2010 together with a different transition between the $2S_{1/2} (F=0)$ and the $2P_{3/2}(F=1)$ levels. From the electron scattering community, two values of the proton charge radius were reported around the same time, and they are 0.8791(79) fm by Bernauer {\it et al.}~\cite{Bernauer10}, and 0.875 (10) fm by Zhan {\it et al.}~\cite{Zhan11} -- both were included in the 2010 CODATA compilation and are in excellent agreement with its recommended value. The muonic hydrogen results~\cite{Pohl10,Antognini13} had not been included into the CODATA compilation until its most recent release~\cite{codata18}.

 

\vspace{0.2in}
\section{Elastic Electron-Proton Scattering}
\label{epscattering}

Electron scattering has proved to be an effective and clean way to probe the internal structure of the nucleon -- as the lepton vertex is well described by QED -- and higher-order contributions are suppressed compared with the leading-order, one-photon-exchange contribution. This has been demonstrated by the Nobel Prize winning electron-proton elastic scattering experiment carried out by 
Robert Hofstadter and collaborators in the 1950s at the Stanford University~\cite{Hofstadter55}~\cite{Hofstadter56} -- in which the root-mean-squared charge radius of the proton -- $0.74 \pm 0.24$ (fm) -- was determined for the first time.
The success of the lepton scattering was further demonstrated by another Nobel Prize awarded to Friedman, Kendall, and Taylor~\cite{Bloom:1969kc,Breidenbach:1969kd}  for leading the DIS experiments with electron beams at SLAC between 1967 to 1973 -- that discovered for the first time the existence of point-like-particles -- quarks inside the proton.  For details about the discovery of quarks, one may refer to the article written by Michael Riordan~\cite{MRiordan92}.

\subsection{Introduction to Electron-Proton Scattering and Proton Electromagnetic Form Factors}
\label{sec:elintro}

\begin{figure}[h]
\includegraphics[scale=0.28]{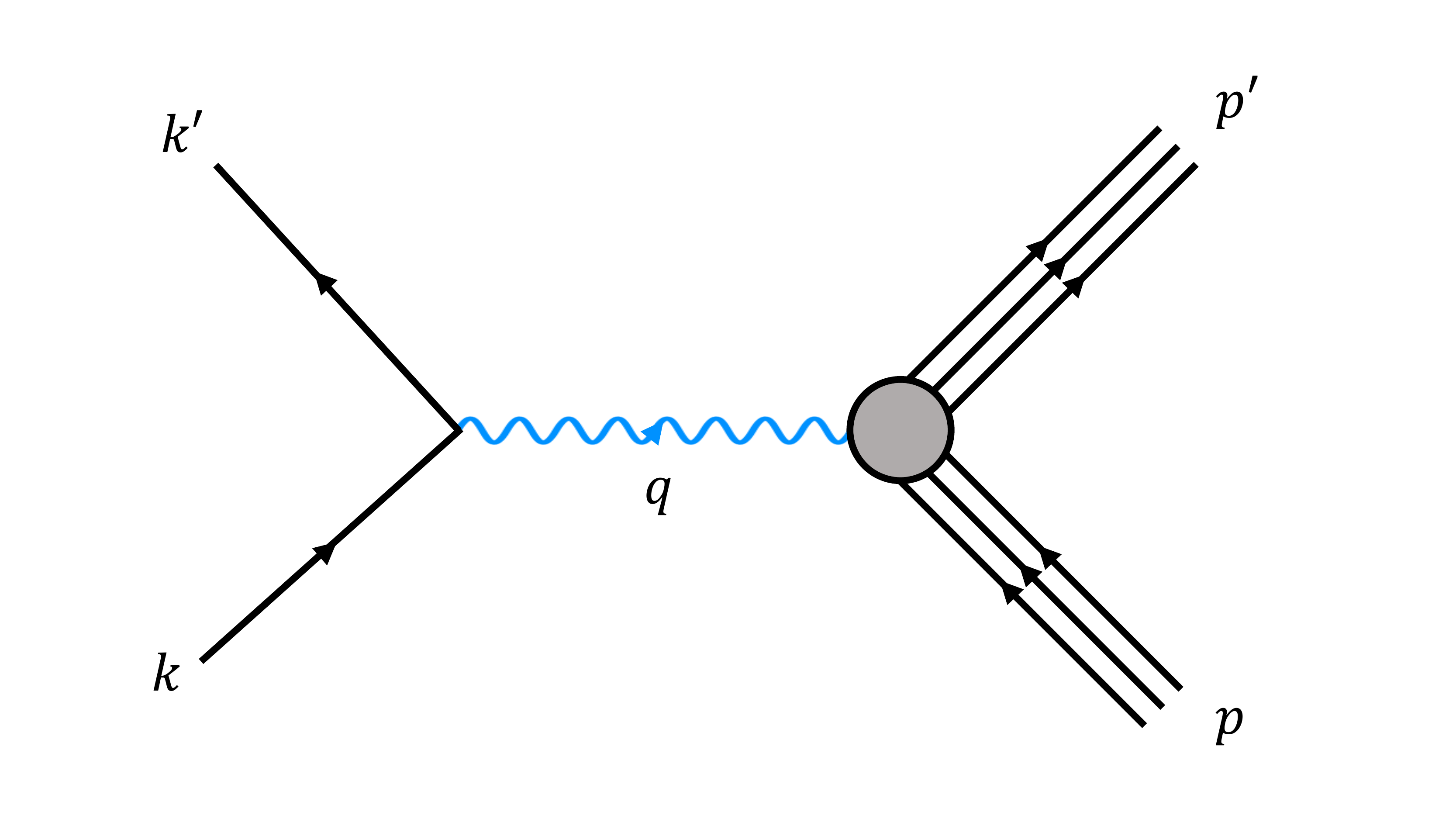}
\vspace{-1cm}
\caption{The one-photon-exchange diagram describing the elastic electron-proton scattering (figure credit: Jingyi Zhou).} 
\label{fig:one-photon}
\end{figure}

To lowest-order in QED, the dominant contribution to the electron-proton elastic scattering is the one-photon-exchange (OPE) Feynman diagram as shown in Fig.~\ref{fig:one-photon}. The 4-momentum of the incoming (scattering) electron is labeled by $k$ ($k'$). The 4-momentum of the target (recoil) proton is labeled by $p$ ($p'$). A virtual photon exchanged between the electron and the proton carries a 4-momentum, $q$, and the corresponding momentum transfer squared  $q^2$, is a Lorentz invariant.  
In electron scattering, the opposite of the four-momentum transfer squared, $Q^2$
($Q^2=-q^2 \ge 0$) is commonly used.

The scattering amplitude for the elastic electron scattering from a hadronic target in OPE based on QED can be written, as
\begin{equation}
{\cal M} = i \frac{e^2}{Q^2} u(k',h) \gamma_\mu u(k, h) 
\langle p', \lambda'| J^\mu_{em}(0) | p, \lambda \rangle,
\label{OPE}
\end{equation}
in which $u$ denote the electron Dirac spinors with $h$ the (conserved) helicity of the electrons, $\lambda$ ($\lambda'$) denote the helicities of initial (final) hadrons, and $\langle p', \lambda'| J^\mu_{em}(0) | p, \lambda \rangle$ the hadron matrix element of the local electromagnetic current operator (taken at space-time point $x = 0$).

For a spin-$\frac{1}{2}$ extended object such as a nucleon, its electromagnetic transition current -- following the requirements of current and parity conservation and covariance under the improper Lorentz group -- can be written as
\begin{equation}
\langle p', \lambda'| J^\mu_{em}(0) | p, \lambda \rangle =  \bar{N}(p',\lambda') \Gamma^\mu N(p, \lambda),
\end{equation}
in which $N$ denote the nucleon spinors, and 
where $\Gamma^{\mu}$ is the virtual photon-proton vertex:
\begin{equation}
\Gamma^{\mu} \equiv F_{1}(q^2)\gamma^{\mu} + F_2(q^2)  
\frac{i \sigma^{\mu\nu}q_{\nu}}{2M}.
\label{eq:pvertex}
\end{equation}
The functions $F_{1}$ and $F_{2}$, two independent quantities which depend on $q^{2} (Q^{2}) $ only, are called the Dirac and Pauli form factors (FF), respectively, and $M$ is the mass of the nucleon.

The electric ($G_E$) and the magnetic ($G_M$) form factor of the nucleon, also called the Sachs' form factors, are two independent linear combinations of $F_1$ and $F_2$, originally proposed by 
Ernst, Sachs and Wali~\cite{Ernst60} as:   
\begin{eqnarray}
G_E &=& F_1 - {\frac{Q^2}{4M^2}}F_2, 
\label{eq:GE} \\
G_M &=& F_1 + F_2.
\label{eq:GM}
\end{eqnarray}
In the limit of $Q^2 =0$, $G_{Ep} (0) =1$, $G_{En} (0) =0$, which are just the charge of the proton, and neutron, respectively; while  $G_{Mp} (0) =\mu_p$, $G_{Mn} (0) =\mu_n$, the proton and neutron magnetic moments, correspondingly. The Pauli FF at $Q^2 = 0$ is given by the anomalous magnetic moment $F_2(0) \equiv \kappa$, with $\mu_p = 1 + \kappa_p$ and $\mu_n = \kappa_n$. In comparison to the $F_1$ and $F_2$ form factors, the $G_E$ and $G_M$ were proposed to have a more intuitive physical interpretation, though $G_E (0) =F_{1}(0)$. Sachs~\cite{Sach62} showed that in the Breit frame $G_E$ and $G_M$ can be interpreted as Fourier transforms of spatial distributions of charge and 
magnetization, under the assumption of considering the nucleon as a non-relativistic static system.  In the Breit frame the incoming electron has a momentum of $\vec{q} / 2$ and
the nucleon initial momentum is $- \vec{q} / 2$; the scattered electron
has a momentum of $- \vec{q} /2$ and the recoil proton has a momentum
of $\vec{q} / 2$. So it
is a special Lorentz frame in which $q^2 = - {\vec{q}}^{\, 2}$, i.e., no energy transfer is involved in this 
particular reference frame, and it also coincides with the electron-proton (e-p) center-of-mass frame for e-p elastic scattering. 
For each $Q^2$ value, there is the corresponding center-of-mass frame (Breit frame) 
in which the form factors are represented as $G_{E,M}(q^2) = G_{E,M}(-{\vec{q}}^{\, 2})$. 
For non-relativistic (n-rel) static systems, the analogy to a  ``classical'' charge density 
distribution has then been introduced in the literature 
through the three-dimensional ($3d$) Fourier transformation of the matrix element of the charge operator in the Breit (B) frame: 
\begin{eqnarray}
\rho_{3d, n-rel}(r) &=& \int {\frac{d^{3}\vec q}{(2\pi)^3}} e^{-i \vec{q}\cdot \vec{r}} 
\frac{\langle p', \lambda| J^0_{em}(0) | p, \lambda \rangle_B}{2M}, \nonumber \\
&=& \int {\frac{d^{3}\vec q}{(2\pi)^3}} e^{-i \vec{q}\cdot \vec{r}} G_{E}(-{\vec{q}}^{\, 2}),
\label{eq:dens3dnr}
\end{eqnarray}
which only depends on $r = |\vec r|$ for a spherical symmetric system.  

It has been pointed out in~\cite{Lorce:2020onh} that for a relativistic (rel) system, a proper kinematical factor has to be introduced, leading to the modified quantity:
\begin{eqnarray}
\rho_{3d, rel}(r) 
&=& \int {\frac{d^{3}\vec q}{(2\pi)^3}} e^{-i \vec{q}\cdot \vec{r}} \frac{\langle p', \lambda| J^0_{em}(0) | p, \lambda \rangle_B}{2 P^0_B}, \nonumber \\
&=& \int {\frac{d^{3}\vec q}{(2\pi)^3}} e^{-i \vec{q}\cdot \vec{r}} \frac{1}{\sqrt{1 + \vec{q}^{\, 2}/(4 M^2)}} G_{E}(-{\vec{q}}^{\, 2}),\quad
\label{eq:dens3dr}
\end{eqnarray}
where $P^0_B$ is the nucleon energy in the Breit frame. 
It was furthermore argued in \cite{Lorce:2020onh} from a phase-space perspective that the quantity $\rho_{3d, rel}(r)$ can be interpreted as an internal charge  quasi-density of the target.  
One notices that such relativistic quasi-density is obtained by the Fourier transform of $G_E(q^2)$ multiplied by the relativistic factor $M/P^0_B = 1/\sqrt{1 + Q^{2}/(4 M^2)}$, as $Q^2 \equiv - q^2 = {\vec{q}}^{\, 2}$ in the Breit frame. 

To arrive at a strict density or probabilistic interpretation, requires the momentum transfer to remain small compared to the inertia of the system. The concept of a rest-frame density is therefore intrinsically limited by the Compton wavelength of the system. This limitation can however be avoided in the infinite-momentum frame (IMF), in which the magnitude of the nucleon's momentum $|{\bf{p}}| \gg M$, i.e. the nucleon is moving at infinite momentum. The IMF is advantageous in discussing deep inelastic scattering process in which the virtual photon interacts with a parton (quark) inside the nucleon. In IMF due to relativistic time dilation - the struck quark is essentially free from interacting with other partons inside the nucleon during the short time when the quark interacts with the virtual photon. Rinehimer and Miller~\cite{PhysRevC.80.015201} studied the connection between the Breit frame and IMF and showed that when the nucleon matrix element of the time component of the electromagnetic current, which gives $G_E/\sqrt{1 + Q^{2}/(4 M^2)}$ in the Breit frame as discussed above, is boosted to the IMF, one obtains the $F_1$ form factor, as was also confirmed by the analysis in  \cite{Lorce:2020onh}.

Miller pointed out~\cite{Miller19}  that the above picture connecting the proton charge density distribution to the Fourier transform of the $G_E$ form factor is not correct, and showed that a three-dimensional charge density, in the strict sense of a probability interpretation, cannot be defined for a nucleon - as a relativistic system of quarks and gluons - because the initial and final state proton wave functions are not the same. Instead, a two-dimensional charge density can be defined, and determined by the Dirac form factor $F_1$, as a matrix element of a density operator between identical initial and final states which are localized in the plane transverse to the direction of the fast moving 
nucleons.  

Jaffe~\cite{Jaffe:2020ebz} looked at this issue from a fundamental aspect -- the interplay between relativity and the uncertainty principle -- and pointed out that any attempt to extract spatial distributions of local properties of a hadronic system that is not much larger than its Compton wavelength would fail. In the case of the proton, its Compton wavelength is about 0.2 fm which is not significantly smaller than its size of $\sim$0.85 fm. Defining the expectation value of the spatial charge density distribution of the proton requires one to localize the proton, which introduces a localization dependence into the relationship between the form factor and the local charge density distribution. Only for systems such as atoms and heavy atomic nuclei -- for which the intrinsic sizes of the systems are much larger than the corresponding Compton wavelength -- the connection between the three-dimensional Fourier transform of the charge form factor and the local charge density distribution is meaningful. Belitsky {\it et al.} also discussed the proton form factors and charge distributions in their seminal paper~\cite{PhysRevD.69.074014} on the development of the concept of quantum phase-space (Wigner) distributions for quarks and gluons in the proton.
 
In the last two decades or more, there have been major developments in three-dimensional imaging of the partonic structure of the nucleon -- motivated to a large extent 
-- by the desire to solve the ``proton spin crisis or puzzle". These developments also shed new light on the electromagnetic structure of the nucleon. It is important to discuss the proton charge distribution and electric and magnetic form factors in the context of these new developments.
In the next section, we briefly introduce the three-dimensional parton distributions first before we discuss the two-dimensional charge density.

\subsection{Three-dimensional parton distributions}

The general framework to describe the partonic structure of the proton is through the generalized transverse momentum dependent parton distributions (GTMDs)~\cite{Meissner09,Lorce11a}, which are obtained by integrating the fully unintegrated generalized quark-quark correlation functions for a nucleon in momentum space over the light-cone energy component of the quark momentum~\cite{Meissner08,Meissner09}.
The thus obtained GTMDs depend on $x$, ${\bf k}_\perp$, and $\Delta$,  where $x$ is the longitudinal momentum fraction of the parton, 
${\bf k}_\perp$, the transverse momentum of the parton, and $\Delta$ is the four-momentum transfer to the nucleon.
The GTMDs are related to the Wigner distributions~\cite{Ji03,PhysRevD.69.074014,Lorce11} via a Fourier transformation between the transverse momentum transfer ${\bf \Delta}_\perp$ and the quark's transverse position ${\bf b}$. 
The five-dimensional Wigner distributions $\rho({\bf b},{\bf k}_\perp,x, \vec{S})$~\cite{Lorce:2011ni}, for a nucleon with polarization $\vec S$, are the quantum mechanical analogues of the classical phase-space distributions, with the five dimensions being $x$, ${\bf  k}_\perp$, and the transverse coordinates ${\bf b}$. 

\begin{figure}[h]
\includegraphics[]{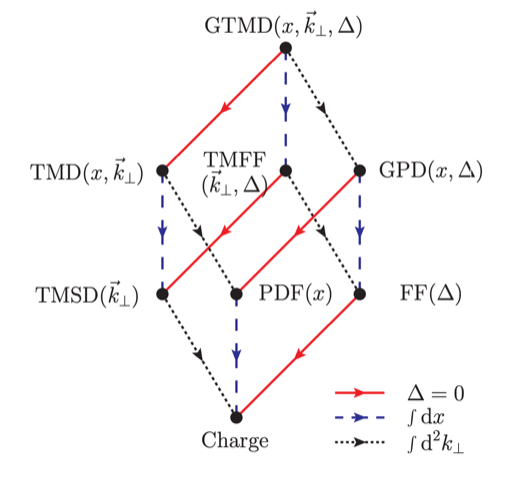}
	\caption{(Color online) Parton distribution family. The figure is from~\cite{Lorce11a}.}
	\label{fig-GTMD}
\end{figure}

As illustrated in Fig.~\ref{fig-GTMD}, one can obtain the generalized parton distributions (GPDs) by integrating the GTMDs over the transverse momentum ${\bf k}_\perp$. The GPDs can be viewed as the generalization of the parton distribution functions (PDFs) and the form factors. On the other hand, one can obtain the transverse momentum dependent parton distributions (TMDs) by setting the momentum transfer $\Delta$ to zero or equivalently by integrating the Wigner distributions over the transverse coordinate ${\bf b}$. The TMDs will reduce to PDFs when the transverse momentum is integrated. In Fig.~\ref{fig-GTMD}, TMFF and TMSD refer to transverse-momentum dependent form factors, and transverse-momentum dependent spin densities, respectively. 
While the most general one-parton information is contained in the GTMDs, which are connected to the Wigner distributions through Fourier transformations, unfortunately, neither the GTMDs nor the Wigner distributions are measurable in experiments. However, there are ways to access GPDs and TMDs experimentally, which we will briefly discuss next.

Deeply virtual Compton scattering  (DVCS)~\cite{Ji:1996nm} was proposed as the experimental tool to probe GPDs. We refer the reader to \cite{Ji:1996ek,Mueller:1998fv,Radyushkin:1996nd,Ji:1996nm}
for the original articles on GPDs and to
\cite{Goeke:2001tz,Diehl:2003ny,Belitsky:2005qn,Boffi:2007yc,Guidal:2013rya,Kumericki:2016ehc} for reviews of the field.   
In the Bj\"{o}rken limit, the DVCS amplitude is described through four off-forward parton distributions~\cite{Ji:1996nm}: $H^q$ and $\tilde{H}^q$ for a quark of flavor $q$, which conserve the nucleon helicity, and $E^q$ and $\tilde{E}^q$ that flip the nucleon helicity. These GPDs are functions of  $x$, $\xi$, and $\Delta^{2}$ -- for example, $H^q(x, \xi,\Delta^2)$, $E^q(x, \xi,\Delta^2)$ -- where $x$ is the average fraction of quark longitudinal momentum, $\xi$ is the average fraction of the longitudinal momentum transfer $\Delta$, and $\Delta^2$ is the squared momentum transfer. 

In the forward limit, $\Delta^{\mu} \rightarrow 0$, $H$ and $\tilde{H}$ are just the quark momentum, and helicity PDFs:
\begin{equation}
H^q(x,0,0) =q(x), \  \tilde{H}^q(x,0,0)=\Delta q(x).
\end{equation}
Furthermore, one can write down the following sum rules relating these new distributions to the quark flavor components of the Dirac and Pauli form factors in a nucleon as:
\begin{eqnarray}
F^q_1(\Delta^2) &=& \int_{-1}^{+1} dx H^q(x, \xi, \Delta^2), 
\label{eq:srf1}\\
F^q_2(\Delta^2) &=& \int_{-1}^{+1} dx E^q(x, \xi, \Delta^2), 
\label{eq:srf2}
\end{eqnarray}
where the $\xi$-independence of these sum rules is a consequence of Lorentz invariance.

Semi-inclusive deep inelastic scattering (SIDIS) processes -- where the scattered lepton together with the leading hadron are detected -- have been proven as effective ways to access TMDs. Within the TMD factorization, SIDIS structure functions are expressed as convolutions of transverse momentum dependent parton distribution functions (TMD-PDFs) and transverse momentum dependent fragmentation functions (TMD-FFs). There are eight leading-twist (twist-two) quark TMD PDFs for a nucleon. If the transverse momentum is integrated, three of them, $q$, $\Delta q$ (also written as $g_1$ or $g_{1L}$ in the literature), and $h_{1}$, will reduce to their collinear limits: the unpolarized, helicity distribution, and transversity distribution PDFs, while the remaining five will vanish. Hence TMDs, especially the spin-dependent ones, contain much richer information than collinear PDFs, and allow for accessing the correlation between quark transverse momentum and quark/nucleon spin. At leading twist, there are also eight gluon TMD-PDFs~\cite{PhysRevD.63.094021,PhysRevD.73.014020}. These TMDs can be probed~\cite{SoLID-EPJPlus,12GeVWP} at the 12-GeV CEBAF at Jefferson Lab in the valence quark region and at the future EIC~\cite{Accardi16} in the sea quark/gluon region. In SIDIS processes where the leading hadron is a spinless hadron such as a pion or kaon, two TMD-FFs describe a quark fragmenting into a hadron at the leading twist -- the unpolarized fragmentation function, and the Collins function~\cite{COLLINS1993161} characterizing the correlation between the transverse spin of the fragmenting quark with the transverse momentum of the leading hadron.
For more details about this subject, we refer to a recent review by Anselmino, Mukherjee, and Vossen~\cite{ANSELMINO2020103806}.

Both the GPDs and the TMDs are three-dimensional parton distributions. They have a much richer spin dependence than PDFs, and have been shown to be related to quark orbital angular momentum in some model dependent or model independent relations. 
One well-known relation is the total quark angular momentum $J_q$ to the gravitational (or generalized) form factors, which are the second moments of the GPDs~\cite{Ji:1996ek}: 
\begin{equation}
J_q \,=\, {1 \over 2} \, \int_{-1}^{+1} d x \, x \, 
\left\{ H^{q}(x,\xi,\Delta^2 = 0) + E^{q}(x,\xi,\Delta^2 = 0) \right\},
\label{eq:ji_spin}
\end{equation}
in which the $\xi$-dependence drops out. The total quark spin contribution $J_q$ decomposes (in a gauge invariant way) as $J_q = {1 \over 2}\Delta\Sigma+L_q$
where 1/2 $\Delta \Sigma$ and $L_q$ are respectively 
the quark helicity and quark orbital contributions to the nucleon spin.
Together with the measurement of quark helicity distributions, one may thus obtain the kinetic quark orbital angular momentum via the GPDs:
\begin{equation}
L_q=\int_{-1}^{+1} dx \left\{xH_{q}(x,0,0)+xE_{q}(x,0,0)-\tilde{H}_{q}(x,0,0) 
\right\}.
\end{equation}

Among the eight TMDs, the pretzelosity TMD $h_{1T}^\perp(x,\bm{k}_\perp)$ is proposed as a quantity to measure quark orbital angular momentum through
\begin{equation}
L_q=-\int dxd^2\bm{k}_\perp\frac{\bm{k}_\perp^2}{2M^2}h_{1T}^{\perp q}(x,\bm{k}_\perp),
\end{equation}
 initially from spectator model calculations~\cite{She:2009jq}. This finding is later proved valid for all spherically symmetric situations~\cite{Lorce:2011kn}. Another TMD -- sensitive to the quark orbital angular momentum -- is the Sivers function $f_{1T}^\perp(x,\bm{k}_\perp)$~\cite{Bacchetta:2011gx}, which is related to the GPD $E$ with a lensing function in a model dependent way~\cite{Burkardt:2003uw}. 
  
\vspace{0.2in} 

\subsection{The nucleon transverse charge densities}

\begin{figure}
\begin{minipage}{0.49\linewidth}
\hspace*{-1.7cm}\includegraphics[width =2\linewidth]{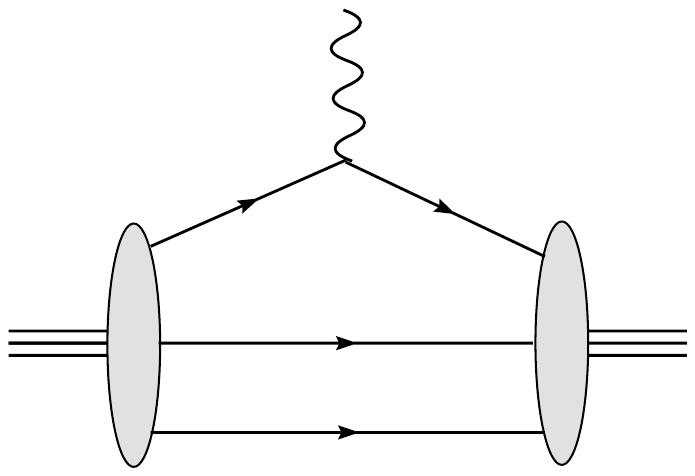}
\end{minipage}\hfill
\begin{minipage}{0.49\linewidth}
\hspace*{-1.7cm}\includegraphics[width =2\linewidth]{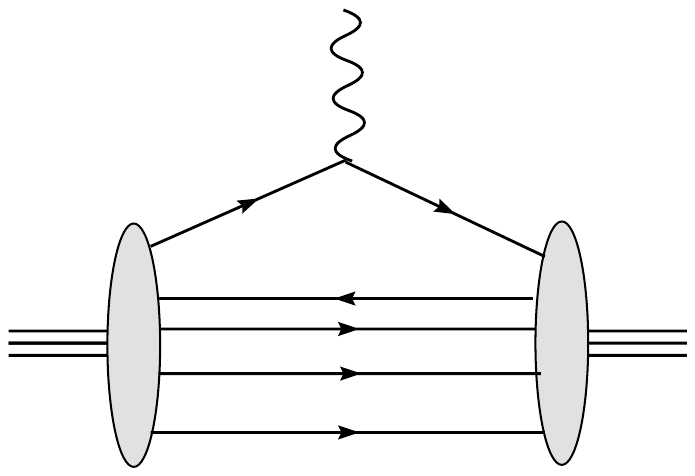}
\end{minipage}
\begin{minipage}{\linewidth}\vspace*{-9cm}
\hspace*{1cm}
\includegraphics[width=\linewidth]{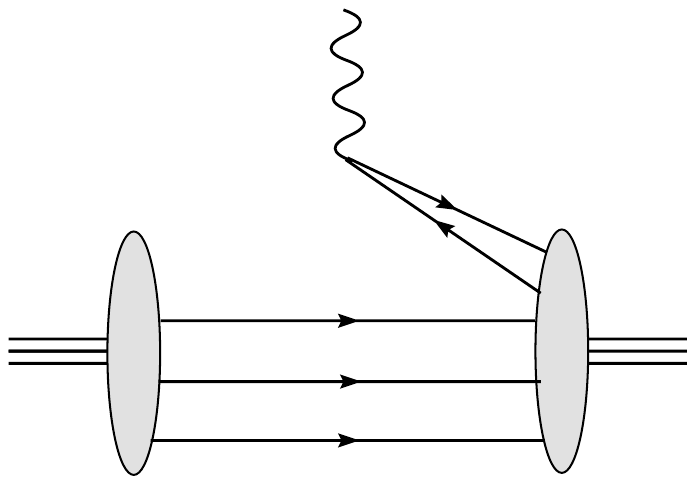}
\end{minipage}
\vspace*{-8cm}
\caption{Coupling of a space-like virtual photon to a relativistic
many-body system, as a proton. Upper panel : diagonal
transition where the photon couples to a quark, in the leading $3q$
Fock component (left), or in a higher $5q$ Fock  component
(right). Lower panel : process where the
photon creates a $q \bar q$ pair leading to a non-diagonal
transition between an initial $3q$ state and a final $5q$ state in
the proton. }
\label{fig:overlap}
\end{figure}

We next discuss in more detail how to define a charge density in a nucleon, and how such density is related to the elastic form factors and generalized parton distributions discussed above.   
For relativistic quantum systems, such as hadrons composed of nearly massless quarks, a proper definition of a charge density requires care as discussed above. 
For such systems, the number of constituents is not constant as a result of virtual pair production. 
 Consider, as an example, a hadron such as the proton, which is probed by a space-like virtual photon, as shown in
 Fig.~\ref{fig:overlap}.
 A sizable fraction of the proton's response when probed by a virtual photon with small (or even intermediate) virtuality is coming from wave function components beyond 
the three valence quark state state~\cite{Sufian:2016hwn}.
 In such a system, the  wave function contains,
 besides the three valence quark Fock component  $| qqq \rangle$,  also components where
 additional $q \bar q$ pairs, so-called sea-quarks, or (transverse) gluons $g$ are excited,
 leading to an infinite tower of $ |qqq q \bar q \rangle$, $ |qqq g \rangle$, ... components.
 When probing such a system using electron scattering,
 the exchanged virtual photon will couple to any quark or anti-quark 
 in the proton as shown in Fig.~\ref{fig:overlap} (upper panel).
In addition, the virtual photon,
 can also produce a $q \bar q$ pair, giving rise e.g. to a transition from a $3q$ state in the initial wave function to a $5q$ state in the final wave function, as shown in Fig.~\ref{fig:overlap} (lower panel). Such processes,
 leading to non-diagonal overlaps between initial and final wave functions, are not positive definite, and do not allow
 for a simple probability interpretation of the density $\rho$ anymore. Only the processes shown in the upper panel of
 Fig.~\ref{fig:overlap}  with the same initial and final wave function yield a positive definite particle density, allowing for a probability interpretation.

This relativistic dynamical effect of pair creation or annihilation
fundamentally hampers the interpretation of density and
any discussion of size and shape of a relativistic quantum system.
An interpretation in terms of the concept of a density requires
suppressing the contributions
shown in the lower panel of Fig.~\ref{fig:overlap}. This is possible when viewing
the hadron from a light-front frame, which allows to describe the hadron
state by an infinite tower of light-front wave functions. Consider the electromagnetic
(e.m.) transition from an initial hadron (with four-momentum $p$) to
a final hadron (with four-momentum $p^\prime$) when viewed from a
light-front moving towards the hadron. Equivalently, this
corresponds with an infinite-momentum frame (IMF) where the hadrons have a large
momentum component along the $z$-axis chosen along the direction of
the hadrons average momentum $P = (p + p^\prime)/2$.
One then defines the light-front plus  (+) component by $P^+ \equiv
(P^0 + P^3)/\sqrt{2}$, which is always a positive quantity for the quark or
anti-quark four-momenta in the hadron. When one views the hadron in a so-called
Drell-Yan frame~\cite{Drell:1969km}, where the virtual photon
four-momentum $\Delta = p^\prime - p$ is purely transverse, satisfying $\Delta^+ = 0$, energy-momentum conservation
will forbid processes where this virtual photon splits into a $q
\bar q$ pair.
Such a choice is possible for a space-like virtual photon, and its
virtuality is then given by  $t \equiv \Delta^2 = - {\bf \Delta}^2_\perp < 0$,
where $\bf \Delta_\perp$ is the transverse photon momentum, lying in the transverse spatial $(x, y)$-plane. 
Here $-t$ or ${\bf \Delta}^2_\perp$ is the same as the virtuality $Q^2$ in elastic e-p scattering. 
In such a frame, the virtual photon only couples to
forward moving partons, i.e. only processes such as those shown in the
upper panel in
Fig.~\ref{fig:overlap}  are allowed. We can then define
a proper density operator through the + component of the four-current by
$J^+ = (J^0 + J^3)/\sqrt{2}$. 
For one quark flavor $q$ it is given by~\cite{Soper77}:
\begin{eqnarray}
J_q^+(z^-, {\bf b}) &=& \bar q(0, z^-, {\bf b}) \gamma^+ q(0, z^-, {\bf b}) \nonumber \\
&=& \sqrt{2} q_+^\dagger(0, z^-, {\bf b}) q_+(0, z^-, {\bf b}), 
\label{eq:reloperator}
\end{eqnarray} 
where the $q_+$ fields are related to the quark fields $q$ through a field redefinition, 
involving the $\pm$ components of the Dirac $\gamma$-matrices as $ q_+ \equiv (1/2) \gamma^- \gamma^+ q$. 
In Eq.~(\ref{eq:reloperator}) light-cone coordinates are used with $a^{\pm} \equiv (a^{0}\pm a^3)/\sqrt{2}$, and both quark fields are taken at equal light-cone time $z^+ = 0$. 
The transverse spatial coordinates are written as two-dimensional vector ${\bf b}$. 
The relativistic density operator $J_q^+$, defined in  Eq.~(\ref{eq:reloperator}), is a positive definite quantity. 
The electromagnetic charge density operator $J_{em}^+$ is then obtained by a sum over quarks weighted by their 
electric charges $e_q$ (in units of $e$) as~:
\begin{eqnarray}
J_{em}^{+}(z^-, {\bf b}) &=& \sum_{q} {e_q {\bar q} (0, z^-, {\bf b}) {\gamma^+} q (0, z^-, {\bf b})}. 
\label{eqn:rhoq}
\end{eqnarray}

One can then examine the transverse structure of the nucleon due to the fact that transverse boosts are independent of interactions in the infinite momentum frame~\cite{Kogut70,Burkardt06}. Transversely localized nucleon states~\cite{Burkardt:2002hr,Diehl02,Diehl03,Soper77} with its transverse 
center-of-mass position ${\bf R}$ being set to 0, can be defined in terms of linear superposition of states of transverse momentum as~\cite{Miller19}
\begin{equation}
|p^+, {\bf R=0},\lambda \rangle \equiv N \int{{\frac{d^2 {\bf p_\perp}}{(2\pi)^2 \sqrt{2p^+}}} | p^+, {\bf p_\perp}, \lambda \rangle},
\label{eqn:wf}
\end{equation}
with $|p^+, {\bf R=0},\lambda \rangle$ being light-cone helicity ($\lambda$) eigenstates~\cite{Soper77}, and $N$ a normalization factor.

Using the density operator of Eq.~(\ref{eq:reloperator}), one can define transverse densities $\rho^q_\lambda$ for a quark of flavor $q$ in a
transversely localized hadron as~\cite{Burkardt:2000za,Burkardt:2002hr,Miller:2007uy}:
\begin{eqnarray}
\rho^q_\lambda(b) &\equiv& \frac{1}{2 P^+} \langle  P^+, {\bf R=0},\lambda | J_q^+(0, {\bf b}) | P^+, {\bf R=0},\lambda \rangle.  
\nonumber \\
\label{eq:dens1a}
\end{eqnarray}
Using the translation operator in transverse spatial direction, 
one can express $J_q^+(0, {\bf b}) = e^{- i {\bf {\hat P_\perp}} \cdot {\bf b}} J_q^+(0) e^{i {\bf {\hat P_\perp}} \cdot {\bf b}}$,  
in terms of the 
local current operator at the origin $J_q^+(0)$. Using Eq.~(\ref{eqn:wf}) then allows to express the quark transverse density of Eq.~(\ref{eq:dens1a}) as:  
\begin{eqnarray}
\rho^q_\lambda(b) &\equiv& \int \frac{d^2 {\bf \Delta_\perp}}{(2
\pi)^2} \, e^{- i \, {\bf \Delta_\perp} \cdot {\bf b}} \, \frac{1}{2 P^+}
\nonumber \\
&& \hspace{.25cm}\times \langle P^+, \frac{{\bf \Delta_\perp}}{2},
\lambda \,|\, J_q^+(0) \,|\, P^+, -\frac{{\bf \Delta_\perp}}{2}, \lambda
\rangle .
\label{eq:dens1}
\end{eqnarray}
In the two-dimensional Fourier transform of Eq.~(\ref{eq:dens1}), the vector
$\bf b$ denotes the quark position (in the transverse plane) from the
transverse center-of-momentum of the hadron. It is the position
variable conjugate to the hadron relative transverse momentum $\bf \Delta_\perp$.
The quantity $\rho^q_\lambda(b)$ has the interpretation of the two-dimensional (transverse) 
density to find a quark of flavor $q$ at distance $b = | {\bf b}|$ from the transverse c.m. of the hadron with
helicity $\lambda$. 

For a quark of flavor $q$ in the proton, 
the matrix element of the $J_q^+$ operator, entering the two-dimensional Fourier-transform in Eq.~(\ref{eq:dens1}), 
can be expressed in terms of the quark flavor contribution $F_1^q$ to the proton Dirac form factor as: 
\begin{equation}
\frac{1}{2 P^+} \langle P^+, \frac{{\bf \Delta_\perp}}{2},\lambda | J_q^+(0) | P^+, -\frac{{\bf \Delta_\perp}}{2}, \lambda \rangle 
= F_1^q (-{\bf \Delta}^2_\perp), 
\end{equation}
which allows to express the density for a quark of flavor $q$ in the proton, using Eq.~(\ref{eq:dens1}), as: 
\begin{eqnarray}
\rho^q(b) &=& \int \frac{d^2 {\bf \Delta_\perp}}{(2
\pi)^2} \, e^{- i \, {\bf \Delta_\perp} \cdot {\bf b}} \, F_1^q (-{\bf \Delta}^2_\perp), \nonumber \\
&=& \int_0^\infty \frac{d  Q}{2 \pi}  Q \, J_0(b \, Q) F_1^q(-Q^2), 
\label{eq:dens1b}
\end{eqnarray}
where in the last line the circular symmetry of the transverse density 
was used to convert the two-dimensional Fourier transform to a one-dimensional integral over 
$Q \equiv |{\bf \Delta}_\perp|$, with $J_n$ denoting the cylindrical Bessel function of order $n$. 
Furthermore, the helicity subscript $\lambda$ has been omitted, as for a spin-1/2 system $\rho_{+ \frac{1}{2}} = \rho_{- \frac{1}{2}}$. 

The two-dimensional electric charge density in a proton is then obtained as sum over the quarks weighted by their electric charges: 
\begin{equation}
\rho(b) = \sum_q e_q \rho^q(b). 
\label{eq:rhopq}
\end{equation}
From the experimentally measured Dirac form factor $F_{1}$ of the proton:  
\begin{eqnarray}
F_{1p} =  \sum_q e_q F_1^q,
\label{eq:f1pq} 
\end{eqnarray}
one obtains:
\begin{eqnarray}
\rho_p(b) &=& \int_0^\infty \frac{d  Q}{2 \pi}  Q \, J_0(b \, Q) F_{1p}(-Q^2).
\label{eq:dens1c}
\end{eqnarray}
A similar formula holds for the neutron with the interchange $\rho^u \leftrightarrow \rho^d$ in Eq.~(\ref{eq:rhopq}) and 
$F_1^u \leftrightarrow F_1^d$ in Eq.~(\ref{eq:f1pq}).
In this way, it was observed in \cite{Miller:2007uy} that
the neutron transverse charge density reveals the well known negative contribution 
at large distances, around 1.5~fm, 
due to the pion cloud, a positive contribution at 
intermediate $b$ values, and a negative core at $b$ values smaller than 
about 0.3~fm.  
One can understand the negative value of the neutron $\rho(b = 0)$ 
from Eq.~(\ref{eq:dens1c}) and the observation that over the whole 
measured $Q^2$ range the neutron Dirac form factor $F_{1n}$ is negative. 

The quark charge densities in Eq.~(\ref{eq:dens1c})
do not fully describe the e.m. structure of the hadron. For a proton  
the densities with $\lambda = \pm 1/2$ yield the same
information, while a spin-1/2 system is described by two independent electromagnetic 
form factors. In general, a
particle of spin $S$ is described by $(2 S + 1)$ e.m. moments. To
fully describe the structure of a hadron one also needs to consider
the charge densities in a transversely polarized hadron state,
denoting the transverse polarization direction by
${\bf S}_\perp$.
The transverse charge densities can be defined through matrix
elements of the density operator $J_q^+$ in eigenstates of transverse spin
~\cite{Carlson:2007xd,Carlson:2008zc,Lorce:2009bs} as:
\begin{eqnarray}
\rho^q_{T s_\perp} ({\bf b}) &\equiv& \int \frac{d^2 {\bf
\Delta}_\perp}{(2 \pi)^2} \, e^{- i \,   {\bf \Delta}_\perp \cdot {\bf b} } \,
\frac{1}{2 P^+}  \nonumber \\
&& \hspace{.25cm} \times \langle P^+, \frac{{\bf \Delta}_\perp}{2},
s_\perp | J_q^+(0) | P^+, \frac{- {\bf \Delta}_\perp}{2}, s_\perp  \rangle, \quad 
\label{eq:dens2a}
\end{eqnarray}
where $s_\perp$ is the hadron spin projection along the transverse spin direction 
${\bf S}_\perp \equiv \cos \phi_S {\bf e_x} + \sin \phi_S {\bf e_y}$, with ${\bf e_x}$ and ${\bf e_y}$ the two unit-vectors in the transverse plane.  

By expressing the transverse spin state in terms of the light-front helicity spinor states as:
\begin{equation} 
| s_\perp = + \frac{1}{2} \rangle = \frac{1}{\sqrt{2}}
\left\{ | \lambda = + \frac{1}{2} \rangle 
+ e^{i \phi_S } \, | \lambda = - \frac{1}{2} \rangle \right\},  
\end{equation}
the matrix element of the $J_q^+$ operator, entering the two-dimensional Fourier-transform in Eq.~(\ref{eq:dens2a}), 
can be expressed in terms of the quark flavor contribution to both the Dirac ($F_1^q$) and Pauli ($F_2^q$) form factors as: 
\begin{eqnarray}
&&\frac{1}{2 P^+} \langle P^+, \frac{{\bf \Delta_\perp}}{2}, s_\perp  | J_q^+(0) | P^+, -\frac{{\bf \Delta_\perp}}{2}, s_\perp  \rangle 
\nonumber \\
&& = F_1^q (-{\bf \Delta}^2_\perp) + \frac{i}{2 M} \left( {\bf S}_\perp \times {\bf \Delta}_\perp \right)_z F_2^q (-{\bf \Delta}^2_\perp).
\end{eqnarray}

\begin{figure}[h]
\includegraphics[width=4cm]{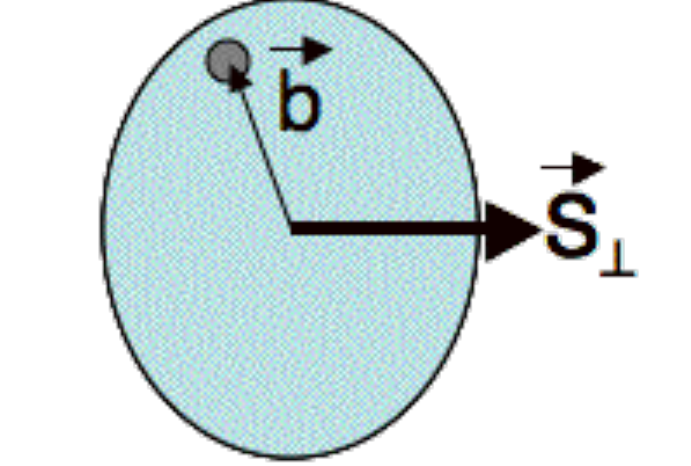}
\caption{Schematic view of the projection of the charge density along the line-of-sight (perpendicular to the figure), for a hadron polarized along the direction of ${\bf S}_\perp$. The position of the (quark) charge inside the hadron is denoted by ${\bf  b}$. }
\label{fig:relativity}
\end{figure}

Taking the weighted sum over the quark charges, the Fourier transform defined as in Eq.~(\ref{eq:dens2a}) can then be worked out as~\cite{Carlson:2007xd}:
\begin{eqnarray}
&&\rho_{T s_\perp}({\bf  b}) = \rho(b) \nonumber \\
&&\hspace{1cm}+ \sin (\phi_b - \phi_S)
\int_0^\infty \frac{d Q}{2 \pi} \frac{Q^2}{2 M} \, J_1(b \, Q) 
F_{2}(-Q^2), \quad 
\label{eq:dens2b}
\end{eqnarray}
where the second term, which describes the deviation from the circular
symmetric unpolarized charge density, depends on the quark position  
${\bf b} = b ( \cos \phi_b {\bf e_x} + \sin \phi_b {\bf e_y} )$. 
Whereas the density $\rho_\lambda$ for a hadron
in a state of definite helicity is circularly symmetric for all spins,
the density $\rho_{T  s_\perp}$ depends also on the
orientation of the position vector ${\bf b}$, relative to the
transverse spin vector ${\bf S}_\perp$,  
 as illustrated in Fig.~\ref{fig:relativity}. 
Therefore, it contains 
information on the hadron shape, projected on a
plane perpendicular to the line-of-sight. 

As the density $\rho_T$ is not circularly symmetric, one can calculate the dipole moment of its distribution as
\begin{eqnarray}
{\bf d} &\equiv& e \int d^2 {\bf b} \, {\bf b} \, \rho_{T s_\perp} ({\bf b}) 
= - \frac{e}{2 M} F_2(0) \left( {\bf S}_\perp \times {\bf e_z} \right).
\label{eq:dens2c}
\end{eqnarray} 
Eq.~(\ref{eq:dens2c}) implies that polarizing the proton along the $x$-axis 
leads to an induced electric dipole moment along the 
$y$-axis which is equal to the value of the 
anomalous magnetic moment, {\it i.e.} $F_2(0)$ (in units $e/2 M$) as first 
noticed in~\cite{Burkardt:2000za}. One can understand this induced electric dipole field pattern from special relativity,  as the nucleon spin along the $x$-axis is the source 
of a magnetic dipole field, denoted by $\vec B$. 
An observer moving towards the nucleon with velocity $\vec v$ will see an
electric dipole field pattern with $\vec E^\prime = - \gamma (\vec v \times
\vec B)$ giving rise to the observed effect.

\begin{center}
\begin{figure}[h]
\includegraphics[width=8.5cm]{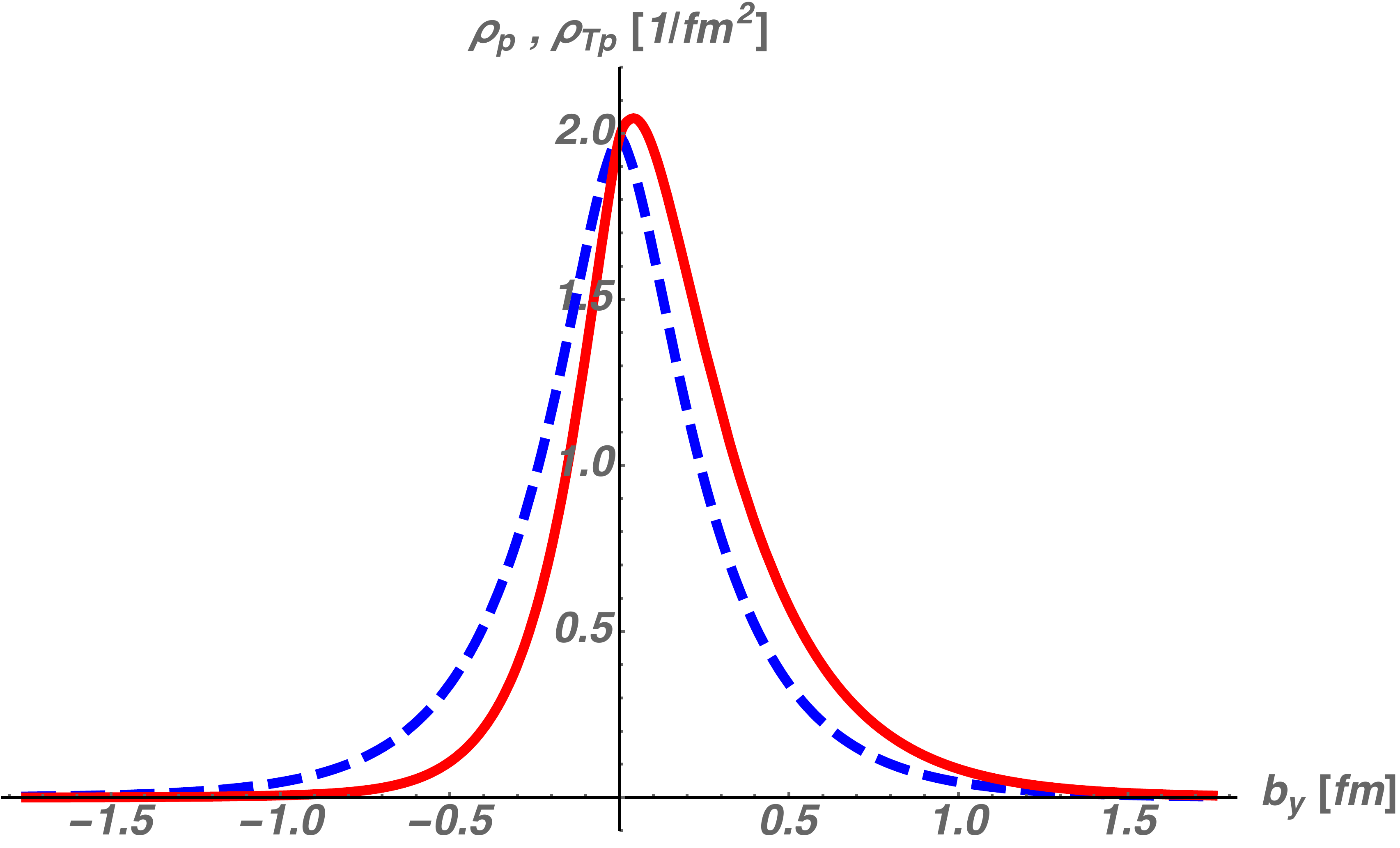}
\includegraphics[width=8.5cm]{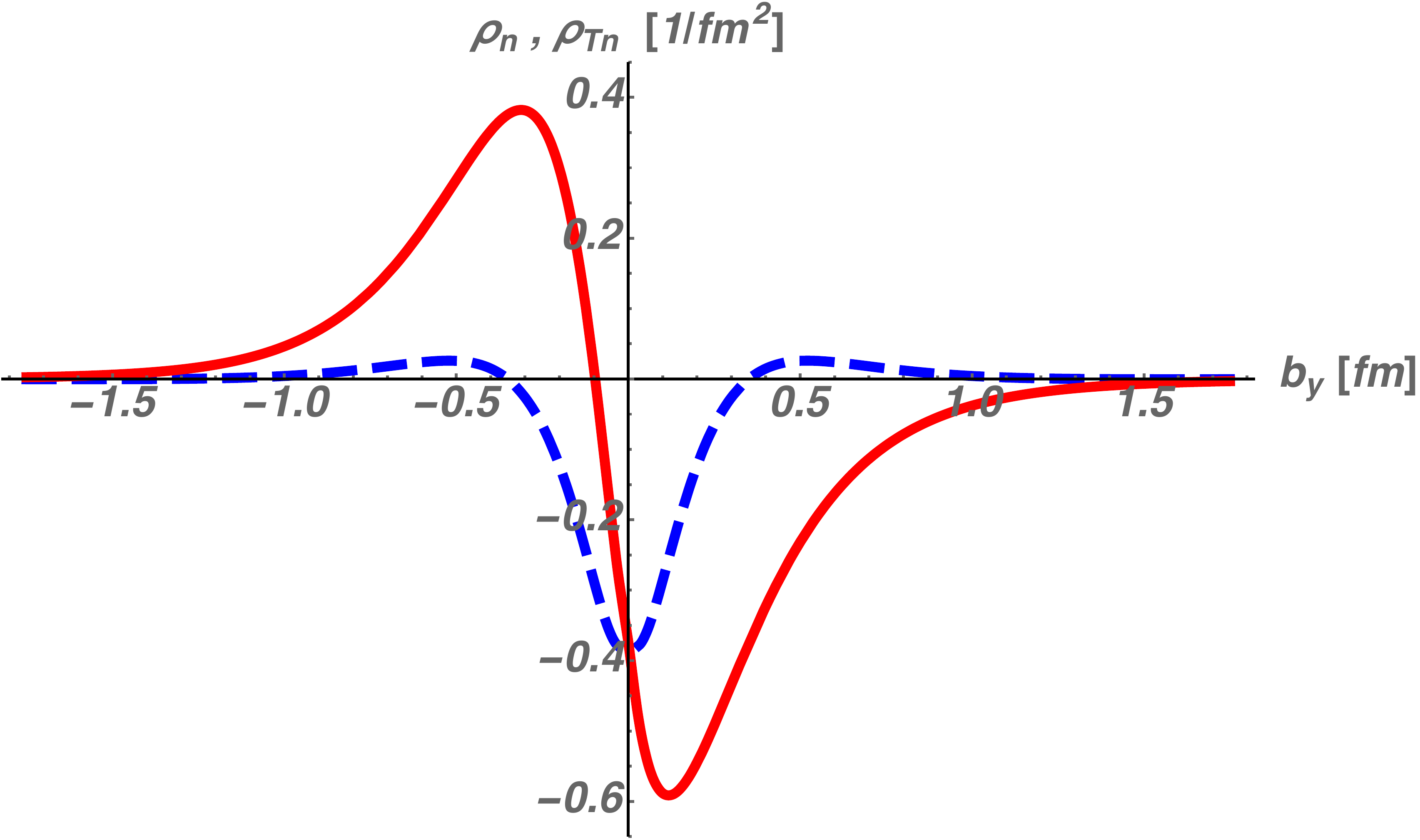}
\caption[fig]{\label{fig:nucdensities} (Color online) Transverse charge densities for proton (left panel) and neutron (right panel). 
The curves show the density along the y-axis for an unpolarized nucleon (dashed curves), and for a nucleon polarized along the x-axis (solid curves). For the nucleon form factors, the empirical parameterization of \cite{Ye18} was used. }
\label{fig:nucdens}
\end{figure}
\end{center}

We show the transverse charge densities in proton and neutron in Fig.~\ref{fig:nucdens} based on the recent parameterization of \cite{Ye18} for the proton and neutron form factors. One notices that for the proton the unpolarized charge density is everywhere positive. For a transversely polarized proton along the x-axis one notices a small displacement of the charge density along the y-axis proportional to the proton's anomalous magnetic moment. For the neutron, the unpolarized density shows the well-known negative contribution at large distances, 
around 1.5~fm, due to the pion cloud, a positive contribution at intermediate b values, and a negative core at b-values smaller than about 0.3 fm as first noticed by \cite{Miller:2007uy}. One can understand the negative value of the neutron $\rho_n(0)$ from the formula similar to Eq.~(\ref{eq:dens1c}) and the observation that over the whole measured $Q^2$ range the neutron Dirac form factor $F_{1n}$ is negative. 
 The corresponding transverse charge density for a neutron polarized along the x-axis gets significantly displaced due to the large (negative) value of the neutron anomalous magnetic moment.
 
\onecolumngrid
\twocolumngrid

The above discussed light-front densities require us to develop some new intuition, as they 
are defined at equal light-front time ($z^+ = 0$) of their constituents. 
When constituents move non-relativistically, it does not make a difference whether they are observed
at equal time ($t = 0$) or equal light-front time ($z^+ = 0$),
since the constituents can only move a negligible small distance during the small time interval that a light-ray needs to connect them.
 This is not the case, however,  for bound systems of relativistic constituents
such as hadrons~\cite{Jarvinen:2004pi, Hoyer:2009ep}. 
For the latter, the transverse density at equal light-front time can be interpreted as a 2-dimensional 
flash photograph of a 3-dimensional object~\cite{Brodsky:2014yha},  reflecting the position of charged constituents at different times, which are (causally) connected by a light-ray.

\subsection{GPDs and quark densities in longitudinal momentum and transverse position in a proton}

\begin{figure}[h]
\includegraphics[width =8.75cm]{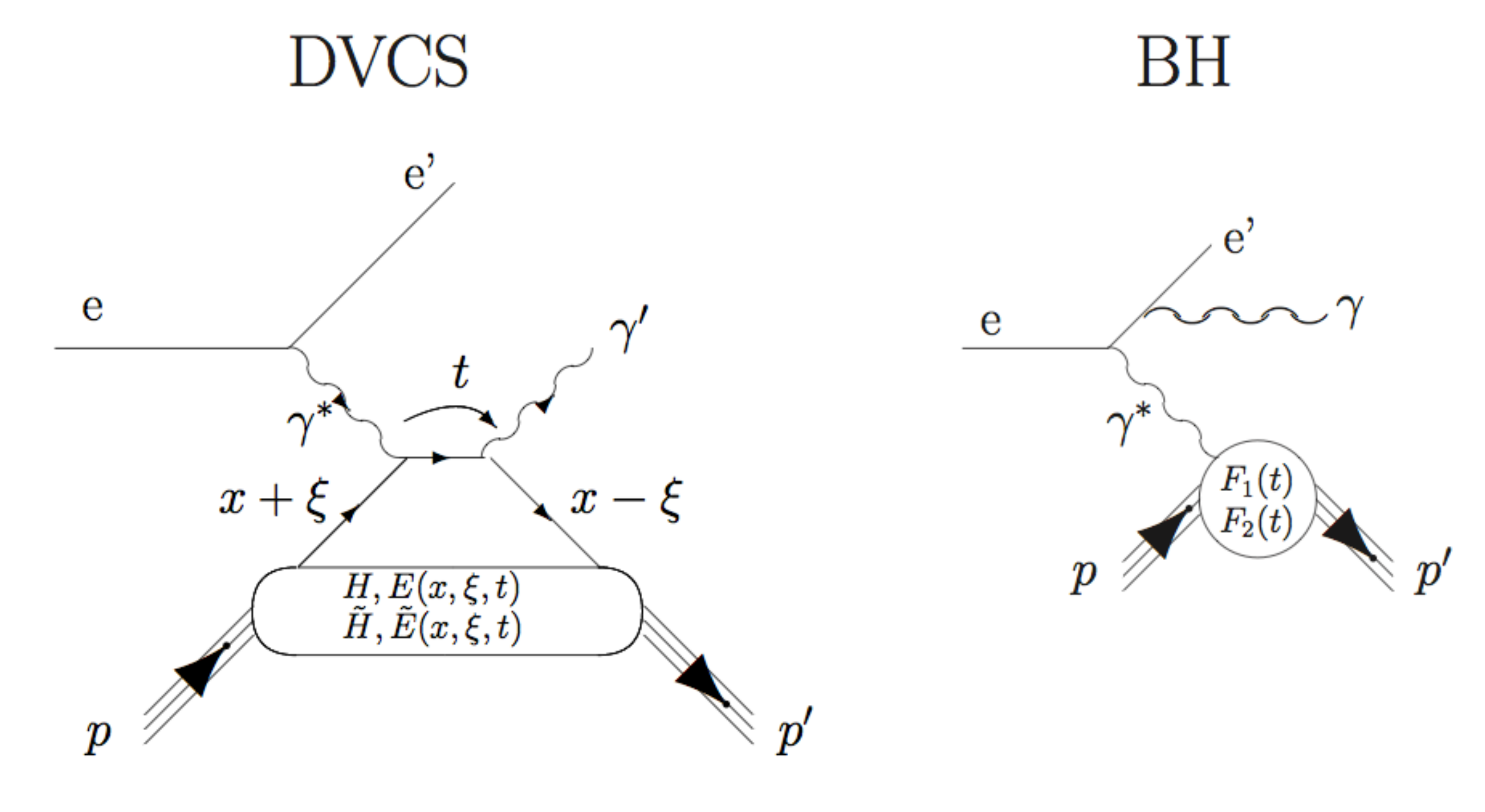}
\caption{Left: handbag diagrams for the deeply virtual Compton scattering (DVCS) as accessed through the $e p \to e p \gamma$ process. Right: competing Bethe-Heitler contribution, in which the nucleon structure enters through its elastic form factors.}
\label{fig:handbag}
\end{figure}

Generalized parton distributions (GPDs) also offer a thorough perspective on the nucleon electromagnetic structure. They correspond to the Fourier transform of bi-local operators between nucleon states of different momenta and can be accessed through the deeply virtual Compton scattering (DVCS) process, with a large virtuality of the intial photon. A factorization theorem, shown through the handbag diagram of  Fig.~\ref{fig:handbag}, allows to parameterize the nucleon structure entering the DVCS process at leading twist-2 through four GPDs which conserve the quark helicity. 
The GPDs entering the DVCS process can be parameterized as the one-dimensional Fourier transform of the bi-local quark operator 
along the light-cone $z^-$ direction. For the vector operator, in light-cone gauge $A^+ = 0$, it takes the form: 
\begin{eqnarray}
 &  & {{P^{+}}\over {2\pi }} \int dz^{-}e^{ixP^{+}z^{-}}\langle p^\prime, \lambda^\prime|\bar{q}(0,-\frac{z^-}{2}, {\bf 0}) \gamma^+ q(0,\frac{z^-}{2},{\bf 0})|p,  \lambda \rangle \nonumber \\
 && = H^{q}(x,\xi ,\Delta^2)\; \bar{N}(p^\prime, \lambda^\prime)\gamma ^{+}N(p, \lambda)
 \nonumber \\
 &&+ E^{q}(x,\xi ,\Delta^2)\; \bar{N}(p^\prime, \lambda^\prime)
 i\sigma ^{+\kappa } \frac{\Delta_\kappa}{2M} N(p, \lambda),
 \label{eq:gpd} 
\end{eqnarray}
where the GPDs $H^q$ and $E^q$ for a quark of flavor $q$ are functions of three variables: $x$, $\xi$, and $t = \Delta^2$. In a fast moving proton consisting of near-collinear partons, $x+\xi$ ($x-\xi$) 
represent the momentum fractions (+ components) of the 
initial (final) quark momenta w.r.t. the average nucleon momentum $P^+$, as shown in Fig.~\ref{fig:handbag}, 
with $\Delta^+ = -2 \xi P^+$, 
while $\Delta^2$ is the momentum transfer to the nucleon.  
In the forward limit, the GPD $H^q$ reduces to the unpolarized PDF as discussed above and illustrated in Fig.~\ref{fig-GTMD}.    
By integrating Eq.~(\ref{eq:gpd}) over the quark momentum fraction $x$, the bi-local operator on the {\it lhs} reduces to a local operator which directly allows to make the connection with the flavor dependent  elastic form factors of the nucleon. This leads to the sum rules given by Eqs.~(\ref{eq:srf1},\ref{eq:srf2}). 
As Eqs.~(\ref{eq:srf1}) and (\ref{eq:srf2}) can be expressed equivalently as integrals from 0 to 1 involving the valence part of the GPDs:
\begin{equation}
H^q_v(x,\xi,\Delta^2) \equiv  
H^q(x,\xi,\Delta^2) + H^q(-x,\xi,\Delta^2),  
\label{eq:Hval}
\end{equation}
the elastic form factors thus 
only constrain the valence parts of the GPDs through the first moment sum rules.  
The second moment of the GPDs $H^q$ and $E^q$ gives the form factors of the energy-momentum tensor. For the combination $H^q + E^q$, its second moment gives the total quark angular momentum contribution to the nucleon spin through Ji's sum rule \cite{Ji:1996ek}, Eq.~(\ref{eq:ji_spin}). 

Analogous to the discussion of Eq.~(\ref{eq:dens1a}), which allowed to define a transverse density by considering the matrix element of the local operator $J^+$ between states localized in transverse space, 
\cite{Burkardt:2000za,Burkardt:2002hr} has generalized this argument to non-local operators entering Eq.~(\ref{eq:gpd}). Defining a non-local operator by making a translation in the transverse plane: 
\begin{eqnarray}
\hat{O}^q(x,{\bf b}) \equiv
\int \frac{dz^{-}}{4 \pi}e^{ixP^{+}z^{-}} \bar{q}(0,-\frac{z^-}{2}, {\bf b}) \gamma^+ q(0,\frac{z^-}{2},{\bf b}), \nonumber \\
\label{eq:bilocal}
\end{eqnarray}
one can define a quark density which is dependent on $x$ and ${\bf b}$ by  considering matrix elements of this operator between nucleon states which have the same large $P^+$ component (corresponding with $\xi = 0$), same helicity, and transverse center-of-mass position ${\bf R}$ set to zero as: 
\begin{equation}
\rho^q(x,{\bf b}) \equiv <P^+, {\bf R=0}, \lambda|\hat{O}^q(x,{\bf b})|P^+,{\bf R=0},\lambda>.
\label{eq:dens1xb}
\end{equation}
Using the defining expression for GPDs of Eq.~(\ref{eq:gpd}), this then leads to:
\begin{equation}
\rho^q(x, {\bf b})=\int
\frac{d^2 {\bf \Delta_\perp}}{(2\pi)^2}e^{- i {\bf \Delta}_\perp \cdot {\bf b}   }
     H^q(x,0,-{\bf \Delta}^2_\perp),
\label{eq:dens2xb}
\end{equation}
which depends on the GPD $H^q$ at $\xi = 0$ and $t = -{\bf \Delta}^2_\perp$.  
The function $\rho^q(x, {\bf b})$ can then be interpreted 
as the number density of quarks of flavor $q$ with {\it longitudinal} momentum 
fraction $x$ at a given {\it transverse} distance ${\bf b}$, relative 
to the transverse c.m. in the proton~\cite{Burkardt:2000za}. 
By integrating Eq.~(\ref{eq:dens2xb}) over quark longitudinal momentum $x$,  
and using Eqs.~(\ref{eq:srf1}) and (\ref{eq:dens1b}), one recovers the 2-dimensional transverse density as:
\begin{equation}
\int_{-1}^{+1} dx \, \rho^q(x,{\bf b}) = \rho^q(b).
\end{equation}
Analogous to the transverse density of Eq.~(\ref{eq:dens2a}) in a nucleon state of transverse spin, 
we can also define a density $\rho^q_{T s_\perp}(x, \bf b)$, depending on $x$ and ${\bf b}$, in a nucleon state of transverse spin $s_\perp$ as:
\begin{equation}
\rho^q_{T s_\perp}(x,{\bf b}) \equiv <P^+, {\bf R=0}, s_\perp|\hat{O}^q(x,{\bf b})|P^+,{\bf R=0},s_\perp>,
\label{eq:dens3xb}
\end{equation}
which can be expressed as 2-dimensional Fourier transform over GPDs as:
\begin{eqnarray}
&&\rho^q_{T s_\perp}(x, {\bf  b}) = \rho^q(x, {\bf b}) \nonumber \\
&&+ \frac{i}{2 M}  \int
\frac{d^2 {\bf \Delta_\perp}}{(2\pi)^2}e^{- i {\bf \Delta}_\perp \cdot {\bf b}   }
\left( {\bf S}_\perp \times {\bf \Delta}_\perp \right)_z E^q(x,0,-{\bf \Delta}^2_\perp). \nonumber \\
\label{eq:dens4xb}
\end{eqnarray}
Integrating Eq.~(\ref{eq:dens4xb}) over quark longitudinal momentum $x$ 
yieds again the corresponding transverse density:
\begin{equation}
\int_{-1}^{+1} dx \, \rho^q_{T s_\perp}(x,{\bf b}) = \rho^q_{T s_\perp}({\bf b}).
\end{equation}

One can define the $x$-dependent squared radius of the quark density in the transverse plane as:
\begin{equation}
\langle b^2 \rangle^q (x) = \frac{ \int d^2 {\bf b} \, 
{\bf b}^2 \, \rho^q(x, {\bf b})}{\int d^2 {\bf b} \,  \rho^q(x, {\bf b})}. 
\label{eq:cr5}
\end{equation}
Inserting Eq.~(\ref{eq:dens2xb}) in Eq.~(\ref{eq:cr5}) 
allows one to express the $x$-dependent mean-squared radius as:
\begin{equation}
\langle b^2 \rangle^q (x)= - 4 \frac{\partial}{\partial {\bf \Delta}^2_\perp} 
\ln H^q(x,0,-{\bf \Delta_\perp}^2) \biggr| _{{\bf \Delta_\perp} = 0}.
\label{eq:crgpd}
\end{equation}

\begin{figure}[h]
\begin{center} 
\includegraphics[width=0.48\textwidth]{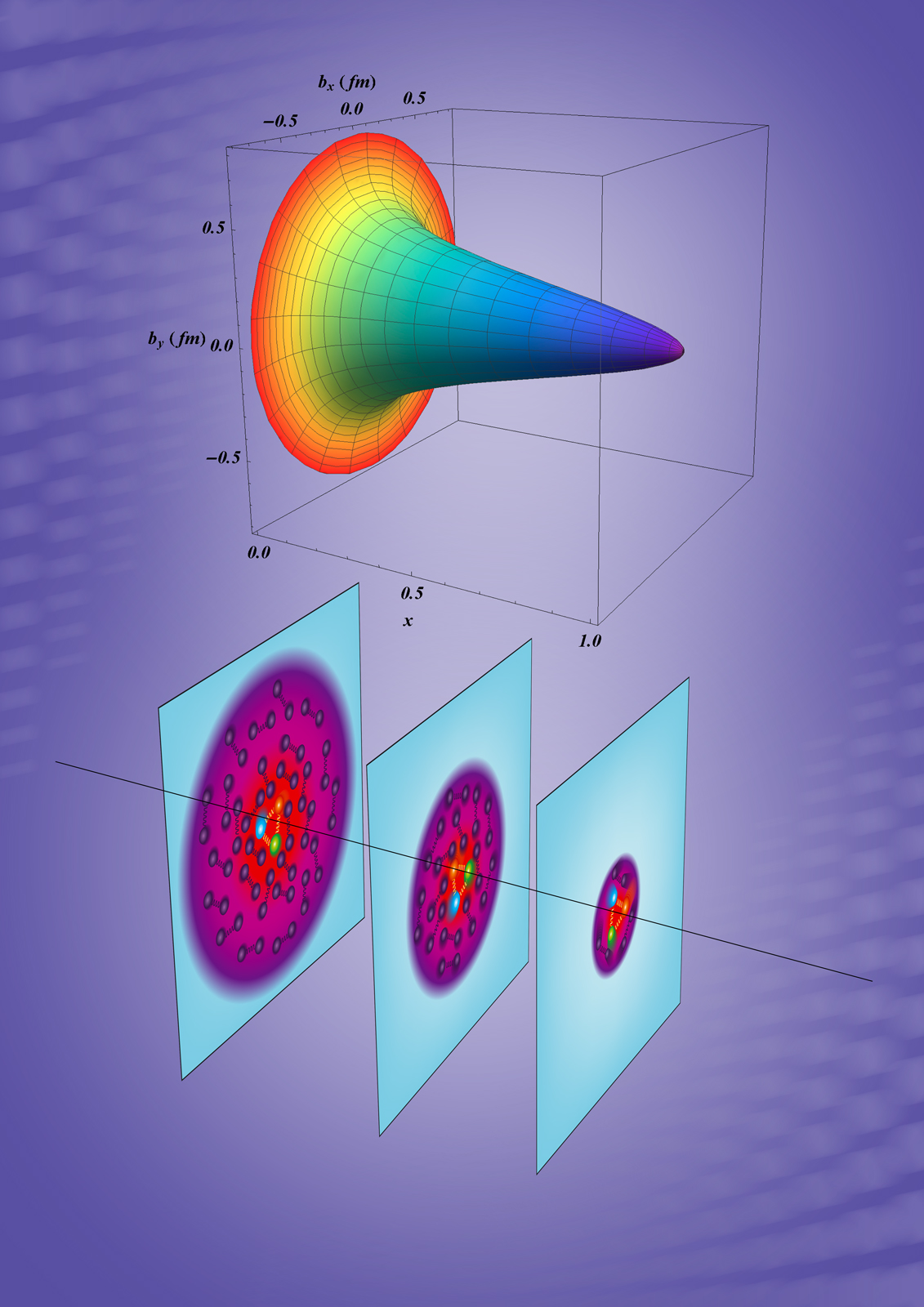}
\caption{(Color online) Top panel: three-dimensional representation of the
function of Eq.~(\ref{eq:bperp1}) fitted to DVCS data,
showing the $x$-dependence of the proton's transverse charge radius. 
Bottom panel: artistic illustration of the corresponding
rising quark density and transverse extent as a function of $x$. Figure from \cite{Dupre:2016mai,Dupre:2017hfs}. 
}
\label{fig:illus}
\end{center}
\end{figure}

The knowledge of the $x$-dependence of the $t$-slope of the GPD $H^q$ thus allows to experimentally access the $x$-dependent mean-squared radius of the quark distributions in a proton.  
Dupr\'e {\it et al.} \cite{Dupre:2016mai,Dupre:2017hfs} have performed a GPD QCD leading-twist and leading-order analysis of $e p\to e p \gamma$ unpolarized cross sections, difference of beam-polarized cross sections, longitudinally polarized target single spin,
and beam-longitudinally polarized target double spin asymmetries measured at HERMES and 
at JLab with the aim of constraining the dominant GPD $H$.  
Assuming the $t$-dependence of the valence GPD $H^q_v(x,0,t)$ as defined in Eq.~(\ref{eq:Hval}) to be exponential of the form:
\begin{eqnarray}
H^q_v(x,0, t) = q_v(x) e^{B(x) t},   
\label{eq:hzeroxi}
\end{eqnarray}
with $q_v(x)$ the corresponding valence PDF, 
the data were found to be consistent with a logarithmic rise of the exponential $t$-slope $B(x)$ with decreasing values of $x$\cite{Dupre:2016mai,Dupre:2017hfs}: 
\begin{equation}
B(x) = a \ln(1/x),
\end{equation}
with $a \simeq 1$~GeV$^{-2}$, consistent with a Regge slope for hadrons. 
Eq.~(\ref{eq:crgpd}) then yields for each quark flavor:
\begin{eqnarray}
\langle b^2 \rangle (x)= 4 B(x). 
\label{eq:bperp1}
\end{eqnarray}
This $x$-dependent profile of the mean-squared radius in the proton is shown in the upper plot of Fig.~\ref{fig:illus}. Its lower panel 
is an artistic view of the tomographic quark content of the proton, with
the charge radius and the density of the quarks increasing as smaller and smaller quark momentum
fractions are probed.

 One can also look at the charge density in the context of the five-dimensional Wigner distributions (two transverse positions and three momentum variables) in the infinite momentum frame. Lorc\'{e} and Pasquini~\cite{Lorce11}  used light-cone variables $k^+$, ${\bf k}$ for the three-momentum of a quark and arrived at a definition of Wigner distributions consistent with relativity. 
 One starts by generalizing the bilocal operator of Eq.~(\ref{eq:bilocal}) by introducing a Wigner operator at equal light-cone time of both quark fields ($z^+ = 0$) as:
 \begin{eqnarray}
\hat{W}^q({\bf b}, {\bf k}, x) &\equiv&
\frac{1}{2}\int \frac{dz^{-} d^2 {\bf z}}{(2 \pi)^3}e^{i(xP^{+}z^{-} - {\bf k} \cdot {\bf z})} \nonumber \\
&\times& \bar{q}(0,-\frac{z^-}{2}, {\bf b} - \frac{{\bf z}}{2}) \gamma^+ q(0,\frac{z^-}{2},
{\bf b}+\frac{{\bf z}}{2}), \quad 
\end{eqnarray}
 Wigner distributions are then defined as matrix elements of the Wigner operators sandwiched between nucleon states with polarization $\vec S$ as:
 \begin{eqnarray}
&& \rho^{q}({\bf b},{\bf k}, x, \vec{S}) \nonumber \\
&&\equiv \int{{\frac{d^{2}{\bf \Delta_\perp}}{(2\pi)^2}} \left< P^+, {\frac{{\bf \Delta}_\perp}{2}}, \vec{S}\left | \hat{W}^{q}({\bf b}, {\bf k},x)\right | P^+, - {\frac{{\bf \Delta}_\perp}{2}},\vec{S}\right>   }, \nonumber \\
 \end{eqnarray}
 which is equivalently given by 
\begin{eqnarray}
&&\rho^{q}({\bf b},{\bf k}, x, \vec{S})
= \int {\frac{d^{2}{\bf \Delta_\perp}}{(2\pi)^2}} e^{- {\bf \Delta}_\perp \cdot {\bf b}} \nonumber \\
&& \hspace{1.5cm} \times \left< P^+, {\frac{{\bf \Delta}_\perp}{2}}, \vec{S}\left | \hat{W}^{q}({\bf 0}, {\bf k},x)\right | P^+, - {\frac{{\bf \Delta}_\perp}{2}},\vec{S}\right>   , \nonumber \\
\label{eq:gtmd}
\end{eqnarray}
which is the two-dimensional Fourier transform of the GTMD for $\Delta^+ = 0$, i.e. $\xi = 0$. 
Although Wigner distributions are not positive definite, and as a result do not have a strict probabilistic interpretation, as Heisenberg's uncertainty principle prevents to determine the quark's position and momentum simultaneously, by integrating out position or momentum variables they reduce to probability distributions. 
By integration Eq.~(\ref{eq:gtmd}) over transverse position ${\bf b}$, they reduce to forward matrix elements for ${\bf \Delta}_\perp = {\bf 0}$, described by TMDs. 
Furthermore, by integrating Eq.~(\ref{eq:gtmd}) over 
{\bf k} one readily obtains the densities $\rho^q(x, {\bf b})$ of Eq.~(\ref{eq:dens2xb}) as 2-dimensional Fourier transforms over GPDs. 
Lastly, by integrating Eq.~(\ref{eq:gtmd}), for a nucleon in a light-front helicity state $\lambda$,
over $x$ and ${\bf k}$, one recovers from Eq.~(\ref{eq:dens1}) the transverse density as:
\begin{equation}
\int dx d^{2} {\bf k} \, \rho^{q} ({\bf b}, {\bf k}, x, \lambda) = \rho^q_\lambda(b),
\end{equation}
and analogously the transverse density of Eq.~(\ref{eq:dens2a}) for a nucleon in a transverse spin state $\vec S_\perp$.
This correspondence demonstrates that the Wigner distributions provide a unified description of the partonic structure of the nucleon, and nucleon electromagnetic form factors occupy a small but important part of their vast space, illustrated in 
Fig.~\ref{fig-GTMD}.

\subsection{Radii of quark distributions in a proton}

As discussed above, to define and reconstruct a 3-dimensional charge distribution from elastic electron scattering measurements of the form factors of a system requires that one is able to localize the object and fix its center-of-mass, with respect to which one defines the charge distribution~\cite{Jaffe:2020ebz}. This is possible for non-relativistic (static) systems for which the typical size is much larger than its Compton wavelength, allowing the probe to localize the charges at distances between both scales. For atomic nuclei, this condition is well satisfied as their Compton wavelength (of order 0.2/A fm) is typically much smaller than their size (of order 1.2 A$^{1/3}$ fm). As example, for the $^{12}$C nucleus, its size of around 2.5~fm is much larger than its Compton wavelength of around 0.02~fm, allowing to localize charges in between these length scales and reconstruct a charge distribution. 
For such systems, one can define a 3-dimensional charge distribution as Fourier transform of the measured electric form factor $G_E$ as given in Eq.~(\ref{eq:dens3dnr}). 
For such charge distribution, one can define a radius through the normalized moment:
\begin{eqnarray}
r_E^2 \equiv \frac{\int d^3 \vec r \, r^2 \, \rho_{\rm 3d, n-rel}(r)}{\int d^3 \vec r \, \rho_{\rm 3d, n-rel}(r)}. 
\label{eq:3dimdens2}
\end{eqnarray}
Inserting the 3-dimensional density defined in Eq.~(\ref{eq:dens3dnr}) allows to express the charge radius as:
\begin{eqnarray}
r_E^2 = - 6 \frac{ G_E^\prime(0)}{G_E(0)},
\label{eq:redef}
\end{eqnarray}
where $ G_E^{\prime}(0) \equiv \frac{d G_E}{d Q^2}\big|_{Q^2 = 0}$, with $Q^2 = \vec q^{\, 2}$.  
One can therefore express the Taylor expansion of $G_E$  at low values of $Q^2$ as:
\begin{eqnarray}
G_{E}(-Q^2) &\equiv& G_E(0)\left\{1 - \frac{1}{6}  r_{E}^2  Q^2  + {\cal O} (Q^4) \right\}, 
\end{eqnarray}
and access the charge radius experimentally from the electric form factor slope at the origin. 

To apply the above concepts to a nucleon becomes problematic as for the nucleon its size (or order 0.85 fm) is not very much larger than its Compton wavelength (of order 0.2 fm), making it not possible to well localize the center-of-mass in three spatial dimensions. Besides for light-quark systems, we have discussed that an interpretation in terms of a positive definite quantity is spoiled in the rest frame due to pair creation processes.  In the previous section, we have reviewed how to properly define density  distributions for a relativistic bound state as a nucleon by going to the infinite momentum frame which allows to localize the hadron in a plane perpendicular to the direction of a fast moving observer and define density distributions in that plane. For the resulting two-dimensional transverse distributions for a quark of flavor $q$ in the proton, one can then define a mean-squared radius as:
\begin{eqnarray}
\langle b^2 \rangle^q  &=& \frac{\int d^2 {\bf b} \, {\bf b}^2 \, \rho^q(b)}{\int d^2 {\bf b} \,  \rho^q(b)} 
= - 4 \frac{F_1^{\prime  q}(0)}{F_1^q(0)} ,
\label{eq:rad1}
\end{eqnarray}
where $ F_1^{\prime  q}(0) \equiv \frac{d F_1^q}{d Q^2}\big|_{Q^2 = 0}$ denotes the slope at the orgin of the corresponding Dirac form factor.  
Note that the radius definition of Eq.~(\ref{eq:rad1}) for each quark flavor is properly 
normalized to the number of valence quarks in the proton: $F_1^u(0) = 2$ and $F_1^d(0) = 1$, yielding:
\begin{eqnarray}
\langle b^2 \rangle^u  = - 2  F_1^{\prime  u}(0), \quad \quad  
\langle b^2 \rangle^d  = - 4  F_1^{\prime  d}(0). 
\label{eq:rad2}
\end{eqnarray} 
We can also obtain the mean-squared radius 
$\langle b^2 \rangle^q$ through an  
average over the $x$-dependent radius $\langle b^2 \rangle^q(x)$ of Eq.~(\ref{eq:cr5}) as:
\begin{eqnarray}
\langle b^2 \rangle^q 
=  \frac{1}{F_1^q(0)} \int_{-1}^1 dx \,q(x) \, \langle b^2 \rangle^q (x).
\label{eq:bperp2}
\end{eqnarray}

To determine the mean-squared radii Eq.~(\ref{eq:rad2}) for each quark flavor, we start by expressing the proton and neutron Dirac form factors, using isospin symmetry, as:
\begin{eqnarray}
F_{1p} &=& e_u F_1^u + e_d F_1^d, \nonumber \\
F_{1n} &=& e_u F_1^d + e_d F_1^u, 
\label{eq:rad3}
\end{eqnarray}
which allows to extract the Dirac form factors for the $u$- and  $d$-quark flavors, which enter the corresponding transverse quark densities, 
as:
 \begin{eqnarray}
F_{1}^u = 2 F_{1p} + F_{1n}, \quad \quad 
F_{1}^d = 2 F_{1n} + F_{1p}. 
\label{eq:rad4}
\end{eqnarray}
Combining Eqs.~(\ref{eq:rad2}) and (\ref{eq:rad3}), this allows to express the proper mean-squared radii for the $u$- and $d$-quark distributions in a proton as:
\begin{eqnarray}
\langle b^2 \rangle^u  &=& -2 \left\{ 2 F^\prime_{1p}(0) + F^\prime_{1n}(0) \right\}, 
\nonumber \\
\langle b^2 \rangle^d  &=& -4 \left\{  F^\prime_{1p}(0) + 2 F^\prime_{1n}(0) \right\}. \quad  
\label{eq:rad6}
\end{eqnarray} 
The last equation allows to express the difference of the mean-squared radii for $d$- and $u$-quark distributions in a proton as:
\begin{equation}
\langle b^2 \rangle^d  - \langle b^2 \rangle^u =  -6 F^\prime_{1n}(0).
\label{eq:rad7}
\end{equation}
In order to empirically determine the mean-squared radii of $u$- and $d$-quark distributions in a proton, we relate 
the derivative of the Dirac form factors to the conventional Sachs form factors $G_E$ and $G_M$, defined through Eqs.~(\ref{eq:GE}, \ref{eq:GM}), which yields: 
\begin{eqnarray} 
F_1^{\prime }(0) = G_E^{\prime }(0) + \frac{\kappa}{4 M^2}.
\label{eq:rad8}
\end{eqnarray}
Following the convention for non-relativistic (static) systems, one can Taylor expand the 
proton and neutron Dirac form factors at low momentum transfer $Q^2$ as:
\begin{eqnarray}
G_{Ep}(- Q^2) &\equiv& 1 - \frac{1}{6}  r_{Ep}^2  Q^2  + {\cal O} (Q^4), 
\label{eq:rad9p} \\
G_{En}(- Q^2) &\equiv& - \frac{1}{6}  r_{En}^2  Q^2  + {\cal O} (Q^4).
\label{eq:rad9n}
\end{eqnarray}
We like to emphasize again that for relativistic bound states as a nucleon where the concept of a 3-dimensional charge distribution is not well defined,  Eqs.~(\ref{eq:rad9p},\ref{eq:rad9n}) are  merely used 
as operational definitions for the form factor slopes at the origin, even though we will refer to these quantities for simplicity as ``radii" in the remainder of this review.   
Eqs.~(\ref{eq:rad9p},\ref{eq:rad9n}) then allow to express for the nucleon 
$(N = p, n)$:
\begin{eqnarray}
-6 F^\prime_{1N}(0) = r_{E N}^2 - \frac{3 \kappa_N}{2 M^2},
\label{eq:rad8}
\end{eqnarray}
where the anomalous magnetic moment contribution is known as the Foldy term. 
\begin{table*}
{\centering
\begin{tabular}{c|c|c|c|c}
\hline  \hline
& $\langle r_E^2 \rangle$  & \quad $- \frac{3 \kappa_N}{2 M^2}$ \quad & $-6 F^\prime_1(0)$  & \quad $\langle b^2 \rangle$ \quad  \\
&&&& \\
&(fm$^2$) & \quad (fm$^2$) \quad & (fm$^2$) & \quad (fm$^2$) \quad \\
\hline \hline
&&&& \\
\quad proton \quad  & $0.717 \pm 0.014$ & 
& \quad $0.598 \pm 0.014$ \quad
& \quad $0.399 \pm 0.009$ \quad \\
(e-p) &  \cite{Cui:2021vgm} & & & \\
&& \quad $-0.1189$ \quad & & \\
\quad proton \quad     & \quad $0.7071 \pm 0.0007$ \quad & 
&  $0.5882 \pm 0.0007$ & \quad $0.3921 \pm 0.0005$ \quad \\
($\mu$H) &  \cite{Antognini13} & & & \\
&&&& \\
\hline 
&&&& \\
\quad neutron \quad & $-0.1161 \pm 0.0022$ & \quad $0.1266$ \quad & \quad $0.0105 \pm 0.0022$ \quad & \quad $0.0070 \pm 0.0015$ \quad \\
 (PDG) &  \cite{Zyla:2020zbs} &&& \\
&&&& \\
\hline \hline
\end{tabular}
\par}
\caption{Empirical values of the proton and neutron radii $r_E^2$, Foldy terms, Dirac slopes $F^\prime_1(0)$, and transverse charge radii $\langle b^2 \rangle$. For the proton, we show the values both using the e-p scattering data analysis of \cite{Cui:2021vgm}, as well as the values from $\mu$H Lamb shift measurements~\cite{Antognini13}.
\label{tab:quarkrad1}}
\end{table*}

The radius of the transverse charge distribution in a proton is then obtained as sum over the radii for the quark distributions weighted by their charges:
\begin{eqnarray}
\langle b^2 \rangle_p = \frac{4}{3} \langle b^2 \rangle^u 
- \frac{1}{3} \langle b^2 \rangle^d = - 4 F^\prime_{1p}(0). 
\end{eqnarray}
For the neutron, assuming isospin symmetry, we define a transverse charge radius
as\footnote{Note that for a neutron, this follows the convention in defining a charge radius for a neutral system, as one cannot use the definition of Eq.~(\ref{eq:rad1}) which is normalized to the total charge.}:
\begin{eqnarray}
\langle b^2 \rangle_n = \frac{2}{3} \langle b^2 \rangle^d 
- \frac{2}{3} \langle b^2 \rangle^u = - 4 F^\prime_{1n}(0). 
\end{eqnarray}

In Table~\ref{tab:quarkrad1}, we show the empirical values of proton and neutron radii  $r_E^2$,  
the Foldy terms, the extracted Dirac slopes  $F^\prime_1(0)$, 
and transverse charge radii $\langle b^2 \rangle$. 
For the proton values for $r_{Ep}^2$ we are showing both the recent analysis of \cite{Cui:2021vgm} based on e-p scattering results, which will be discussed in Section~\ref{sec:scatt}, Eq.~(\ref{eq:rpcui}), as well as the extracted value from the $\mu$H Lamb shift measurements, 
which will be discussed in Section~\ref{sec:spec}, Eq.~(\ref{eq:rpLS}). 
Anticipating the discussion in Section~\ref{sec:spec}, the quantity entering the hydrogen spectroscopy Lamb shift experiments is also given by the slope $G^\prime_{Ep}(0)$. Therefore, it is important and meaningful to compare the proton charge radius values obtained by these two experimental techniques. 
We see from Table~\ref{tab:quarkrad1} that the extracted mean-squared radii $\langle b^2 \rangle$ are consistent between both analyses, showing that the transverse charge distribution in a proton has a rms radius around 0.63~fm, as seen by an observer moving with a light-front.  
For the neutron, one notices that its Dirac slope  value $F^\prime_{1n}(0)$ is the result of a large cancellation between the $r_{En}^2$ term and the Foldy term, which have opposite signs, resulting in a value of $F^\prime_{1n}(0)$ around 10\% of the size of each contribution.   
As the Foldy term for the neutron is slightly larger in absolute value than the $r_{En}^2$ term, the resulting positive value of $- 6 F^\prime_{1n}(0)$ combined with Eq.~(\ref{eq:rad7}) results in a slightly larger mean-squared radius for the $d$-quarks in a proton in comparison with the $u$-quarks in the proton, confirming the observation of \cite{Cates:2011pz} based on a flavor decomposition of proton and neutron form factors. 

In Table~\ref{tab:quarkrad2}, we show the extracted values of the mean-squared radii for $u$- and $d$-quark distributions in the proton, using the neutron PDG value for $r_{En}^2$, and both analyses for the proton as shown in Table~\ref{tab:quarkrad1}.  
For the more accurate values  extracted from the $\mu$H Lamb shift measurements,
one obtains a precision of 
0.3\% (0.7\%) on the mean-squared radii for the $u$ ($d$)-quark distributions. Using the values in Table~\ref{tab:quarkrad1}, we notice that the neutron $F^\prime_{1n}(0)$ term contributes 1\% (4\%) to the mean-squared radii for the $u$ ($d$)-quark distributions respectively in Eq.~(\ref{eq:rad6}). 
One also notices that the uncertainty on the neutron $F^\prime_{1n}(0)$ value is at present the limiting uncertainty in the extraction of the mean-squared radius value for the $d$-quark distribution. 

\begin{table}
{\centering
\begin{tabular}{c|c|c}
\hline \hline
& $\langle b^2 \rangle^u$  & $\langle b^2 \rangle^d$    \\
& (fm$^2$) & (fm$^2$) \\
\hline \hline
&& \\
proton (e-p)  & \quad $0.402 \pm 0.009$ \quad 
& \quad $ 0.413 \pm 0.010$ \quad \\
&& \\
\hline
&& \\
proton ($\mu$H)  & \quad $ 0.396 \pm 0.001$ \quad & 
\quad $ 0.406 \pm 0.003$ \quad \\
&& \\
 \hline \hline
\end{tabular}
\par}
\caption{Extracted values of the mean-squared radii for $u$- and $d$-quark distributions in the proton, using the neutron PDG value for $r_{En}^2$ as given in Table~\ref{tab:quarkrad1}, and for the proton values for $r_{Ep}^2$ from both the analysis of \cite{Cui:2021vgm} based on e-p scattering results, as well as the extracted value from the $\mu$H Lamb shift measurements~\cite{Antognini13}.
\label{tab:quarkrad2}}
\end{table}

In the next sections, we will discuss unpolarized and polarized electron-proton elastic scatterings and the method to extract the proton electric form factor and the 
proton charge radius value based on the definition of Eq.~(\ref{eq:rad9p}) as slope at the origin of the form factor $G_E$.  
Likewise, one can also define a magnetic radius as slope at the origin of the form factor $G_{MN}$ for the nucleon ($N = p, n$):
\begin{eqnarray}
G_{MN}(- Q^2) &\equiv& \mu_N \left\{ 1 - \frac{1}{6}  r_{MN}^2  Q^2  + {\cal O} (Q^4) \right\}, 
\label{eq:radM} 
\end{eqnarray}
where $\mu_N$ is the nucleon magnetic moment, $\mu_p = 2.79$ and 
$\mu_n = - 1.91$, in the units of the nucleon magneton.

Ideally, to extract the proton charge radius value $r_{Ep}$, one needs to extract the proton electric form factor, $G_E$ all the way down to $Q^2 =0$, and then determine its slope. In practice, it is not possible to measure $G_E$ at $Q^2 \sim 0$,  which corresponds to near $0^\circ$ scatterings. Therefore some type of extrapolation is unavoidable which may introduce systematic uncertainties associated with the determination of $r_{Ep}$ as discussed below. 

A theoretical determination of the proton radius starting from QCD requires a nonperturbative framework. The only ab-initio tool so far is lattice QCD. The standard procedure in lattice QCD is to compute the electric form factor for finite values of the momentum transfer and then perform a fit to determine the slope at zero momentum transfer, e.g. through a popular dipole fit or a $z$-expansion fit. However, on a finite lattice, the smallest nonzero momentum is $2 \pi/L$ with $L$ is the spatial size of the lattice. Therefore, to reach very small momentum transfers is challenging as it requires very large lattices. 
Furthermore, although electromagnetic form factors have been studied within lattice QCD for many years, it is only recently that they have been extracted using simulations with physical values of the light quark masses.

\onecolumngrid
\begin{center}
\begin{figure}[h]
\includegraphics[width=0.43\textwidth]{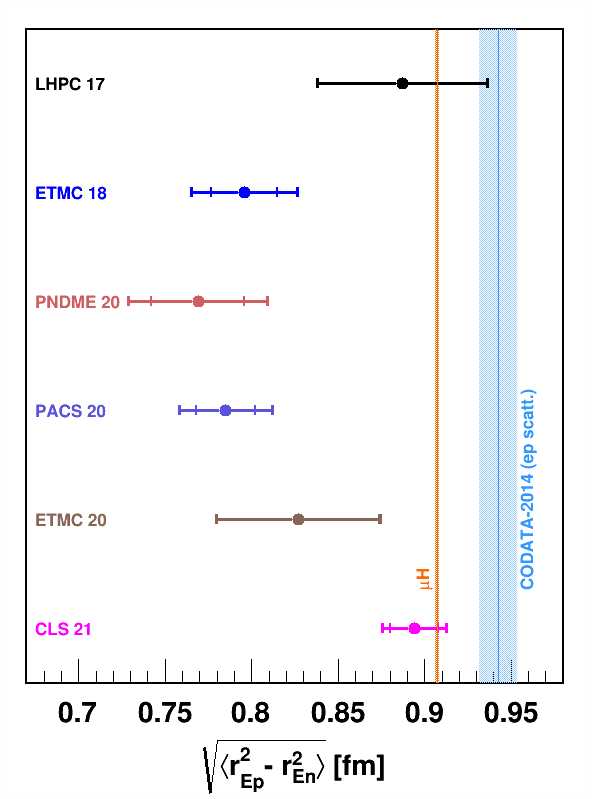}
\includegraphics[width=0.43\textwidth]{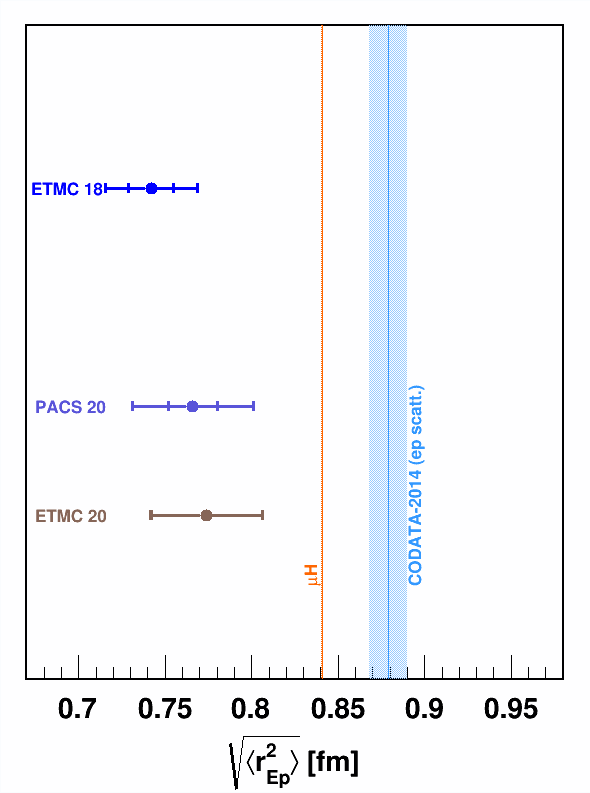}
\caption{(Color online) Compilation of recent lattice QCD results for the isovector charge radius (left panel) and the proton charge radius (right panel), obtained from ensembles at the physical pion mass.  Results shown are from 
LHPC~\cite{Hasan:2017wwt}; 
ETMC, both using a form factor fit  ETMC~18~\cite{Alexandrou:2018sjm}, 
as well as the direct calculation of the radius ETMC~20, avoiding an extrapolation through a form factor fit~\cite{Alexandrou:2020aja}; 
PNDME~\cite{Jang:2019jkn}; 
PACS~\cite{Shintani:2018ozy};
CLS~\cite{Djukanovic:2021cgp}.
Inner error bars display the statistical errors, whereas outer error bars display  the full error. The vertical bands show the empirical result extracted from muonic hydrogen spectroscopy and the CODATA-2014 recommended value, as discussed in Sections
\ref{sec:scatt} and \ref{sec:spec} 
(figure credit: Jingyi Zhou).}
\vspace{.5cm}
\label{fig:lattice}
\end{figure}
\end{center}
\twocolumngrid

In Fig.~\ref{fig:lattice}, we show a compilation of recent lattice QCD results for both the isovector charge radius 
$(\langle r_{Ep}^2 \rangle - \langle r_{En}^2 \rangle)^{1/2}$, as well as the proton charge radius, obtained from ensembles at or near the physical pion mass. For the isovector radius only the connected quark diagrams, in which the photon couples to the quarks connected to either the intial or final nucleon, contribute.  The 
proton charge radius also requires the much harder calculation of the contribution from disconnected diagrams, in which the photon couples to a $q \bar q$ loop, which interacts with the quarks in the intial and final proton through gluon exchanges. 
Although the disconnected contribution to the proton electric form factor at low momentum transfer is found to be in the 1~\% range~\cite{Alexandrou:2018sjm}, its omission would result in an uncontrolled systematic error. Such systematics need to be under control for precision comparisons  of the proton charge radius at the 1\% level or better.

Improving on the precision of the lattice extractions of the proton charge radius also 
requires to reduce the model error induced by a form factor fit, which is done in most of the lattice results so far. To this end, a first step was taken 
in the lattice study of 
~\cite{Alexandrou:2020aja}, which 
has explored a direct method to extract the proton radius that does not depend on fitting the form factor, displayed by ETMC~20 in Fig.~\ref{fig:lattice}.

The lattice calculations have made important progress in recent years, by controlling excited state contaminations, and by performing calculations at the physical point. One notices however from Fig.~\ref{fig:lattice} that further improvements are called for to reach the precision level obtained in the empirical extractions.

\subsection{Unpolarized electron-proton elastic scattering}

The differential cross-section based on OPE for elastic 
electron-nucleon scattering can be written in terms of the two electromagnetic form factors: $G_E$ 
and $G_M$ as:
\begin{eqnarray}
{\frac{d\sigma}{d\Omega}}_{lab} 
&=& {\frac{\alpha^2}{4E^2 \sin^{4}{
{\frac{\theta}{2}}}}}
{\frac{E'}{E}} \nonumber \\
&\times& \left\{{\frac{G^2_{E} + \tau G^{2}_M}{1+\tau}}
\cos^{2}{\frac{\theta}{2}} 
+ 2 \tau G^{2}_M \sin^{2}{{\frac{\theta}{2}}} \right\},
\label{eq:unpol}
\end{eqnarray}
where $E$ is the incident electron energy; $E'$ the energy of the scattered electron, and $\theta$ the electron scattering angle, respectively; $\alpha$ is the fine
structure constant, $\tau \equiv {\frac{Q^2}{4M^2}}$, and where the mass of the electron is neglected. 

In the remaining of this paper, we will focus on the proton electric and magnetic form factors. To separately determine the proton electric and magnetic form factor for each $Q^2$ value, ideally one would need to perform two measurements with independent combinations of the $G_E$ and $G_M$ at the corresponding $Q^2$ value, with one of the measurements involving polarizations which we will discuss further on.  However, polarization experiments only became possible in recent decades. Historically, the Rosenbluth technique ~\cite{Rosenbluth50} had been used extensively which allows for the separation of these two form factors by performing unpolarized differential cross section measurements only. To see how this works, one can rewrite Eq.~(\ref{eq:unpol}) as: 
\begin{eqnarray}
{\frac{d\sigma}{d\Omega}}_{lab} &=& \sigma_M  \frac{1}{1+\tau} \left\{{G^2_E} + 
{\frac{\tau}{\epsilon}}{G^2_M}\right\},
\label{eqn:rosenbluth}
\end{eqnarray}
where $\epsilon={(1+2(1+\tau) \tan^{2}{\frac{\theta}{2}})}^{-1}$ 
is the virtual photon longitudinal polarization, and
$\sigma_{M}$ is the Mott cross section describing the scattering
from a pointlike spinless target (where we included the recoil factor $E'/E$):
\begin{equation}
\sigma_{M} = {\frac{\alpha^2  \cos^{2}{\frac{\theta}{2}}}{4 E^{2}
    \sin^{4}{\frac{\theta}{2}}}} \left( \frac{E'}{E} \right).
\end{equation}

The Rosenbluth separation technique works in the following way: at a fixed $Q^2$ value, one can take a series of measurements by varying the incident electron beam energy and the scattering angle. According to Eq.~(\ref{eqn:rosenbluth}), one can then fit the measured reduced cross section $G_M^2 + \epsilon/\tau G_E^2$ as a function of ${\epsilon}$. Then from the slope and the intercept of the fit, one can determine $G_E^2$ and $G_M^2$. There are limitations to the Rosenbluth method: at low $Q^2$, due to the kinematic suppression, the extraction of the proton magnetic form factor is problematic while at high $Q^2$, the magnetic contribution dominates the cross section and the extraction of the proton $G_E$ becomes difficult. This is due to the fact that $G_E$ only yields a small $\epsilon$-dependent contribution to the cross section, which makes its extraction sensitive to small $\epsilon$-dependent corrections. Notwithstanding these limitations, this method was used extensively prior to any polarization measurements. 

An improved Rosenbluth separation, also known as the super Rosenbluth method allows for significantly improved extraction of the proton electromagnetic form factor ratio at higher values of $Q^2$ with precision comparable to those from polarization techniques.   In the super Rosenbluth method, the recoil proton is detected to overcome issues at large values of $Q^2$ associated with detecting the scattered electrons. 
The first such experiment was reported by Qattan {\it et al.}~\cite{PhysRevLett.94.142301}, in which
at large and fixed values of $Q^2$, the change in the differential cross section $d\sigma/d\Omega_{p}$ over the measured $\epsilon$ range was less than a factor of 2, 50 times smaller compared with detecting electrons. Further, the proton momentum was constant at fixed $Q^2$, while the electron momentum varied by a factor of 10. Finally the QED radiative correction effects were also suppressed when the recoil proton was detected~\cite{PhysRevLett.94.142301}.

\subsection{Double-Polarization Elastic Electron-Proton Scattering}
 
To overcome the aforementioned limitations associated with the Rosenbluth technique, an independent combination of the proton electric and magnetic form factors can be obtained by a double polarization measurement from electron-proton elastic scattering in addition to unpolarized differential cross section measurements, therefore allowing for the separation of these two form factors.

Polarization degrees of freedom in electron scattering have proven to
be very useful in extracting information about small amplitude
processes by isolating terms sensitive to the interference of the
small amplitude with a much larger amplitude. The ability to
selectively isolate certain combinations of amplitudes that either polarized
electron beams or polarized targets (recoil polarimeters) 
have historically provided when
used in isolation, is significantly enhanced by using them together. Double polarization measurements in the context of electron-proton scattering refer to the following two cases: (i) longitudinally polarized electrons scattering from a polarized proton target; (ii) longitudinally polarized electrons scattering from an unpolarized proton target with the recoil proton polarization measured by a polarimeter. In this paper, we will not review the technical aspects of polarized electron beams, polarized proton targets, nor the recoil proton polarimeters. We refer interested readers to review articles~\cite{Gao03,Perdrisat:2006hj} instead.

\vspace{0.1in}
\subsubsection{Spin-Dependent Asymmetry from $\vec{p}(\vec{e},e')p$}
 
\begin{figure}
\includegraphics[scale=0.3]{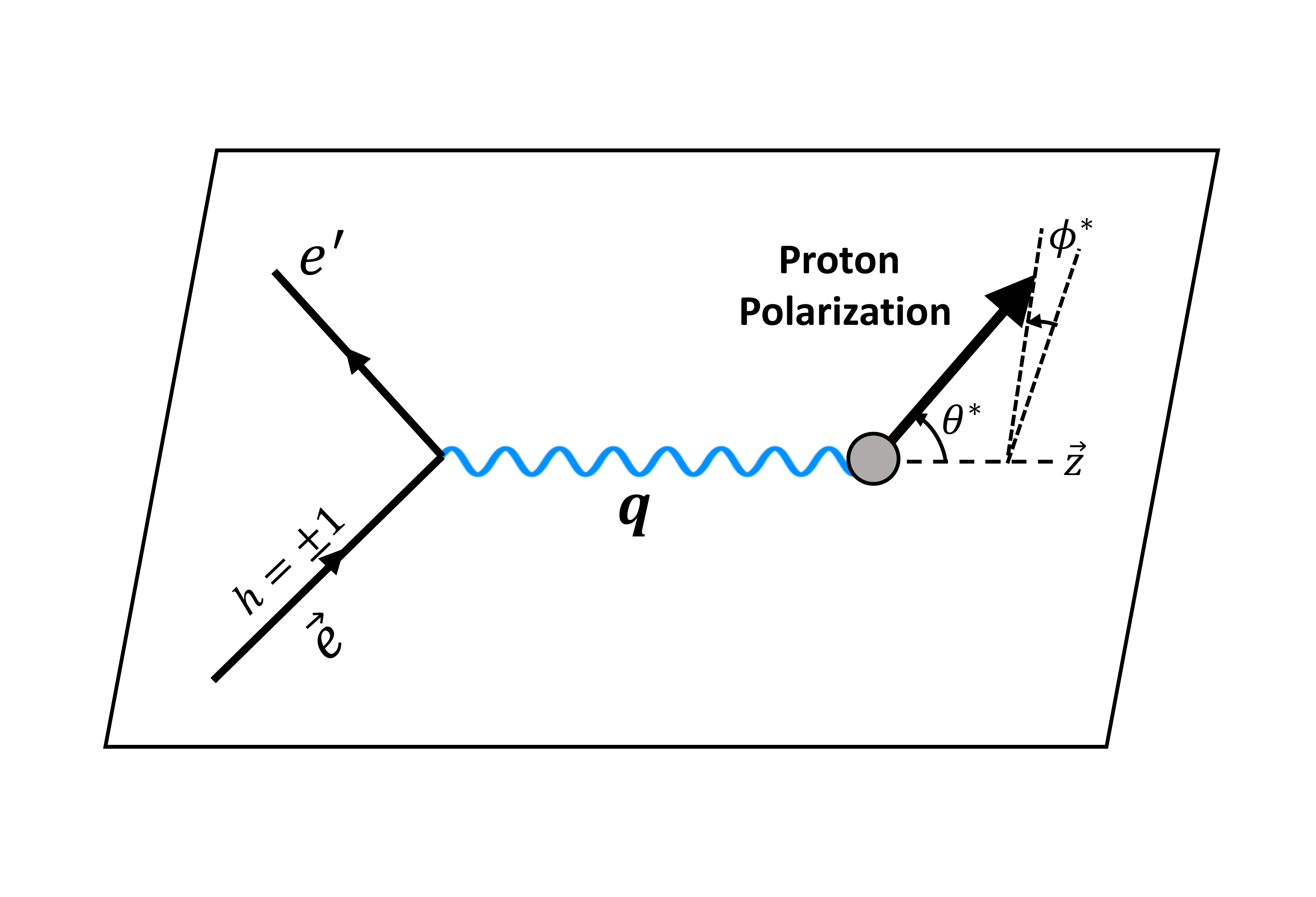}
\caption[fig]{\label{fig:one-photon-spin} (Color online) The one-photon-exchange diagram for spin-dependent electron-proton scattering (figure credit: Jingyi Zhou).} 
\end{figure}

The one-photon-exchange diagram for spin-dependent electron-nucleon scattering is shown in Fig.~\ref{fig:one-photon-spin}. In this picture the incident electron is longitudinally polarized with helicity of $h=\pm 1$, corresponding to an electron's spin being parallel or anti-parallel to its momentum direction, respectively.  The target proton spin vector is shown by a thick arrow, with  $\theta^{*}$ and $\phi^{*}$ as its polar and azimuthal angles defined with respect to the three-momentum transfer vector
${\bf q}$ of the virtual photon. The scattering plane is defined as the $x,z$ plane with $\hat{z}={\bf q}/|{\bf q}|$ and 
$\hat{y}=({\bf k}\times{\bf k'})/(|{\bf k}||{\bf k'}|$), with ${\bf k}$ and ${\bf k'}$ being the incident and scattered electron three-momentum vector, respectively.

The spin-dependent asymmetry $A$ is defined as
    $A = (\sigma^{h+}-\sigma^{h-})/(\sigma^{h+}+\sigma^{h-})$,
where $\sigma^{h^{\pm}}$ denotes the differential cross sections for the two 
different helicities of the polarized electron beam.

For longitudinally polarized electrons scattering from a polarized proton 
target, the differential cross section can be written \cite{Donnelly86} as:

\begin{equation}
{\frac{d\sigma}{d\Omega}} = \Sigma + h  \Delta \>,
\end{equation}
where $\Sigma$ is the unpolarized differential cross section 
given by Eq.~(\ref{eq:unpol}), and $\Delta$ is the spin-dependent differential cross
section given by:
\begin{equation}
\Delta =  \sigma_{Mott} [v_{z} 
\cos \theta^{*} {G^2_{M}} + v_{x}
\sin\theta^{*} \cos\phi^{*} G_{M} G_{E}]\>,
\end{equation}
where
\begin{eqnarray}
v_{z} &=&-2 \tau \tan{\frac{\theta}{2}} \sqrt{{\frac{1}{1+\tau}}+{\tan^{2}{\frac{\theta}{2}}}}, \\
v_{x} &=-&2 \tan{\frac{\theta}{2}} \sqrt{{\frac{\tau}{1+\tau}}},
\end{eqnarray}
are kinematic factors.
The spin-dependent asymmetry $A$ is defined in terms of the polarized
and unpolarized cross-sections as: 
\begin{equation}
A = {\frac{\Delta}{\Sigma}} =  {\frac{ v_{z} \cos\theta^{*}
G^2_{M} + v_{x} \sin\theta^{*}
\cos\phi^{*} G_{M} G_{E}}{ ({\epsilon G^2_{E}}  + \tau
 {G^2_{M}})/[\epsilon (1+\tau)]}} \>.
\label{eqn:asymmetry}
\end{equation}
The experimental asymmetry $A_{exp}$, is related to the spin-dependent
asymmetry of Eq.~(\ref{eqn:asymmetry})  by the relation 
\begin{equation}
A_{exp} = P_{b} P_{t} A \>,
\end{equation}
where $P_{b}$ and $P_{t}$ are the beam and target polarization,
respectively. A determination of the ratio 
$G_{E} / G_{M}$, independent of the knowledge
of the beam and target polarization 
can be precisely obtained by measuring the so-called super ratio 
\begin{equation}
R = {\frac{A_1}{A_2}} = {\frac{v_{z} \cos\theta_{1}^{*}
{G^2_{M}} + v_{x} \sin\theta_{1}^{*}
\cos\phi_{1}^{*} G_{M} G_{E}} {v_{z} \cos\theta_{2}^{*}
{G^2_{M}} +  v_{x}
\sin\theta_{2}^{*} \cos\phi_{2}^{*} G_{M} G_{E}}} \> , 
\end{equation}
where $A_1$ and $A_2$ are elastic electron-proton scattering asymmetries
measured at an identical value of $Q^2$ simultaneously, but at two different proton 
spin orientations relative to ${\bf q}$, corresponding to $(\theta_{1}^{*}, \phi_{1}^{*})$ and
$(\theta_{2}^{*}, \phi_{2}^{*})$, respectively. However, the proton spin direction is fixed in the laboratory frame, therefore it is feasible if one has a symmetric detection system.
For a symmetric detector configuration with respect to the incident 
electron momentum direction, the
$A_{1}$ and $A_{2}$ can be measured simultaneously by forming
two independent asymmetries with respect to either
the electron beam helicity or the target spin orientation 
in the beam-left and beam-right sector of the detector system, respectively.
Thus, the proton form factor ratio can be determined with high systematic accuracy using 
this technique because it is insensitive to the uncertainties in determining the beam and the target polarizations.
 Such a technique was pioneered~\cite{Crawford07} in the BLAST experiment~\cite{Hasell11} at the former MIT-Bates linear accelerator center, and the proton electric to magnetic form factor ratio was extracted in the $Q^2$ range from 0.15 to 0.65 (GeV/c)$^2$.

\subsubsection{Recoil Proton Polarization from $p(\vec{e},e'\vec{p})$}

 \begin{figure}[h]
\includegraphics[scale=0.32]{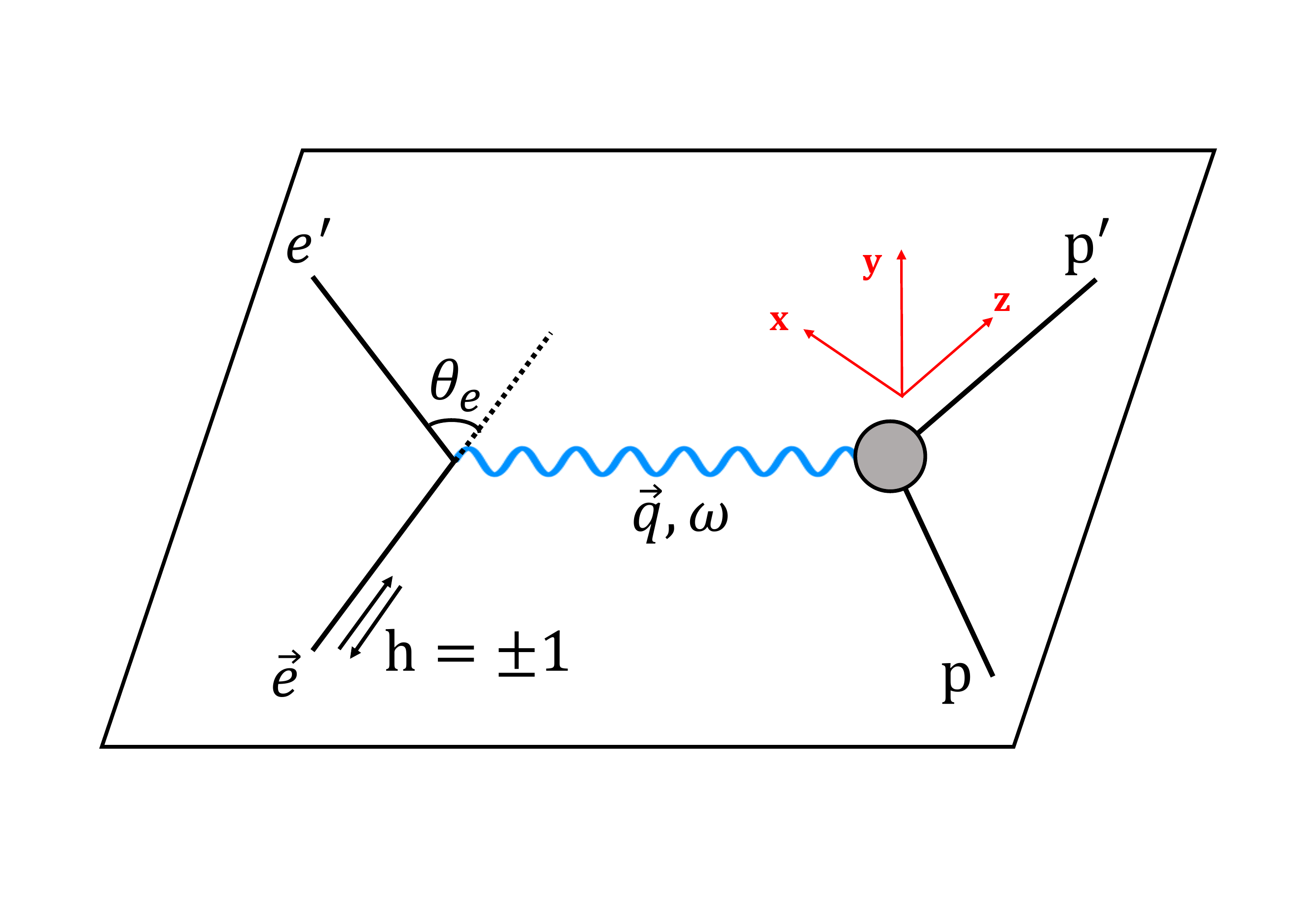}
\caption[fig]{\label{recoil} (Color online) The one-photon-exchange diagram for polarization transfer from longitudinally polarized electron to unpolarized proton (figure credit: Jingyi Zhou).} 
\end{figure}

A different type of double-polarization measurement in $e-p$ elastic scattering involves a longitudinally polarized electron beam, an unpolarized proton target, and a recoil proton polarimeter. Such experiments are called polarization transfer measurements -- the polarization from the incident electron beam is transferred to the recoil protons -- and the polarization of the final state protons is measured using a recoil proton polarimeter as illustrated in Fig.~\ref{recoil}. Such a polarimeter relies on secondary scatterings -- recoil protons from e-p scattering off an analyzer such as CH$_2$ -- and spin-orbital interaction of protons and nuclei, and spin-dependent proton-proton interaction which give rise to azimuthal angular dependence in the distribution of the scattered protons. By analyzing such azimuthal angular dependence, one can determine the recoil proton polarization components in the reaction plane ($x-z$ plane in Fig.~\ref{recoil}). Such secondary scatterings take place at the focal plane of the spectrometer and the polarimeter is also called focal-plane polarimeter (FPP). In order to determine the proton electric to magnetic form factor ratio at the target from the proton polarization components measured at the focal plane, an involved spin transport process is needed because the proton spin rotates as it goes through various magnetic components inside a magnetic spectrometer. The proton polarization measured by FPP, $\vec{P}_{\rm fpp}$, and the proton polarization at the target, $\vec{P}$, are related through a 3-dimensional spin rotation matrix. The elements of the spin rotation matrix can be calculated from a detailed modeling of the magnetic spectrometer including all spectrometer magnets (dipole, quadrupoles), fringe fields, and dipole field gradient, etc.  For details about such polarimeters, we refer interested readers to the review article by Perdrisat, Punjabi, and Vanderhaeghen~\cite{Perdrisat:2006hj}.

In the one-photon exchange Born approximation, the scattering of longitudinally 
polarized electrons results in a transfer of polarization to the 
recoil proton with only two nonzero components, $P_x$ perpendicular to, 
and $P_z$ parallel to the proton momentum in the scattering plane as illustrated in Fig.~\ref{recoil} 
\cite{Arnold81}. 
The form factor ratio can be
determined from a simultaneous measurement of the two recoil polarization
components in the scattering plane as
\begin{equation}
{\frac{G_E}{G_M}} = - {\frac{P_{x}}{P_{z}}} {\frac{E+E'}{2M}} 
\tan(\theta/2),
\end{equation}
in terms of the incident and scattered electron energies $E$ and $E'$ 
respectively, and electron scattering angle $\theta$.

\subsection{Two-Photon-Exchange  Contribution to Electron-Proton  Scattering}

\begin{figure}[h]
\includegraphics[scale=0.8]{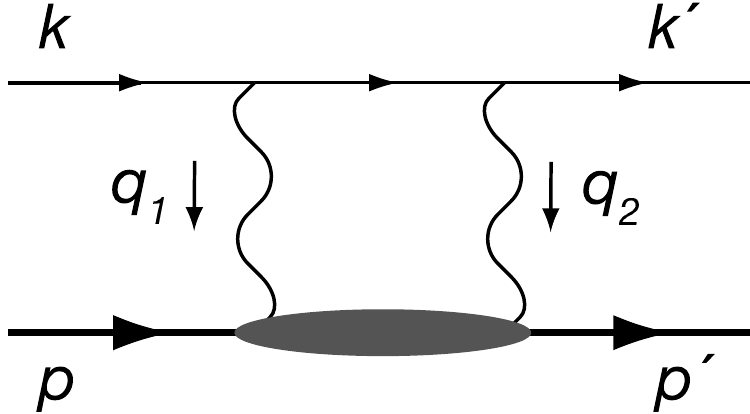}
\caption[fig]{\label{TPEdiagram} (Color online) The two-photon-exchange diagram for elastic electron-proton scattering. The blob denotes the doubly-virtual Compton subprocess on the proton.} 
\end{figure}

Note so far all our discussions are based on the dominant one-photon-exchange (OPE) Born diagram contribution in electron-proton scattering as higher orders contributions are suppressed due to the smallness of the fine-structure constant, $\alpha \simeq 1/137$.  The next-to-leading-order contribution is the two-photon-exchange (TPE) contribution, as shown in Fig.~\ref{TPEdiagram}, which 
is proportional to the doubly-virtual Compton subprocess on the proton side. 

The TPE contribution became a strong  interest after a drastic difference was reported on the proton $G_E / G_M$ ratio measured directly using a recoil proton polarimeter~\cite{Jones:1999rz} from those using Rosenbluth separation.  The data from~\cite{Jones:1999rz} and the subsequent recoil polarization   experiments~\cite{Gayou:2001qd,Punjabi:2005wq,Puckett:2010ac} show very intriguing behavior at higher
$Q^2$, i.e., $G_{Ep}$ falls off much faster than $G_{Mp}$ as a function 
of $Q^2$, while the two form factors extracted from unpolarized differential cross section measurements using the Rosenbluth separation method show a similar $Q^2$ dependence. The near constant behavior of the proton $G_{Ep}/G_{Mp}$ ratio extracted from unpolarized measurements was confirmed, and extended to a higher $Q^2$ value near 5.5 (GeV/c)$^2$ by another experiment in a different experimental hall at Jefferson Lab~\cite{Christy:2004rc}. 
The first explanations of such puzzling behavior pointed towards hard TPE processes between the electron and the proton, which become relevant once experiments aim to access terms which contribute at or below the percent level to the scattering cross section as is the case in the Rosenbluth method at larger $Q^2$ values~\cite{Guichon:2003qm, Blunden:2003sp}. 
This unexpected behavior triggered intensive experimental and theoretical studies of the TPE effect in electron-proton scattering in the last two decades as its effect is expected to be different in unpolarized cross section measurements compared to recoil polarization experiments; see~\cite{Carlson:2007sp, Arrington:2011dn} for some early reviews of this field. 

To account for two- and multi-photon exchange effects in a model independent way, requires one to generalize the amplitude of Eq.~(\ref{OPE}) describing the elastic e-p scattering.  
Neglecting the electron mass, the elastic e-p scattering amplitude, following the notations introduced in Section~\ref{sec:elintro},  
can be expressed through three independent structures~\cite{Guichon:2003qm}~:
\begin{eqnarray}
{\cal M}_{h, \, \lambda' \lambda} \,&=&\, 
i (e^2 / Q^{2}) \, \bar{u}(k', h)\gamma _{\mu }u(k, h)\,  \nonumber \\
&&\hspace{-1.75cm} \times \, 
{\bar N}(p', \lambda')\left( \tilde{G}_{M}\, \gamma ^{\mu }
-\tilde{F}_{2}\frac{P^{\mu }}{M}
+\tilde{F}_{3}\frac{\gamma \cdot K 
P^{\mu }}{M^{2}}\right) N(p, \lambda), \nonumber \\
\label{eq:TPE}
\end{eqnarray}
in which $K \equiv (k + k')/2$, and where the functions 
$\tilde G_M$, $\tilde F_2$, and $\tilde F_3$ are complex functions of $\epsilon$ and $Q^2$. In the OPE approximation, the functions $\tilde G_M$ and  $\tilde F_2$ reduce to the $Q^2$ dependent form factors $G_M$ and $F_2$ respectively, while the function $\tilde F_3$ vanishes. When accounting for the (very small) electron helicity flip effects,  which are proportional to its mass, the analysis has been extended in \cite{Gorchtein:2004ac}, where it was shown that three more amplitudes are needed to fully describe the e-p scattering amplitude. Based on such general analysis, the TPE corrections to both unpolarized and polarization observables have been expressed in~\cite{Guichon:2003qm} in terms of the amplitudes $\tilde G_M$, $\tilde F_2$, and $\tilde F_3$. In that work it was shown that by adding a TPE contribution of the size expected from perturbation theory, it is possible to simultaneously account for the relatively large correction to the unpolarized observable when extracting the $G_{Ep}/G_{Mp}$ ratio at larger $Q^2$, while maintaining a small correction in the polarization observables.     

To use electron scattering as a precision tool, it is clearly indispensable to arrive at a better quantitative understanding of TPE processes, and a lot of activities have taken place over the past two decades or are planned in the near future.  Firstly, there exist observables which provide us with very clear indications of the size of TPE effects, as they would be exactly zero in the absence of two- or multiphoton-exchange contributions. Such observables are normal single-spin asymmetries (SSA) of electron-nucleon scattering, where either the electron spin or the nucleon spin is polarized normal to the scattering plane. Because such SSAs are proportional to the imaginary part of a product of two amplitudes, they are zero for real (nonabsorptive) processes such as OPE. At leading order in the 
fine-structure constant, they result from the product of the OPE amplitude and the imaginary part of the TPE amplitude. 
For the target normal SSA, they were predicted to be in the (sub) percent range some time ago~\cite{DeRujula:1972te}. A measurement  of the normal SSA for the elastic 
electron-$^3$He scattering, by the JLab Hall A Coll., 
has extracted a SSA for the elastic electron-neutron subprocess in the percent range~\cite{Zhang:2015kna}.  
For the experiments with polarized beams, the corresponding normal SSAs were predicted to be in the range of a few to hundred ppm for electron beam energies in the GeV range~\cite{Afanasev:2002gr, Gorchtein:2004ac, Pasquini:2004pv}. Although such beam normal spin asymmetries are small, being proportional to the electron mass, the parity-violation programs at the major electron laboratories have reached precisions on asymmetries with longitudinal polarized electron beams well  below the ppm level, and the next generations of such experiments are designed to reach precisions at the sub-ppb level~\cite{Kumar:2013yoa}. The beam normal SSA, which is due to TPE and thus parity conserving,  
has been measured over the past two decades as a spinoff in the parity-violating electron scattering programs at MIT-BATES (SAMPLE Coll.)~\cite{Wells:2000rx}, 
at MAMI (A4 Coll.)~\cite{Maas:2004pd, BalaguerRios:2012uk,Gou:2020viq}, and at JLab (G0 Coll.~\cite{Armstrong:2007vm, Androic:2011rh}, HAPPEX/PREX Coll.~\cite{Abrahamyan:2012cg}, and Qweak Coll.~\cite{Androic:2020rkw}).  The resulting beam normal SSA ranges from a few ppm in the forward angular range to around a hundred ppm in the backward angular range, in qualitative agreement with theoretical TPE expectations. 

While the nonzero normal SSAs in elastic electron-nucleon scattering quantify the 
imaginary parts of the TPE amplitudes, measurements of their real parts have also been performed by several dedicated experiments over the past few years. In particular, the deviation from unity of the  elastic scattering cross-section ratio $R_{2 \gamma} \equiv e^+ p / e^- p$ is proportional to the real part of the product of OPE and TPE amplitudes. Recent measurements of $R_{2 \gamma}$, for $Q^2$ up to $2$~GeV$^2$,  
have been performed at VEPP-3~\cite{Rachek:2014fam},  
by the CLAS Coll. at JLab~\cite{Adikaram:2014ykv, Rimal:2016toz}, and by the  
OLYMPUS Coll. at  DESY~\cite{Henderson:2016dea}. These experiments show that $R_{2 \gamma}$ 
ranges,  for the kinematic region corresponding with $Q^2 = 0.5 - 1$~GeV$^2$  
and virtual photon polarization parameter $\epsilon = 0.8 - 0.9$,  
from a value $R_{2 \gamma} \approx 0.99$~\cite{Henderson:2016dea}, showing a deviation from unity within $2 - 3~ \sigma$ (statistical and uncorrelated systematic errors),  
to a value $R_{2 \gamma} = 1.02 - 1.03$ for $Q^2 \approx 1.5$~GeV$^2$  and $\epsilon \approx 0.45$~\cite{Rachek:2014fam, Rimal:2016toz}. 
Furthermore, the  GEp2gamma Coll. at JLab~\cite{Meziane:2010xc} has performed a pioneering measurement of the deviation from the OPE prediction for both  double-polarization components $P_x$ and $P_z$ of the ${\vec e} p \to e \vec p$ process at 
$Q^2 = 2.5$~GeV$^2$. While for $P_x$ the TPE corrections were found to be negligeable, for $P_z$ it has found a deviation from the OPE result at the 4$\sigma $ level at $\epsilon = 0.8$~\cite{Meziane:2010xc}. 
In combination with the unpolarized data, these measurements of the $\epsilon$ dependence of both  double-polarization observables in the ${\vec e} p \to e \vec p$ process at a fixed value of $Q^2$ have been used 
in \cite{Guttmann:2010au}
to provide a first disentanglement of  the three TPE amplitudes describing  elastic  e-p scattering for massless electrons, as given by Eq.~(\ref{eq:TPE}).

While the TPE effects have been shown by experiments to be of the size needed to bring the form factor ratio results from unpolarized measurements closer to those from the recoil polarization experiments, further quantitative studies are needed to reach a conclusive statement, especially in the larger $Q^2$ range. On the theoretical side, various dispersion theoretical approaches have been developed in recent years, 
see \cite{Ahmed:2020uso,Borisyuk:2015xma,Tomalak:2017shs} and older references therein, which relate the TPE amplitudes at intermediate $Q^2$ values to empirical input on the electromagnetic structure of the nucleon and its excitations, while at very large $Q^2$ approaches based on perturbative QCD have been proposed~\cite{Chen:2004tw,Borisyuk:2008db,Kivel:2009eg,Kivel:2012vs}.  
Further experiments investigating the TPE effect at larger values of $Q^2$ will be highly desirable to further test and constrain the TPE model descriptions.  

In low $Q^2$ region, the TPE effect can be predicted with less model dependence~\cite{Hill:2012rh,Tomalak:2016vbf}. Especially in the forward angular range, relevant for the proton electric charge radius determination from elastic e-p scattering, it is found to be understood at the level of precision of current experiments. The TPE effect increases for the backward angular range, where a better understanding is required for improving the extraction of the proton magnetic radius.  

\subsection{Radiative Corrections in Electron Scattering}

Besides the TPE correction corresponding with two hard photons in Fig.~\ref{TPEdiagram}, another important aspect associated with lepton scattering, especially with electron scattering is the so-called radiative correction (RC) effect to the OPE picture. RC refers to effects from various types of radiation and soft-photon exchanges in electron scattering which need to be corrected before one can extract information such as the proton electric and magnetic form factors defined in the OPE picture.  A few examples can be the initial state electron radiates a photon prior to the scattering, or the final state electron radiates a photon before it is detected in the detector. Similar pictures can be applied to the proton side, though such radiative effects are suppressed because the proton mass is significantly larger than that of an electron. A different way to look at the proton side is that such RC effect in principle can be included in the definition of the proton electric and magnetic form factors.  Another important RC contribution is due to the QED vacuum polarization, which refers to the fact that a virtual photon can fluctuate into an electron-positron pair before they are absorbed, and the vertex correction on electron and proton sides. Furthermore, the radiative corrections conventionally also include a part of the TPE correction, in which one of the photons in the box diagram of Fig.~\ref{TPEdiagram} has a soft four-momentum. These are just examples of leading-order RC contributions, which are at the next-to-leading order compared to the leading-order OPE in electron-proton scattering. 
Two classic review articles on this subject still widely used and cited are by Mo and Tsai~\cite{Mo69} and by Maximon~\cite{Maximon69}. In recent years, there have been renewed interests in performing and pushing the state-of-the-art calculations on RC for various lepton-nucleon scattering processes not only due to the demand from the experimental side to improve precision, but also due to the need for other processes such as semi-inclusive deep-inelastic-scattering to probe partonic three-dimensional momentum distributions and fragmentation functions. The effect of RC is also experiment specific for which, we refer readers to specific experiments that are discussed in this review for further details.  

\subsection{The Extraction of the Proton Charge Radius from Proton Electric Form Factor}

The proton charge radius can be extracted from the experimentally determined proton electric form factor values. According to Eq.~(\ref{eq:rad9p}), the proton rms charge radius is directly related to the $G_{Ep}$ $Q^2$-slope at $Q^2=0$. Experimentally this is of course not possible due to the requirement of conducting electron-proton elastic scattering at zero-degree scattering angle. Therefore, while it is important to reach as low a $Q^2$ value as possible, it is inevitable that one needs to extrapolate from the measured values of $Q^2$ down to zero. Furthermore, it is also important for any scattering experiment to cover a sufficient range of $Q^2$, i.e. to have a good leverage in $Q^2$ coverage.  When $Q^2$ is sufficiently close to zero, the slope becomes rather flat because $G_E$ would converge to 1, which is just the net charge of the proton as expected. Therefore, it is important to experimentally cover a $Q^2$ range in which one can capture whatever a $Q^2$ dependence nature calls for, and at the same time still be as close to $Q^2=0$ as practically possible. 

Given the aforementioned limitations, it is important to develop ways that allow for an extraction of the proton charge radius in a robust way. Such a study was carried out by Yan {\it et al.}~\cite{Yan18}. Below we briefly describe this study. Pseudo-data sets on the proton electric form factor are generated for a particular experiment or a planned measurement according to various proton electromagnetic form factor parametrizations/models in the literature.  These parametrizations/models in general describe the existing data on the proton form factors well.  One then smears the generated pseudo data sets according to the experimental resolutions, and any other relevant experimental aspects such as the statistical and systematic uncertainties. The way to take into account the experimental systematic uncertainties is quite elaborate and we refer interested readers to the original paper~\cite{Yan18} for more details. 
One then fits the smeared data sets to various functional forms and extracts for each functional form the corresponding proton charge radius value, $r_{Ep}$, and its uncertainty, $\delta r_{Ep}$. The bias is defined as the difference between the input $r_{Ep}$ value from the parameterization/model used to generate the pseudo-data set in the first place, and the $r_{Ep}$ obtained from the fit. The goodness of a fit is to consider both the bias and the variance from the fit by using the root-mean-square error (RMSE) defined as $RMSE = \sqrt{bias^2+\sigma^2}$. 

The functional forms studied by Yan {\it et al.}~\cite{Yan18} include monopole, dipole,  Gaussian, multi-parameter polynomial expansion of $Q^2$, multi-parameter rational function of $Q^2$, continuous fractional (CF) expansion of $Q^2$, and also the multi-parameter polynomial expansion of $z$, defined as:
\begin{equation}
z = {\frac{\sqrt{t_{cut}+Q^2} - \sqrt{t_{cut}-t_0}}{\sqrt{t_{cut}+Q^2} + \sqrt{t_{cut}-t_0}}},
\label{eq:zdef}
\end{equation}
where $t_{cut} = 4 m^2_\pi$ corresponds to the threshold for the lowest $2\pi$ intermediate state in the timelike region, with $m_\pi$ being the mass of $\pi^0$, and $t_0$ is a free parameter set to zero in ~\cite{Yan18}. So the full functional form is expressed as:
\begin{equation}
f_{polyz}(Q^2) = p_{0} G_{E}(Q^2) = p_{0} ( 1 + \sum_{i=1}^{N} {p_{i}z^{i}}).
\label{eq:polyz}
\end{equation}  

The CF expansion form is expressed as: 
\begin{equation}
f_{CF}(Q^2) = p_{0} G_{E}(Q^2) = p_{0} {\frac{1}{1+{\frac{p_{1}Q^2}{1+{\frac{p_{2}Q^2}{1+...}}}}}}.
\label{eq:cf}
\end{equation}  

The multi-parameter rational function of $Q^2$ is written as:
\begin{equation}
f_{rational}(Q^2) = p_{0} G_{E}(Q^2) = p_{0} \frac{1+\sum_{i=1}^{N}{p^{(a)}_{i}Q^{2i}}}{1+\sum_{j=1}^{M}{p^{(b)}_{j}Q^{2j}}}.
\label{eq:rat}
\end{equation}

In all these functional forms of Eqs.~(\ref{eq:polyz},\ref{eq:cf},\ref{eq:rat}), the $p_0$ is a floating normalization parameter. For the PRad experiment~\cite{Xiong19} in its entire data range, the study found that the $(N=M=1) = (1, 1)$ rational function, the two-parameter continued fraction, and the second-order polynomial expansion in $z$ can all extract the proton charge radius in a robust way with small variance independent of the model or parameterization used for generating the pseudo data. The published $r_{Ep}$ result~\cite{Xiong19} from the PRad experiment is based on fits to rational (1,1) function.  While in \cite{Yan18} the case study was presented for the PRad experiment, the approach can be applied to any lepton scattering experiment to extract the proton charge radius. 

\vspace{0.2in}
\section{Atomic Hydrogen Spectroscopy}
\label{sec:atspec}

The proton charge radius is an important input to QED calculations of bound states such as ordinary atomic hydrogen and muonic hydrogen. High precision spectroscopic measurements, combined with the state-of-the-art QED calculations, can determine the proton charge radius. In this section, we provide a brief discussion
and focus on aspects most relevant to the finite size of the proton due to our interest in the determination of the proton charge radius. We follow closely the review paper by Eides, Grotch and Shelyuto~\cite{Eides01}, 
to which we refer for a comprehensive discussion of the QED calculations including various higher-order effects.

The energy levels for one-lepton atoms can be obtained in the first approximation by solving the non-relativistic Schr\"{o}dinger equation for an electron in the field of an infinitely heavy Coulomb center with a charge $Z$ in units of the proton charge. The  energy levels are written as:
\begin{equation}
E_n = -{\frac{m(Z \alpha)^2}{2n^2}}, 
\end{equation}
where $n = 1, 2, 3, ...$ is the principal quantum number,  $\alpha$ the fine structure constant, and $m$ is the mass of the lepton.  
Considering the Coulomb source still to be infinitely heavy, solving the Dirac equation for a lepton in such a Coulomb field, one obtains the following Dirac spectrum:
\begin{equation}
E_{nj} = m f(n,j),
\end{equation}
where
\begin{widetext}
\begin{eqnarray}
f(n,j) &=& \left [ 1 +{\frac{(Z\alpha)^2}{\left({\sqrt{(j+{\frac{1}{2}})^2 - (Z\alpha)^2} +n - j -{\frac{1}{2}} }\right)^2}} \right ]^{-1/2} \\ \nonumber
& \approx & 1 - {\frac{(Z\alpha)^2}{2n^2}} - {\frac{(Z\alpha)^4}{2n^3}} \left ( {\frac{1}{j+1/2}} - {\frac{3}{4n}}   \right ) -{\frac{(Z\alpha)^6}{8n^3}} \left [ {\frac{1}{(j+ 1/2)^3}} + {\frac{3}{n(j+1/2)^2}} +{\frac{5}{2n^3}} - {\frac{6}{n^2(j+1/2)}}  \right ] + ..., 
\end{eqnarray}
 \end{widetext}
 where $j=1/2, 3/2,....n-1/2$ is the total angular momentum of the state.  Compared with the nonrelativistic Schr\"{o}dinger spectrum -- where all levels with the same $n$ are degenerate -- the energy levels in the Dirac spectrum with the same principal quantum number $n$ but different $j$ are no longer degenerate. However,  energy levels with the same $n$ and $j$, but different $l= j \pm 1/2$ remain degenerate. Such degeneracy is lifted when one takes into account the finite size of the proton, recoil contributions and most importantly the QED loop corrections, and the corresponding energy shifts are called the Lamb shifts.  
 Details on calculating the QED radiative corrections, recoil and radiative-recoil corrections can be found in~\cite{Eides01}.

Below we briefly review the leading relativistic corrections with exact mass dependence in the external field approximation following~\cite{Eides01}.
For a non-relativistic system of two particles with Coulomb interaction such as a hydrogen atom, the Hamiltonian in its center-of-mass system can be written as:
\begin{equation}
H_{0}={\frac{{\bf p}^2}{2m}} + {\frac{{\bf p}^2}{2M}} -{\frac{Z\alpha}{r}},
\end{equation}
where ${\bf p}$ is the momentum, and in the case of hydrogen (muonic hydrogen), $Z=1$, and where $m$ and $M$ are the masses of the electron (muon), and the proton, respectively. In the remaining of this section, we will focus on hydrogenlike atoms only. 
For a nonrelativistic loosely bound system such as a hydrogen atom, expansions over $\alpha^2$ correspond to expansions over $v^2/c^2$. Therefore, an effective Hamiltonian including terms of the first order in  $v^2/c^2$ would provide proper corrections of relative order $\alpha^2$ to the nonrelativistic energy levels.  Breit~\cite{Breit29,Breit30,Breit32} proposed such a potential realizing that all corrections to the nonrelativistic two-particle Hamiltonian of the first order in $v^2/c^2$ can be written as the sum of the free relativistic Hamiltonian of each of the particles and the relativistic one-photon exchange between the two.  
Barker and Glover~\cite{Barker55} derived the following Breit potential from the one-photon-exchange amplitude using the Foldy-Wouthuysen transformation~\cite{Foldy50}:
\begin{eqnarray}
V_{Br}  &=& {\frac{\pi \alpha}{2}}\left( {\frac{1}{m^2}}+{\frac{1}{M^2}} \right) \delta^{3}({\bf r}) \nonumber \\
&-& {\frac{\alpha}{2mMr}}\left( {\bf p}^2+{\frac{{\bf r}({\bf r}\cdot {\bf p})
\cdot {\bf p}}{r^2}} \right) \nonumber \\
 &+& {\frac{\alpha}{r^3}}\left({\frac{1}{4m^2}}+{\frac{1}{2mM}}\right)[{\bf r} \times {\bf p}] \cdot {\vec{\sigma}}.
\end{eqnarray}
In the above potential, the hyperfine structure is not considered, i.e., terms which depend on the proton spin are omitted.  The corrections to the energy levels up to order of $\alpha^4$ can be calculated from the total Breit Hamiltonian of $H_{Br}=H_0 + V_{Br}$, where the interaction potential is the sum of the Coulomb and the Breit Potential. These corrections are just the first-order matrix elements of the Breit interaction between the eigenfunctions of the Coulomb Hamiltonian $H_0$, and the result is
\begin{widetext}
\begin{equation}
E^{tot}_{nj} = (m+M) - {\frac{m_r \alpha^2}{2 n^2}} - {\frac{m_r \alpha^4}{2n^3}} \left( {\frac{1}{j+1/2}} - {\frac{3}{4n}} + {\frac{m_r}{4n(m+M)}} \right)  + {\frac{\alpha^4 m_{r}^3}{2n^3 M^2}} \left( {\frac{1}{j+1/2}} - {\frac{1}{l+1/2}} \right) (1-\delta_{l0}), 
\label{eq:breit}
\end{equation}
\end{widetext}
where $m_r = mM/(m+M)$ is the reduced mass of the hydrogenlike atom. One can see that the last term in Eq.~(\ref{eq:breit}) breaks the degeneracy in the Dirac spectrum between states with the same $j$ and $l=j\pm 1/2$, and contributes to the classical Lamb shift defined as $E(2P_{1/2}) - E(2S_{1/2})$. However, due to the smallness of the electron to proton mass ratio, the contribution of this term is extremely small in the hydrogen case and the leading contribution to the Lamb shift is the QED radiative correction.

In the discussion so far, the proton has been treated as a point-like charge.  Eq.~(\ref{eq:pvertex})  provides the photon-proton vertex operator involving the Dirac ($F_1$) and Pauli ($F_2$) form factors of the proton.  
To calculate the finite size contribution to the hydrogen atom energy levels, amounts to evaluate the zero component of Eq.~(\ref{eq:pvertex}) between nucleon spinors, normalized as $N^\dagger N = 1$. An elementary calculation yields the spin independent term at low momentum transfer ${\bf q} \equiv {\bf p}' - {\bf p}$ ~\cite{Eides01}, see e.g. \cite{Miller19} for an explicit derivation, as:  
\begin{eqnarray}
N(p',\lambda) \Gamma^0 N(p, \lambda) 
&=& \left(1 - \frac{{\bf q}^2}{8 M^2} \right) G_E(- {\bf q}^2) + {\cal O}\left( \frac{1}{M^4}\right)\nonumber \\
&\approx& 1 - {\bf q}^2 \left[ {\frac{1}{8M^2}} + \frac{1}{6} \langle r_{Ep}^2 \rangle  \right],
\label{eq:atomsize}
\end{eqnarray}
where in the last line we have used the low-momentum expansion of the proton electric form factor $G_E$ in terms of the proton charge radius $\langle r_{Ep}^2 \rangle$, defined through Eq.~(\ref{eq:rad9p}).  
For a point-like proton, the only term that survives in Eq.~(\ref{eq:atomsize}) is the first term in the square brackets, which leads to the well-known local Darwin term in the lepton-proton interaction~\cite{Barker55} that gives rise to the term proportional to $\delta_{l0}$ in Eq.~(\ref{eq:breit}). 
Note that the leading relativistic correction factor in front of $G_E$ in Eq.~(\ref{eq:atomsize}) is the same 
as the one appearing in Eq.~(\ref{eq:dens3dr}). The established convention is not to include it in the definition of $G_E$, but to include it separately. Therefore, the leading nuclear (proton) structure contribution to the energy shift is determined by the slope of the conventionally defined nuclear (proton) form factor $G_E$. 
The corresponding perturbative potential which corrects the Coulomb potential of a point charge to account for the finite proton size is therefore given by~\cite{Eides01} 
\begin{equation}
\delta V_{\rm fin. size} = {\frac{2\pi \alpha} {3}} \langle r_{Ep}^2 \rangle. 
\end{equation}
The associated energy level shift is then
\begin{eqnarray}
\Delta E_{\rm fin. size} &=& {\frac{2\pi \alpha}{3}}
\langle r_{Ep}^2 \rangle |{\psi}_{nl}(0)|^2 ,
\nonumber \\
&=& {\frac{2 {\alpha}^4}{3n^3}} m^3_r 
\langle r_{Ep}^2 \rangle \delta_{l0}.
\label{eq:finsize}
\end{eqnarray}
One notices from Eq.~(\ref{eq:finsize}) that  the radius entering the finite size correction to the $S$-levels of the hydrogen atom is the proton charge radius, obtained from the form factor $G_E$ as  measured in electron-proton scattering experiments. This consistency between the proton charge radius determined from spectroscopic experiments of hydrogenlike atoms and from electron scattering experiments has also been emphasized  recently in~\cite{Miller19}.  

While the Lamb shift of hydrogenlike atoms is dominated by the QED radiative effects of the lepton, the contribution from the proton charge radius is the leading term due to the finite size of the proton. By measuring Lamb shifts or other transitions between energy levels involving at least one $S$-state of hydrogenlike atoms precisely and utilizing the state-of-the-art QED calculations, one can determine the proton charge radius value. 
In the case of muonic hydrogen, the proton charge radius effect is $6.4 \times 10^6$ times larger compared to that of ordinary hydrogen atoms for the same $nS$ level, due to the $m_r^3$ dependence. For the $2P-2S$ Lamb shift in muonic hydrogen, the term due to the proton charge radius amounts to around -3.7~meV, and contributes to about 2\% of the overall Lamb shift~\cite{Eides01}. This large relative contribution is the important reason why muonic hydrogen spectroscopic measurements are significantly more precise in extracting the proton charge radius than those from ordinary hydrogen atoms. 

\begin{figure}
\includegraphics[width = 0.35\textwidth]{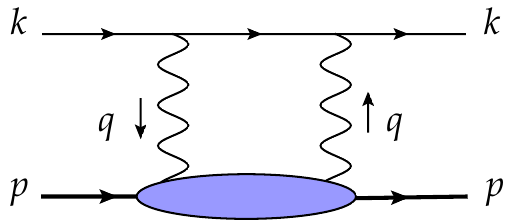}
\caption[fig]{\label{fig:lambbox} (Color online) The box diagram for the $\mathcal O(\alpha^5)$ corrections to $l = 0$ energy levels in muonic hydrogen. The blob denotes all possible hadronic intermediate states.   
}
\end{figure}

In order to extract the proton radius from muonic hydrogen spectroscopic measurements accurately, it is important to also calculate the proton structure corrections of next order in $\alpha$, i.e. 
$\mathcal O(\alpha^5)$. These proton structure corrections, which arise from 
the two-photon exchange (TPE) diagram shown in Fig.~\ref{fig:lambbox}, in which both photons in the loop carry the same four-momentum, are  known as the polarizability correction.
They have been evaluated using different approaches: chiral effective field theory, see~\cite{Hagelstein:2015egb} and references therein for a review of the ongoing activity in this field; within non-relativistic QED~\cite{Pineda:2002as,PhysRevLett.107.160402,Hill:2012rh, Dye:2016uep}; or by connecting them model-independently to other data through dispersive frameworks~\cite{PhysRevA.60.3593,PhysRevA.84.020102,Birse12}. 
Using dispersion relations, with input from forward proton structure functions and a subtraction function, this $\mathcal O(\alpha^5)$ proton structure correction to the $2P - 2S$ Lamb shift in muonic hydrogen has been estimated as~\cite{Antognini:2013rsa,PhysRevA.84.020102,Birse12}:
\begin{equation}
    \Delta E_{TPE}(2P - 2S) = 0.0332 (20) \; {\rm meV}.
    \label{eq:tpeLS}
    \end{equation}
This value for the TPE correction is presently used in the extraction of the proton charge radius from the muonic hydrogen Lamb shift measurements as discussed in Section~\ref{sec:spec}.

\section{Modern lepton scattering experiments}
\label{sec:scatt}

{\subsection{Mainz 2010}
\label{Mainz2010}}

Bernauer {\it et al.}~\cite{Bernauer10} carried out a unpolarized electron-proton elastic scattering experiment at the Mainz accelerator facility MAMI and extracted the proton charge and the magnetic radii. The experiment utilized electron beam energies up to 855 (180, 315, 450, 585, 720 and 855) MeV and three high-resolution magnetic spectrometers with one serving as a relative luminosity monitor at a fixed laboratory angle. The other two spectrometers were moved as a function of electron scattering angle during the experiment to provide the kinematic coverage and also redundancy in the coverage. In total the experiment measured over 1400 differential cross sections covering a $Q^2$ range of 0.004 to 1 (GeV/c)$^2$  and achieved a statistical precision better than 0.2\% for these cross section measurements.  To extract the proton electric and magnetic form factors,  least square fits to models of $G_{Ep}$ and $G_{Mp}$ were carried out to the 1400 cross section data points covering all $Q^2$ and scattering angles of the experiment. The proton form factors up to $Q^2 =0.6$ (GeV/c)$^2$ were extracted from this approach. The authors  carried out detailed studies of model dependence in extracting the proton form factors using various form factor models and parameterizations. The experiment extracted the following for the proton charge and magnetic radii:
\begin{eqnarray}
\langle r^2_{Ep} \rangle^{1/2} &=& 0.879(5)_{stat}(4)_{syst}(2)
_{model}(4)_{group} \ {\rm fm}, \nonumber \\
\langle r^2_{Mp} \rangle^{1/2} &=& 0.777(13)_{stat}(9)_{syst}(5)
_{model}(2)_{group}\  {\rm fm}, \nonumber 
\end{eqnarray}
where the uncertainty labeled as ``group" is assigned to account for the difference between the radius values obtained using two groups of models for the form factors in the fits, namely the spline and the polynomial groups. Details can be found in ~\cite{Bernauer10,Bernauer:2013tpr}. The result on the proton charge radius from this electron scattering experiment was consistent with the CODATA06~\cite{Mohr08} value at the time of the publication, but 5 standard deviations larger than the value from the muonic hydrogen Lamb shift measurement~\cite{Pohl10}.  The magnetic radius obtained is smaller than those from previous fits of electron scattering data, but consistent with the result of 0.778(29)~\cite{volotka05} fm from hyperfine splitting in hydrogen.

\subsection{JLab recoil polarization experiment}

The Jefferson Lab experiment E08-007~\cite{Zhan11} carried out a high-precision measurement of the polarization transfer from electron-proton elastic scattering using a recoil proton polarimeter covering a momentum transfer squared $Q^2$ region between 0.3 to 0.7 (GeV/c)$^2$. The experiment was performed in Hall A and utilized a longitudinally polarized electron beam with polarization higher than 80\% at 1.2 GeV, beam currents between 4 and 15 $\mu$-A, and a 6-cm long unpolarized liquid hydrogen target. There are two high-resolution magnetic spectrometers (HRS) in Hall A placed on each side of the electron beam line.  In E08-007, the recoil proton was detected in the left HRS with its polarization being measured by a focal plane polarimeter, in coincidence with the scattered electron which was measured in a large acceptance spectrometer (``BigBite"). The experiment extracted the proton electric to magnetic form factor ratio $\mu_{p}G_{Ep}/G_{Mp}$ with a total uncertainty of about 1\%. Using these results together with a few other proton form factor ratio measurements from Jefferson Lab~\cite{Puckett:2010ac,Ron11,Paolone10}, a global fit of the proton form factors~\cite{Arrington07} was updated. This updated global analysis did not include the Mainz data~\cite{Bernauer10}, and gave the following values for the proton electric and the magnetic charge radii:
\begin{eqnarray}
\langle r^2_{Ep} \rangle^{1/2} &=& 0.875 \pm 0.010\ {\rm fm}, \nonumber \\
\langle r^2_{Mp} \rangle^{1/2} &=& 0.867 \pm 0.020 \  {\rm fm} \nonumber. 
\end{eqnarray}

The proton charge radius value from this updated global analysis is in excellent agreement with the value from the Mainz electron-proton scattering experiment~\cite{Bernauer10}, and also with the CODATA 2006 value~\cite{Mohr08} which is based mostly from ordinary hydrogen spectroscopic measurements. It is in disagreement with the muonic hydrogen result~\cite{Pohl10}. The magnetic radius value from this global analysis is more than 5 standard deviations (larger) away from the Mainz value~\cite{Bernauer10}.

\subsection{Mainz ISR measurements}

\begin{figure}
\includegraphics[angle=-90,scale=0.23]{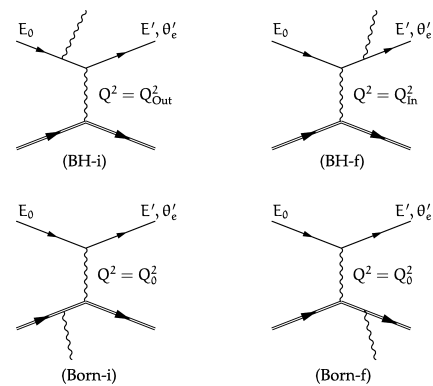}
\caption[fig]{\label{ISI} (Color online) Feynman diagrams showing electron-proton scattering with electron or proton radiates a real photon in the initial state or final state. In the electron case, the two diagrams are labeled as Bethe-Heitler (BH-i) and (BH-f), while for the proton: (Born-i) and (Born-f), where i and f stand for the initial-state and final-state radiation, respectively. 
 The figure is from ~\cite{MIHOVILOVIC2017194}.} 
\end{figure}

Following the Mainz experiment by Bernauer {\it et al.}~\cite{Bernauer10}, another electron-proton elastic scattering experiment at Mainz was carried out using the same three-spectrometer setup but reached lower values of $Q^2$ (0.001 to 0.004 GeV/c$^2$) using the technique of initial-state radiation (ISR)~\cite{MIHOVILOVIC2017194}. For electron-scattering experiments, the lowest $Q^2$ value that is achievable is determined by the lowest electron beam energy the associated accelerator can deliver, and the most forward electron scattering angle the corresponding detector can reach. The ISR technique is adopted to overcome such limits by utilizing the information within the radiative tail of the elastic peak.  The technique works the following as depicted by Fig.~\ref{ISI}. The incoming electron can radiate a real photon before the scattering takes place, as such the corresponding $Q^2$ value for the e-p scattering would be lower than what's limited by the accelerator and the detector because the incident electron energy is lower than its original value delivered by the accelerator and before the initial state radiation of the real photon by the incoming electron. This is the diagram labeled as Bethe-Heitler (BH-i) in Fig.~\ref{ISI}. Such an ISR technique was proposed and used successfully in particle physics experiments previously~\cite{Arbuzov1998,PhysRevD.69.011103}. One of the challenges of such an ISR experiment is to separate the contribution from the diagram labeled as (BH-f), where the scattered electron radiates a real photon as only scattered electrons are measured (inclusive measurement). Furthermore, although contributions from diagrams involving the proton initial-state and final-state radiation are suppressed due to the proton mass, they need to be included as they also contribute to the radiative tail of the elastic scattering, as well as higher order radiative effects. For details on how to account for these effects, see~\cite{MIHOVILOVIC2017194}, which extracted a proton charge radius value of  $\langle r^2_{Ep} \rangle^{1/2}  = 0.810 \pm 0.035_{stat.} \pm 0.074_{syst.} \pm 0.003_{mod.}$ fm, with the last uncertainty accounting for higher moments in parameterizing the proton electric form factor. The collaboration reported a follow-up result through a comprehensive reinterpretation of the existing cross section data from this first ISR e-p scattering experiment by improving the description of the radiative tail. They obtained  $\langle r^2_{Ep} \rangle^{1/2}  = 0.870 \pm 0.014_{stat.} \pm 0.024_{syst.} \pm 0.003_{mod.}$ fm with major improvements in both the statistical and systematic uncertainties, see~\cite{mihovilovi2019proton}.

\subsection{The PRad experiment at JLab}

The proton charge radius (PRad) experiment~\cite{Xiong19} at Jefferson Lab was designed with a number of important points in mind: (i) an experiment that is different from previous e-p scattering experiments, therefore having different systematics; (ii) the reach of unprecedentedly low values of $Q^2$; (iii) the ability to measure precisely e-p elastic scattering cross sections by accurate control of the integrated luminosity; (iv) minimizing changes during the experiment and taking all the data using a fixed experimental apparatus.       

The PRad experiment innovated electron-scattering mesurements in the following ways. Instead of using a magnetic spectrometer, which usually limits the forward most scattering angles due to its physical size, the PRad experiment uses a two-dimensional large-area, granular, high-resolution electromagnetic calorimeter with a hole at the center for the electron beam to pass through. As such it allows the access to significantly smaller scattering angles, e.g. $\sim 0.7^\circ$,  compared with experiments using magnetic spectrometers.  To overcome major background issues associated with small-angle scattering,  the PRad experiment uses a windowless, cryogenically cooled, hydrogen gas flowing target, i.e. an internal target for the first time at Jefferson Lab, an external beam facility so that the electron beam would not see any target window. 
In order to have an excellent control of the integrated luminosity for the electron-proton elastic scattering cross section measurements, M{\o}ller scattering, a well-known QED process, is used as a reference process and is measured simultaneously during the e-p scattering. Lastly, to improve the scattering angle ($Q^2$) determination, a large plane of Gas Electron Multiplier (GEM) detectors is used. The GEM detector used in PRad was the largest ever used in any experiment at the time.

\begin{center}
\begin{figure}[h]
\includegraphics[angle=0,scale=0.4]{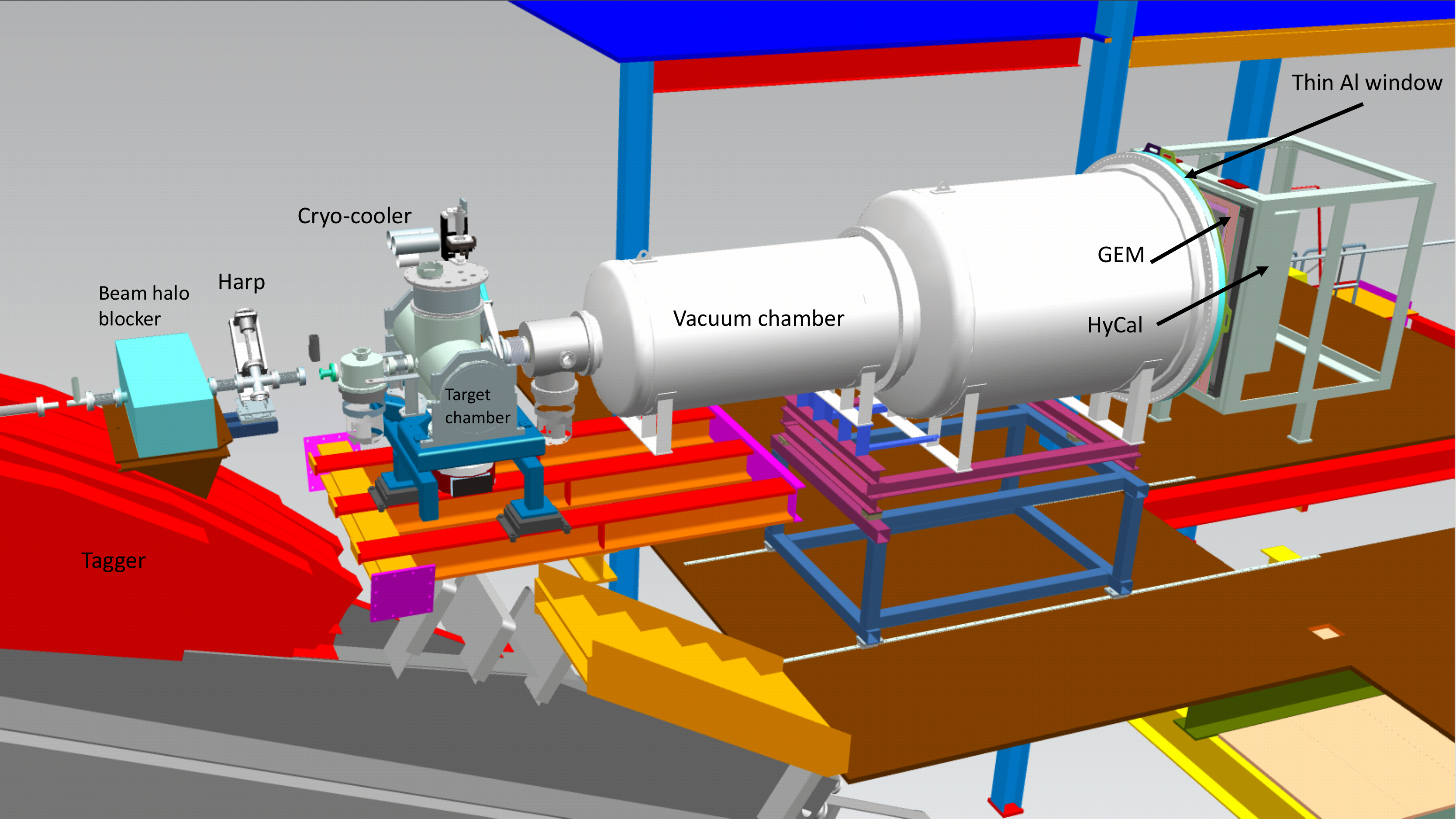}
\caption[fig]{\label{PRad-setup} (Color online) The schematics of the PRad experiment in Hall B at Jefferson Lab. In this figure, the electron beam is from left to right (figure credit: Eugene  Pasyuk and others~\cite{PRad-figure}). }
\end{figure}
\end{center}

The schematics of the PRad experiment is shown in Fig.~\ref{PRad-setup}, in which the electron beam is from left to right. PRad was the first experiment to complete its data taking in June 2016 after the Continuous Electron Beam Accelerator Facility (CEBAF) -- consisting of a polarized electron source, an injector and a pair of superconducting
radio frequency (RF) linear accelerators --  energy upgrade from 6 GeV to 12 GeV at Jefferson Lab was completed. Two values of electron beam energies were used in the PRad experiment, 1.1 and 2.143 GeV. For the 1.1 GeV data set, most of the data were obtained at a beam current of 15 nA with the rest at 10 nA, while for the 2.143 GeV data, the nominal beam current was 55 nA.

The cryogenically cooled hydrogen gas at a temperature of about 20K and a flow rate of 600 sccm was flowing into the target cell through an inlet at its midpoint. 
The target cell was machined from a single block of C101 copper with outer dimensions of 7.5 cm by 7.5 cm by 4.0 cm, and two 6.3 cm diameter holes along the axis of the beam line. The holes were covered at both ends by 7.5 $\mu$m thick kapton foils, held in place by aluminum end caps with 2-mm holes at the center of each kapton foil to allow the electron beam to pass through and also the gas to flow out from both ends. 
As such, it is different from a typical storage cell used in internal gas  targets~\cite{Steffens_2003,PhysRevA.73.020703}, which is an open-ended one-piece cylinder with circular or elliptical shape made of aluminum or copper. 
The target areal density for the PRad experiment is $\sim 2 \times 10^{18}$ atoms/cm$^2$.  The target cell was housed inside the target chamber, which was being pumped by four turbo molecular pumps -- two with pumping speed each of 3000 l/s mounted directly under the target chamber, and the remaining two each with a pumping speed of 1400 l/s mounted upstream and  downstream of the target chamber, respectively during the experiment -- in order to maintain the required beam-line vacuum $\sim 10^{-5}$ Torr, upstream of the target chamber. During the PRad experiment, the pressure inside the target cell was about 0.5 Torr, and $\sim$ 2.3 mTorr inside the target chamber.  There is a gate valve separating the downstream turbo pump from the target chamber. In addition to flowing hydrogen gas into the target cell for production data taking, additional data were taken during the experiment for background subtraction such as in the so-called ``empty'' target configuration with hydrogen flowing into the target chamber through a side inlet, while the gas inlet to the target cell being valved off. This configuration was to emulate the background gas inside the target chamber during the production data taking.  Additional background related data with other configurations were also taken in order to understand possible background sources located upstream of the target cell. More details can be found in~\cite{Xiong_thesis,PRad-target-nim}. 

\onecolumngrid
\twocolumngrid

To minimize the background from the air, the scattered electrons travel through a 5-meter long, two-stage vacuum chamber constructed specifically for the PRad experiment with the downstream end sealed by a large-area chamber window that is made of aluminum with a thickness of 1.6 mm. The electromagnetic calorimeter used in the PRad experiment is a hybrid calorimeter (HyCal) consisting of 1156 PWO$_4$ inner crystal modules augmented by 576 lead glass modules. The HyCal was built for experiments to carry out precision measurements of the neutral pion radiative decay width~\cite{PhysRevLett.106.162303,Larin506}. For the PRad experiment, it provides a scattering angular coverage for electron-proton scattering from about 0.7$^\circ$ to 7.5$^\circ$, which corresponds to a $Q^2$ range of $2 \times 10^{-4}$ to 0.06 (GeV/c)$^2$. However, to reach the resolution of $Q^2$ required for the PRad experimental precision, a large plane consisting of two large-area Gas Electron Multiplier (GEM) detectors with a small overlap region in the middle with a hole for the beam to pass through is installed in front of the HyCal. The position resolution of the GEM detector is 72 $\mu$m, which represents more than a factor of 20 improvement if only HyCal were used. Upstream of the PRad target chamber, there is the Hall B photon tagger~\cite{SOBER2000263} which was used before the experiment production data taking for calibrating the HyCal.  Additional details can be found in ~\cite{Xiong_thesis}.  

\begin{figure}
\includegraphics[scale=0.44]{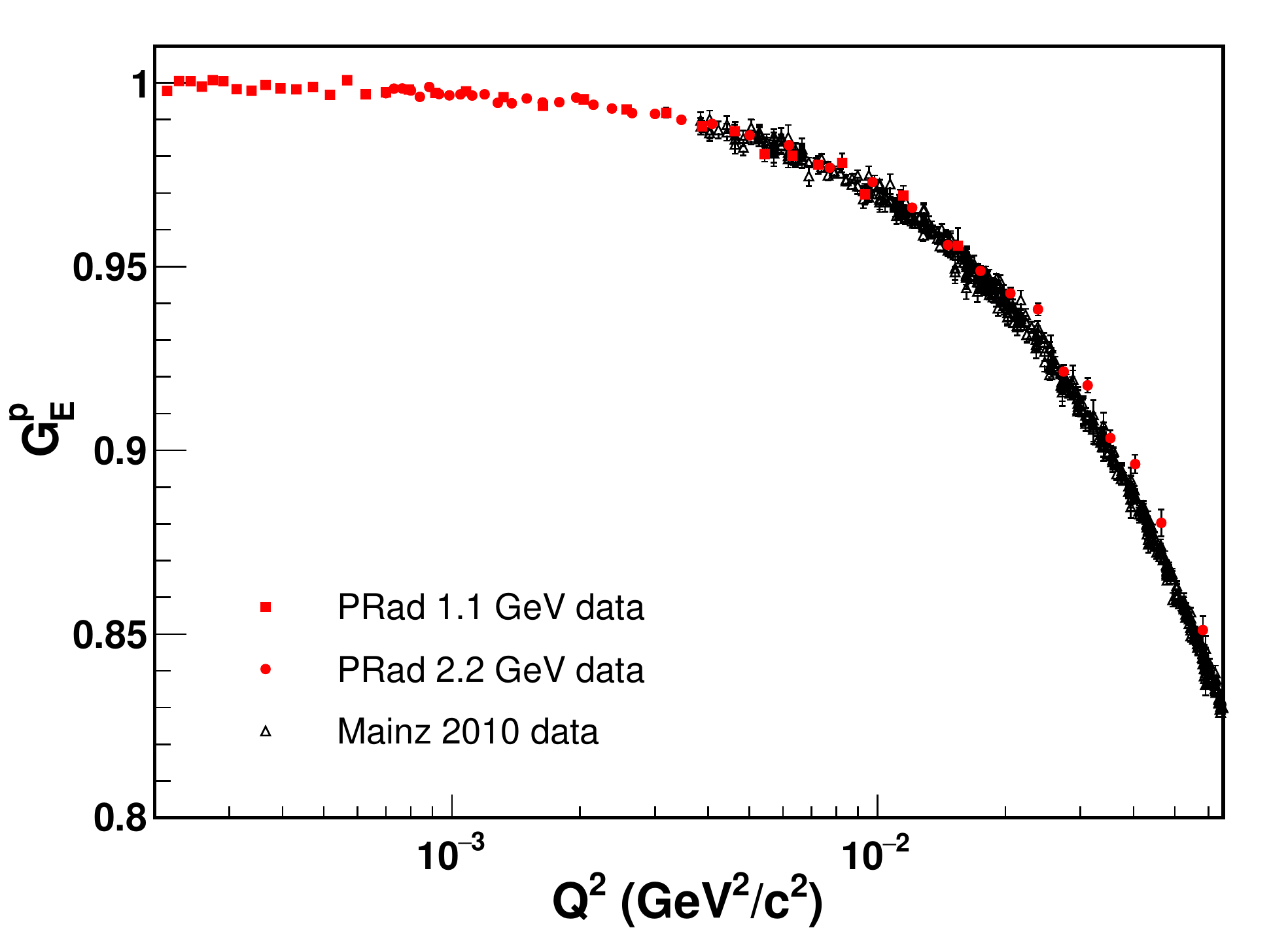}
\caption[fig]{\label{fig:PRadGEp} (Color online) The proton electric form factor $G^p_E$ from the PRad experiment together with those from the Mainz experiment~\cite{Bernauer10} in the overlap $Q^2$ region (figure credit: Weizhi Xiong). } 
\end{figure}

\begin{figure}
\includegraphics[scale=0.44]{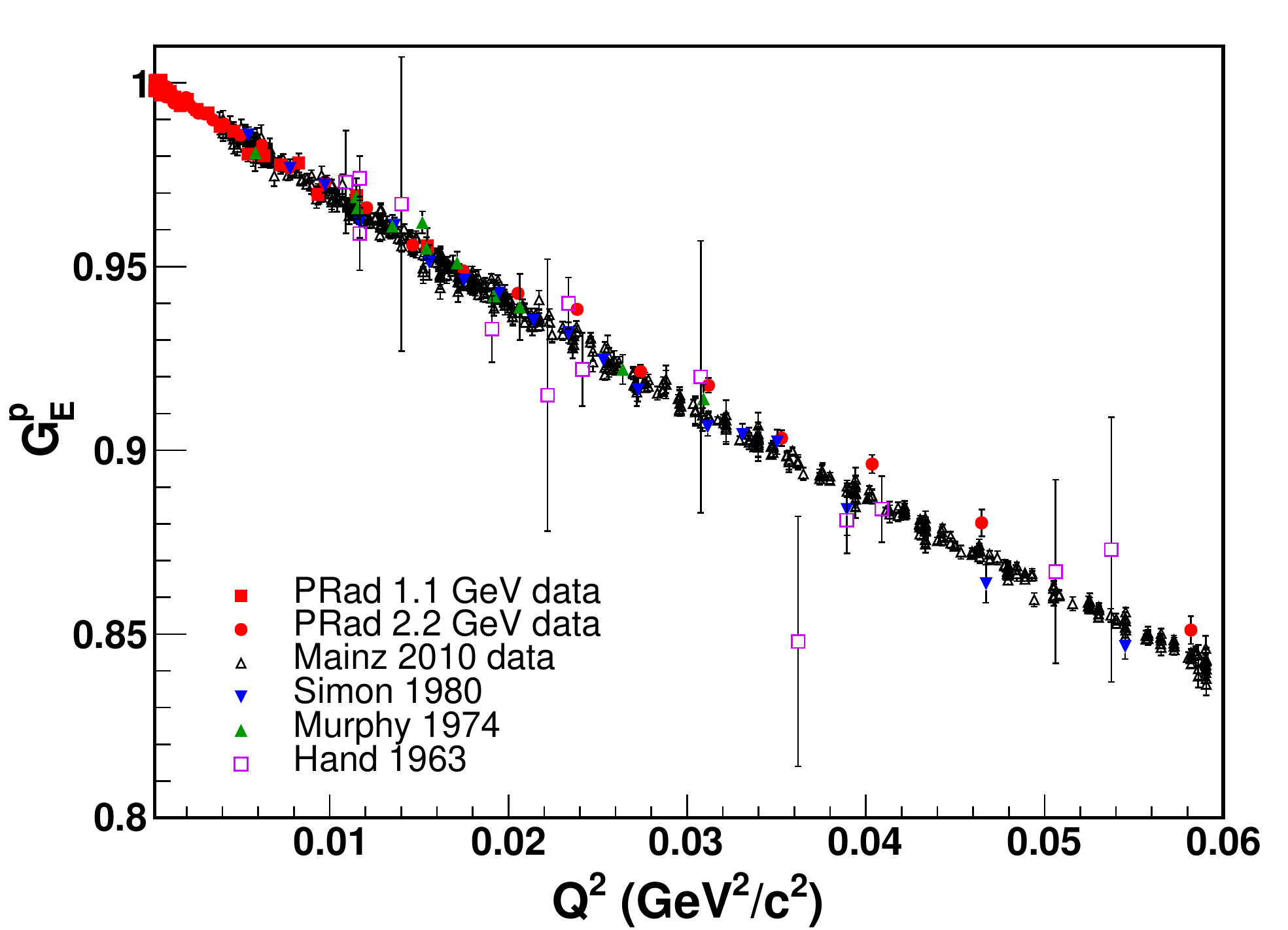}
\caption[fig]{\label{GEp-all} (Color online) The proton electric form factor $G^p_E$ from the PRad experiment together with those from ~\cite{Bernauer10,Hand63,Murphy:1974zz,Simon80} in the overlap $Q^2$ region, on linear scale (figure credit: Weizhi Xiong).} 
\end{figure}

The proton electric form factor values in the $Q^2$ range of $2 \times 10^{-4}$ to 0.06 (GeV/c)$^2$ have been extracted from the PRad experiment and they are presented in Fig~\ref{fig:PRadGEp} with statistical uncertainties only.  The systematic uncertainties range from $\sim 0.1\%$ to 0.6\% (relative) for the entire PRad data set~\cite{Xiong_thesis}.  The Mainz data~\cite{Bernauer10} shown are from ~\cite{Griffioen16} in which statistical, and point-to-point systematic uncertainties -- with an additional 15\% inflation of point-to-point systematic uncertainty -- have been included. In Fig.~\ref{GEp-all}, additional $G^p_E$ data from~\cite{Hand63,Murphy:1974zz,Simon80} are also shown. Other than the data by ~\cite{Hand63} which has rather larger uncertainties, the PRad results are systematically higher than other data on $G^p_E$ in the higher end of the $Q^2$ range covered by the PRad experiment,  $\sim$ 0.03 (GeV/c)$^2$ and higher. 

\begin{center}
\begin{figure}[h]
\centering
\includegraphics[scale=0.33]{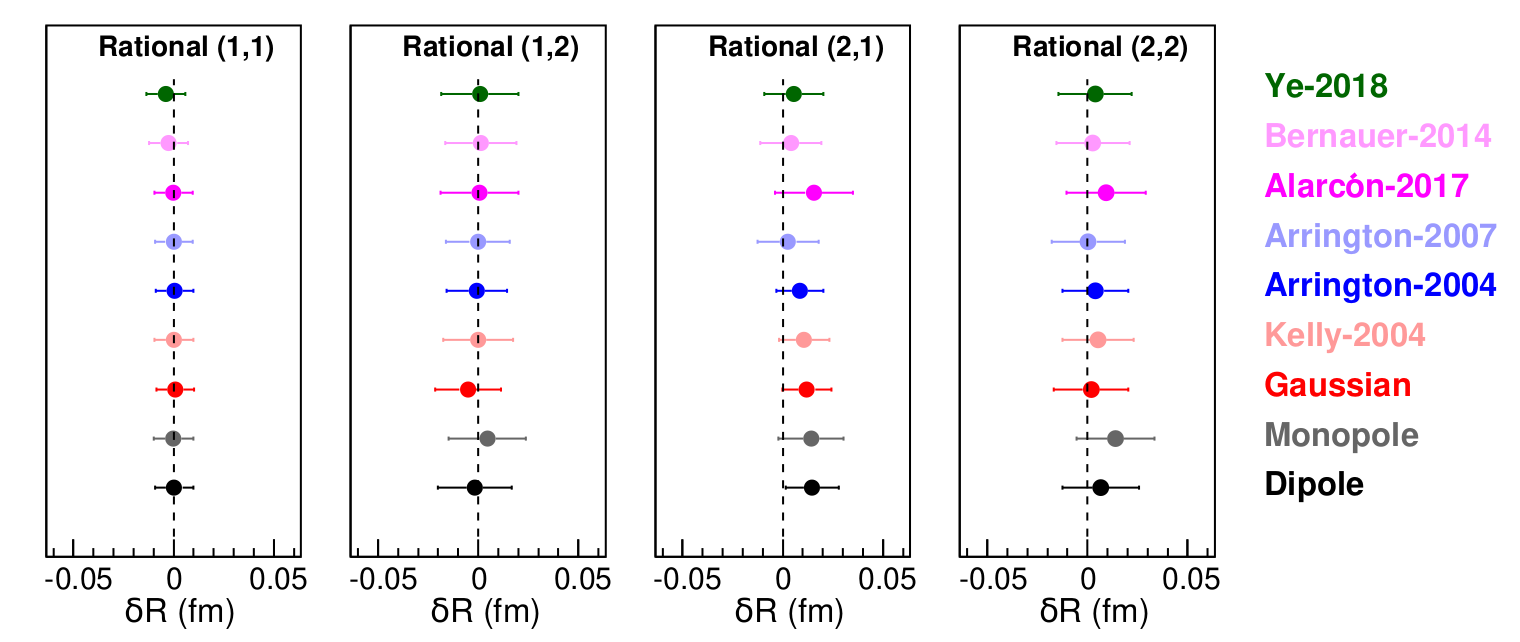}
\caption[fig]{\label{Yan-fig5} (Color online) Sample fits using rational functions of pseudo-data generated with nine proton form factor models including the projected PRad statistical and systematic uncertainties. The figure is from~\cite{Yan18}. } 
\end{figure}
\end{center}

Yan {\it et al.} ~\cite{Yan18} studied how to extract the proton charge radius in the low $Q^2$ region from the measured $G^p_E$ values in a robust way and demonstrated that the rational (1,1) function, defined in Eq.~(\ref{eq:rat}) is such a function and the best choice for the PRad data. Fig.~\ref{Yan-fig5} shows fits using various rational functions of pseudo-data generated with nine proton form factor models including the projected PRad statistical and systematic uncertainties.  Apart from monopole, dipole and Gaussian functional forms, the proton form factor parameterizations and fits from~\cite{Kelly04,Arrington07,Bernauer:2013tpr,Ye18,ALARCON2017} have been used. Additional details including fits of other functional forms can be found in~\cite{Yan18}.   
The PRad collaboration adopted the rational (1,1)  functional form to fit the data with two individual normalization parameters, $n_1$ and $n_2$, corresponding to the two separate beam energy values for which the data were taken, while keeping the rest of the rational (1,1) parameters the same, i.e. $n_{1}{\frac{1+p_{1}Q^2}{1+p_{2}Q^2}}$, and $n_{2}{\frac{1+p_{1}Q^2}{1+p_{2}Q^2}}$. At $Q^2=0$, this normalization parameter is just the proton charge, which should be 1.
The results from the fit are given by:
\begin{eqnarray}
 \langle r^2_{Ep} \rangle^{1/2} &= &0.831 \pm 0.007 (\rm stat.)  \pm 0.012 (\rm syst.) \ {\rm fm}, \nonumber \\
 n_{1} & = &1.0002 \pm 0.0002 (\rm stat.) \pm 0.0020 (\rm syst.), \\
 n_{2} & = & 0.9983 \pm 0.0002 (\rm stat.) \pm 0.0013 (\rm syst.), \nonumber
 \end{eqnarray}  
showing that the two normalization values obtained are consistent with 1.

\begin{center}
\begin{figure}[h]
\vspace{-1.75cm}
 \includegraphics[angle=-90,width=0.8\textwidth]{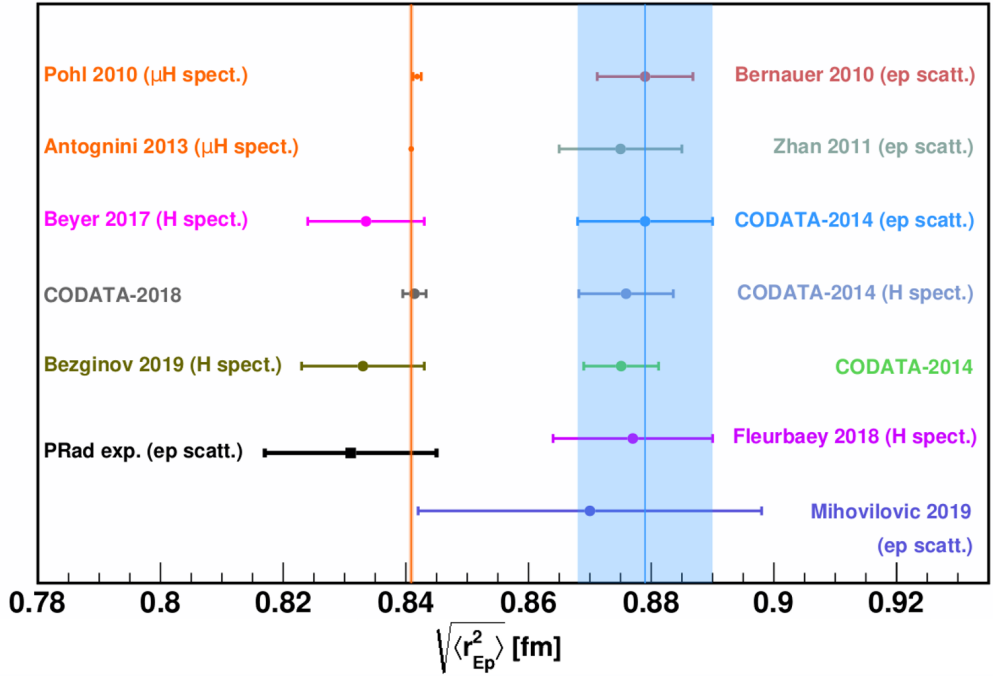}
\vspace{-0.75cm}
  \caption{\small{(Color online) The proton charge radius $\langle r^2_{Ep} \rangle^{1/2}$ as extracted from electron scattering and spectroscopic experiments since 2010 and before 2020 together with CODATA-2014 and CODATA-2018 recommended values (figure credit: Jingyi Zhou).~\label{PRad}}}
\end{figure}
\end{center}

The PRad result on the proton charge radius is smaller than the two latest $\langle r^2_{Ep} \rangle^{1/2}$ values extracted from electron scattering experiments~\cite{Bernauer10,Zhan11}, but consistent with the $\langle r^2_{Ep} \rangle^{1/2}$ values from the muonic hydrogen spectroscopic measurements~\cite{Pohl10,Antognini13}. While this result is also consistent with two recent hydrogen spectroscopic measurements ~\cite{Beyer17,Bezginov19}, it is not consistent with ~\cite{Fleurbaey18}. These latest hydrogen spectroscopic measurements will be discussed in later sections. Fig.~\ref{PRad} shows the PRad result on the proton charge radius together with these recent results from the hydrogen spectrocopic measurements, and the muonic hydrogen results. Also shown are the latest CODATA-2018 value released in 2019~\cite{codata18}, CODATA-2014 values~\cite{codata14}, results from~\cite{Bernauer10,Zhan11} and also the result from the Mainz ISR experiment~\cite{mihovilovi2019proton}.  One interesting observation is that among the most precise measurements from hydrogen spectroscopic and electron scattering measurements in recent years~\cite{Xiong19,Beyer17,Bezginov19,Fleurbaey18} before 2020,  three of these experiments reported a value that is smaller than those from the muonic results, though they are all consistent within experimental uncertainties. Improving the precision of such measurements will be crucial to investigate whether there might be a substantiated difference between results from muonic versus electronic systems.

\onecolumngrid
\twocolumngrid

\onecolumngrid
\twocolumngrid

\subsection{Proton charge radius from modern analyses of proton electric form factor data}

In addition to new experiments, numerous analyses have been carried out in recent years including global analyses in order to understand the difference between the $\langle r^2_{Ep} \rangle^{1/2}$ values determined from electron scattering experiments, especially the modern precision electron-proton scattering experiment at Mainz~\cite{Bernauer10}, and the muonic hydrogen results~\cite{Pohl10,Antognini13}. Some of these analyses obtain results consistent with the precise values from muonic hydrogen, while others are in agreement with larger values of $r_{Ep}$. Below we describe some of these analyses.  

Hill and Paz~\cite{PhysRevD.82.113005} carried out a model-independent determination of the  proton charge radius from electron scattering by first performing a conformal mapping of the domain of analyticity onto the unit circle in terms of $z(t,t_{cut},t_0)$ defined in Eq.~(\ref{eq:zdef}), where $t= q^2$, $t_{cut}=4m^{2}_{\pi}$, and $t_0$ is a free parameter mapping onto $z=0$.  The form factor $G_E(q^2)$ can then be written as a function of $z$ and a $z$ expansion can be carried out with the advantage that higher-order terms in $z$ are suppressed. Using electron-proton scattering data sets, a proton charge radius value of $\langle r^2_{Ep} \rangle^{1/2} = 0.870 \pm 0.023 \pm 0.012$ fm is obtained (see Ref.~\cite{PhysRevD.82.113005} for details).

Lorenz, Hammer and  Meissner~\cite{Lorenz12} analyzed the 2010 Mainz data using a dispersive approach to ensure analyticity and unitarity in the description of the nucleon form factors. In their analysis they have included the world data on the proton and also the neutron, and have obtained a charge radius value of  $\langle r^2_{Ep} \rangle^{1/2} = 0.84 \pm 0.01$ fm, that is consistent with the result from muonic hydrogen.
Lorenz and Meissner~\cite{Lorenz14} later also reanalyzed the Mainz data using a fit function based on conformal mapping, and showed that the extracted value for the proton charge radius -- with a larger statistical uncertainty than that from ~\cite{Bernauer10} -- is in agreement with the value from muonic hydrogen spectroscopic measurements, and also their previous dispersive analysis.  Lorenz {\it et al.}~\cite{PhysRevD.91.014023} calculated the TPE corrections to the electron-proton scattering, and applied these corrections to the Mainz data~\cite{Bernauer10}. They also investigated the impact on the extraction of the proton form factors from the inclusion of physical constraints and the extraction of $\langle r^2_{Ep} \rangle^{1/2}$ due to the enforcement of a realistic spectral function, which dominates the latter.  Very recently, a further improvement of the dispersive description has been presented in \cite{Lin:2021umk} using an improved two-pion continuum based on a Roy-Steiner analysis of pion-nucleon scattering, resulting in a value $\langle r^2_{Ep} \rangle^{1/2} = 0.838 \pm 0.005 \pm 0.004$ fm, where the first error is due to the fitting procedure and the second is from the spectral function. 

Lee, Arrington and Hill carried out a comprehensive global analysis~\cite{PhysRevD.92.013013} of the world electron-proton elastic scattering data with a focus on the Mainz measurements~\cite{Bernauer10}. This study involves enforcing model-independent constraints from form factor analyticity, and systematic studies of possible systematic effects. The extracted proton radius from this improved analysis of the Mainz data is $\langle r^2_{Ep} \rangle^{1/2} =0.895(20)$ fm, while $\langle r^2_{Ep} \rangle^{1/2} =0.916(24)$ fm from analyzing the world data without including the Mainz data.   Arrington and Sick~\cite{Arrington15} carried out a global examination of the elastic electron-proton scattering data and recommended a proton charge radius value of 0.879(11) fm. 

Griffioen, Carlson and Maddox~\cite{Griffioen16} analyzed the Mainz data set~\cite{Bernauer10} using a continued fraction functional form to map the $G_E$ assuming it is monotonically falling and inflectionless. They obtained a proton charge radius value of 0.840(16) fm, consistent with the mounic hydrogen result after rescaling different data sets on a level that is smaller than the original normalization uncertainties, and also inflating the point-to-point systematic uncertainty by 15\%. 

A proton charge radius value consistent with muonic hydrogen results was also obtained by Higinbotham {\it et al.} ~\cite{PhysRevC.93.055207} from analyzing data in the low momentum transfer region from Mainz in the 1980s~\cite{Simon80} and Saskatoon in 1974~\cite{PhysRevC.9.2125,PhysRevC.10.2111} using a stepwise regression of Maclaurin series and applying the {\it F}-test and the Akaike information criterion. Including the Mainz results on $G_{Ep}$~\cite{Bernauer:2013tpr}, the same analysis favors a radius that is consistent with the muonic hydrogen results, though their result is more sensitive to the range of the data included in the analysis. 

Horbatsch and Hessels~\cite{PhysRevC.93.015204} also analyzed the Mainz data~\cite{Bernauer10} and obtained 
$\langle r^2_{Ep} \rangle^{1/2}$ values ranging at least from 0.84 to 0.89 fm using two single-parameter form factor models with one being a dipole form, and the other a linear fit to a conformal-mapping variable.  

Sick and Trautmann~\cite{PhysRevC.95.012501} argued that the smaller values of $\langle r^2_{Ep} \rangle^{1/2}$  from~\cite{Griffioen16,PhysRevC.93.055207,PhysRevC.93.015204} are due to the neglect of higher moments in these analyses.  Kraus~{\it et al.}~\cite{PhysRevC.90.045206} found that fits of the proton charge form factor with truncated polynomials give too small values for the proton charge radius.   
In a later paper by Horbatsch {\it et al.}~\cite{PhysRevC.95.035203}, a $\langle r^2_{Ep} \rangle^{1/2}$ value of 0.855(11) fm was obtained with the higher moments fixed to the values based on Chiral Perturbation theory.  

Alarc\'on, Higinbotham, Weiss, and Ye~\cite{PhysRevC.99.044303} used a new theoretical frame work that combines chiral effective field theory and dispersion analysis. The behavior of the spacelike form factor in the finite $Q^2$ region correlates with its derivative at $Q^2=0$ due to the analyticity in the momentum transfer. In this approach predictions for spacelike form factors are made with the proton charge radius as a free parameter.  
 By comparing the predictions for different values of the proton radius with a descriptive global fit~\cite{PhysRevD.92.013013} of the spacelike form factor data, the authors of ~\cite{PhysRevC.99.044303} extracted a proton radius value of 0.844(7) fm, that is consistent with the muonic hydrogen results. A more recent analysis by 
 Alarc\'on, Higinbotham, and Weiss~\cite{PhysRevC.102.035203}  using the aforementioned method to extract both the proton magnetic and charge radius from the Mainz A1 data~\cite{Bernauer10} and obtained  $\langle r^2_{Mp} \rangle^{1/2}=0.850\pm0.001 ({\rm fit}\  68\%) \pm 0.010$ (theory full range) fm, and $\langle r^2_{Ep} \rangle^{1/2}=0.842 \pm0.002~{\rm (fit)} \pm 0.010$  (theory) fm. Including the PRad data~\cite{Xiong19} into their fit, they found no change in the extracted radius values within uncertainties.

Sick~\cite{sick2018proton} carried out a detailed study to reduce the model dependence associated with the required extrapolation in determining ${\frac{dG_{E}}{dQ^2}}(Q^2=0)$ to extract $\langle r^2_{Ep} \rangle^{1/2}$.  The approach takes into account the fact that $G_{Ep}$ in region lower than experimentally measured momentum transfer values is closely related to the charge density $\rho(r)$ at large values of $r$, which is constrained using form factor data at finite values of $Q^2$ to reduce model dependence in extrapolation. While corrections for relativistic effects are applied in this analysis, it is however not possible to rigorously define an accurate 3-dimensional charge density for the proton as has been discussed above. 
Using different form factor parameterizations of the data prior to 2010, Sick obtains a $\langle r^2_{Ep} \rangle^{1/2}$ value of 0.887(12) fm, that is consistent with the Mainz result~\cite{Bernauer10}, but inconsistent with the muonic hydrogen results~\cite{Pohl10,Antognini13}. 

Zhou {\it et al.}~\cite{PhysRevC.99.055202} adopted a flexible approach within a Bayesian paradigm which does not make any parametric assumptions for $G_{Ep}$, but with two physical constraints -- a normalization constraint for $G_{Ep}(0)$, and $G_{Ep}$ being monotonically decreasing as $Q^2$ increases.  The value of the proton charge radius extracted from the Mainz data is found to be sensitive to the $Q^2$ range of the data used in this analysis.

Horbatsch~\cite{HORBATSCH2020135373} analyzed the PRad data on the proton $G_E$ following a proposal by Hagelstein and Pascalutsa~\cite{HAGELSTEIN2019134825} by taking the logarithm to yield a $Q^2$ dependent radius function. This analysis shows that the PRad data is in agreement with theoretical predictions from dispersively improved chiral perturbation theory.  

Atac {\it et al.}~\cite{Atac2021} extracted both the proton and the neutron charge radius from a global analysis of the world proton and neutron form factor data by carrying out a flavor separation of the Dirac form factor $F_1$ assuming isospin symmetry. The u- and d-quark root-mean-squared transverse radii are subsequently determined from a fit to the slope of the corresponding flavor-dependent Dirac form factors, from which both the proton and the neutron charge radii are reconstructed. In this analysis, a proton charge radius value of $0.852\pm0.002_{(stat.)}\pm0.009_{(syst.)}$ fm is obtained, which is consistent with the muonic hydrogen results as well as the latest result from the PRad experiment~\cite{Xiong19}. Excluding the PRad data, a $\langle r^2_{Ep} \rangle^{1/2}$ value of 0.857(13) fm is extracted, consistent with the value including the PRad data but with a larger uncertainty. 

Borisyuk and Kobushkin~\cite{Borisyuk_2020} reanalyzed the Mainz data~\cite{Bernauer10} and found that the radius value obtained under certain conditions can be consistent with the muonic hydrogen results.

Cui {\it et al.}~\cite{Cui:2021vgm} extracted values of $\langle r^2_{Ep} \rangle^{1/2}$ using the electron-proton scattering data from the PRad experiment at JLab~\cite{Xiong19} and the A1 experiment at Mainz~\cite{Bernauer10} using a statistical sampling approach based on the Schlessinger Point Method (SPM). The SPM method, with an important feature that no specific functional form is assumed for the interpolation, is used in this analysis for the interpolation and extrapolation of smooth functions to minimize biases associated with assumed forms. The authors obtained a radius value of 
$
\langle r^2_{Ep} \rangle^{1/2} =0.838 \pm 0.005_{\rm stat} \ {\rm fm} 
$
from the PRad experiment, and a value of 
$
\langle r^2_{Ep} \rangle^{1/2} =0.856 \pm 0.014_{\rm stat} \ {\rm fm}
$
from the Mainz A1 experiment including data up to a $Q^2$ value of $0.014$ (GeV/c)$^2$. Combining these two values, Cui {\it et al.} finds a proton charge radius value of 
\begin{equation}
\langle r^2_{Ep} \rangle^{1/2} =0.847 \pm 0.008_{\rm stat} \ {\rm fm},
\label{eq:rpcui}
\end{equation}
from the two most recent experiments~\cite{Bernauer10,Xiong19} measuring the unpolarized electron-proton elastic scattering cross sections, that is consistent with the muonic hydrogen results~\cite{Pohl10,Antognini13}, as well as the most recent ordinary hydrogen spectroscopy results  ~\cite{Grinin1061,Bezginov19} for the proton charge radius.

\begin{center}
\begin{figure}[h]
\vspace{-1cm}
\includegraphics[angle=-90,scale=0.59]{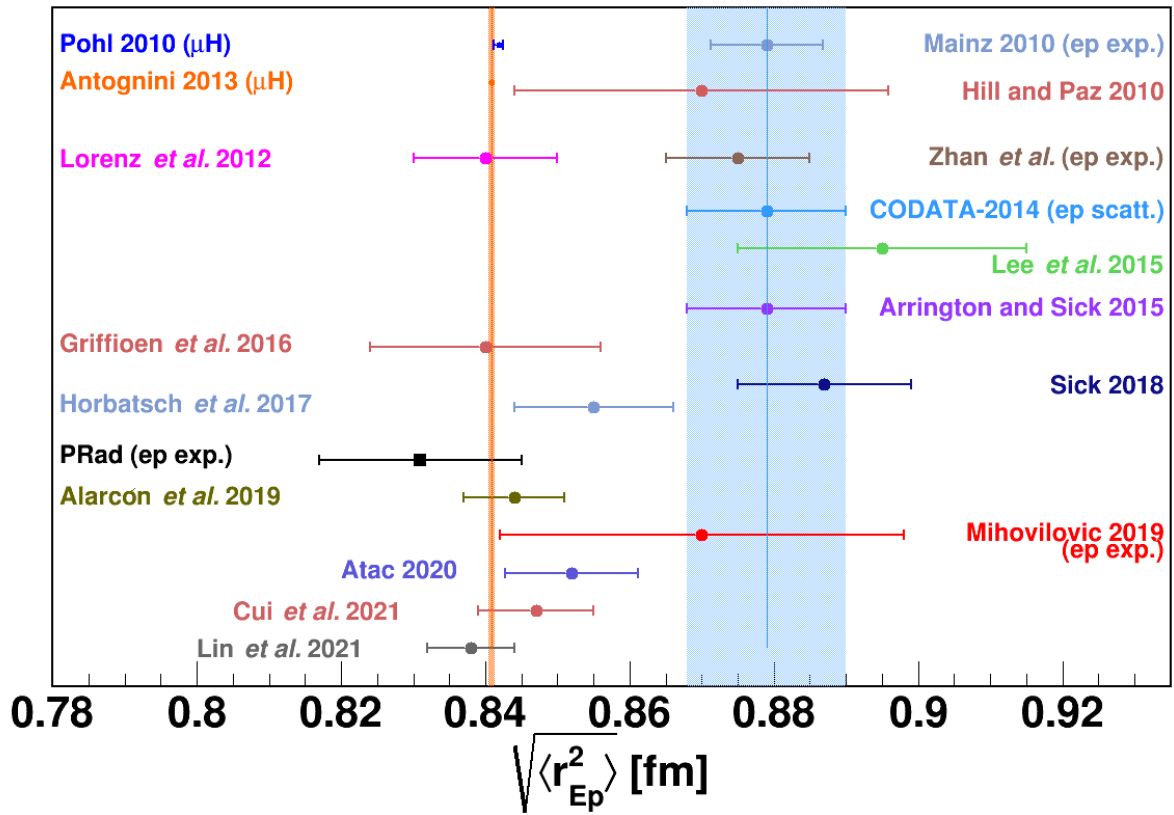}
\vspace{-0.5cm}
\caption[fig]{\label{fig:epdata-analysis} (Color online) The proton charge radius values determined from electron scattering  experiments since 2010 together with the results from the various analyses of electron-proton scattering data (see text) (figure credit: Jingyi Zhou).} 
\end{figure}
\end{center}

Fig.~\ref{fig:epdata-analysis} shows proton charge radius   results from electron-proton scattering experiments since 2010 and the extracted $\langle r^2_{Ep} \rangle^{1/2}$ values from some of the various analyses described above. Also included are the muonic hydrogen results as well as the CODATA-2014 recommended value. While the results of some of these analyses are consistent with muonic hydrogen results on the $\langle r^2_{Ep} \rangle^{1/2}$, others are consistent with the CODATA-2014 recommended value based on electron scattering  data, and few are in between.  There is no conclusive statement one can draw regarding the proton charge radius puzzle from these analyses of electron-proton scattering data. New and further improved measurements from lepton scattering are highly desirable, which we describe in Section VII.

 \onecolumngrid
\twocolumngrid

\section{Modern spectroscopic measurements}
\label{sec:spec}

\subsection{Muonic hydrogen spectroscopic experiments}

\begin{center}
\begin{figure}[h]
\includegraphics[width=0.9\textwidth]{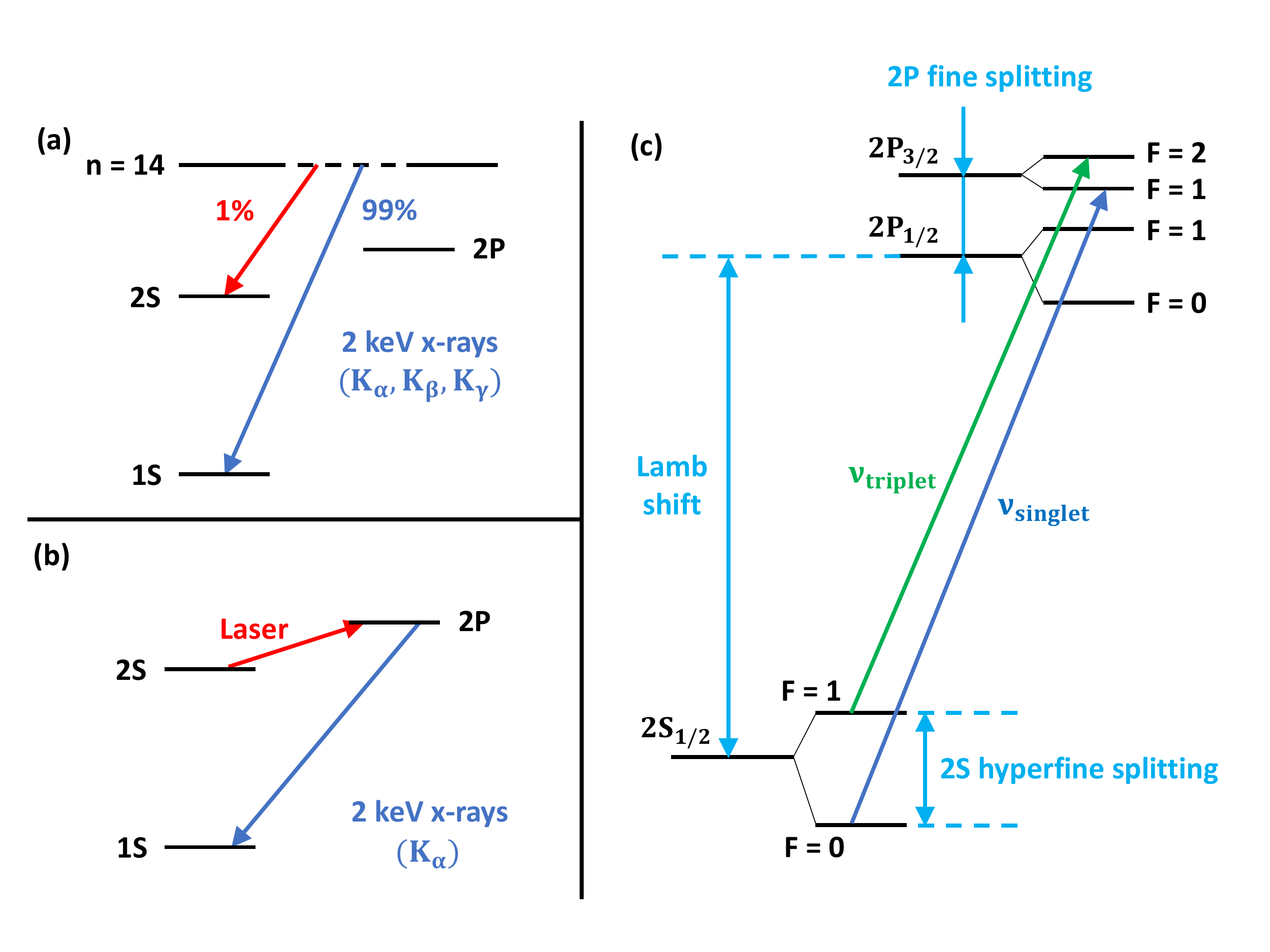}
\caption[fig]{\label{fig:muon} (Color online) The muonic hydrogen energy levels relevant to the proton charge radius measurement (figure credit: Jingyi Zhou).} 
\end{figure}
\end{center}

The first determination of the proton charge radius using muonic hydrogen atoms was carried out by Pohl {\it et al.} ~\cite{Pohl10} at the Paul Scherrer Institute (PSI) by measuring the transition frequency between the $2S_{1/2}^{F=1}$ and the $2P_{3/2}^{F=2}$ states at wavelengths around 6.01 $\mu$m using pulsed laser spectroscopy, 
see Fig.~\ref{fig:muon}. The muonic hydrogen atoms were produced by stopping negative muons in a hydrogen gas target with a pressure of 1 hPa (1 mbar) at the $\pi$E5 beam-line of the proton accelerator at PSI. The muonic atoms produced are in the $n \approx 14$ excited state, which then decay with about 1\% probability to the $2S$ metastable state, while the majority (99\%) decay to the $1S$ ground state. The lifetime of the long-lived $2S$ state at 1 hPa pressure is 1 $\mu$s. A 5-ns pulsed laser with a wavelength tunable around 6 $\mu$m is incident and illuminating the target volume about 0.9 $\mu$s after the muons reach the target. The laser wavelength is scanned through the resonance of the $2S \rightarrow 2P$ transition. Upon the excitation, the $2P$ state with a lifetime of 8.5 ps will decay to the $1S$  state via emission of the 1.9-keV$K_\alpha$ x-ray. Therefore, in this pulsed muonic atom laser spectroscopic measurement, the resonance curve is recorded by the coincidence of the 1.9-keV x-ray and the laser pulse as a function of the laser wavelength.  A coincidence time window of 0.9 to 0.975 $\mu$s is chosen, i.e. 0.9 $\mu$s after the muons enter the H$_2$ target, and the 75-ns window corresponds to the confinement time of the laser light within the optics surrounding the target.

The resonance frequency for the transition between the $2S_{1/2}^{F=1}$ and the $2P_{3/2}^{F=2}$ states was measured to be  49881.88 (76) GHz~\cite{Pohl10}, which gave a proton charge radius value of $r_{Ep} = 0.84184(67)$ fm based on the state-of-the-art QED calculations.  In a follow-up paper by the CREMA collaboration~\cite{Antognini13}, the tunable laser wavelength was scanned from 5.5 to 6.0 $\mu$m, and in addition to the original transition between the $2S_{1/2}^{F=1}$ (triplet) and the $2P_{3/2}^{F=2}$ states, a second transition between $2S_{1/2}^{F=0}$ (singlet) and the $2P_{3/2}^{F=1}$ states  was also measured. The corresponding resonance frequencies were determined to be
\begin{eqnarray}
\nu_t &=& 49881.35 \, (57)_{\rm stat.}(30)_{\rm syst.} \; {\rm GHz}, 
\nonumber \\
\nu_s &=& 54611.16 \, (1.00)_{\rm stat.}(30)_{\rm syst.} \; {\rm GHz}. \nonumber 
\end{eqnarray}
From these two transitions, the Lamb shift (LS) and the hyperfine splitting (HFS) can be independently determined and they are:
\begin{eqnarray}
\Delta E^{exp}_{LS} &=& 202.3706 \, (23) 
\; {\rm meV}, 
\label{eq:muHLSexp} \\
\Delta E^{exp}_{HFS} &=& 22.8089 \, (51) 
\; {\rm meV}.  \nonumber
\end{eqnarray}
Relating the state-of-the-art theory calculations of the Lamb shift~\cite{PhysRevA.53.2092,JENTSCHURA2011500,Borie:2012zz,Karshenboim10,PhysRevA.60.3593,Eides01,PhysRevA.85.032509} to the proton $r_{Ep}$, one obtains (in meV):
\begin{eqnarray}
\Delta E^{th}_{LS}(2P-2S) &=& 206.0336 \, (15) \nonumber \\ &-& 5.2275 \, (10) \, r_{Ep}^{2} + \Delta E_{TPE}, 
\label{eq:muHLSeq}
\end{eqnarray}
where the last term is due to the two-photon-exchange proton polarizability contribution discussed in Section~\ref{sec:atspec}. Using the estimate of Eq.~(\ref{eq:tpeLS}) for the latter, 
the extracted value for the proton charge radius is:
\begin{equation}
r_{Ep} = 0.84087(26)_{\rm exp}(29)_{\rm th} \ {\rm fm}  = 0.84087 (39) \; {\rm fm}. 
\label{eq:rpLS}
\end{equation}
This result is not only consistent with the earlier result from the muonic hydrogen spectroscopic measurement~\cite{Pohl10}, but also represents the most precise value for the proton charge radius. Both these results have been included in the 2018 CODATA compilation~\cite{codata18} and dominate its recommended value for the proton charge radius.

One notices from Eq.~(\ref{eq:tpeLS}) that the uncertainty ($\delta$) of the present 
TPE estimate for the muonic hydrogen $2P - 2S$ Lamb shift, $\delta( \Delta E_{TPE}) = 2.0~\mu$eV, is 
comparable to the present experimental Lamb shift precision,    
$\delta( \Delta E^{exp}_{LS}) = 2.3~\mu$eV, see Eq.~(\ref{eq:muHLSexp}).
A further improvement on the proton charge radius extraction from muonic hydrogen spectroscopy results therefore hinges upon further improving the TPE estimates.

\onecolumngrid
\twocolumngrid

\subsection{Ordinary Hydrogen spectroscopic experiments}

Since the release of the first muonic hydrogen spectroscopic determination of the proton charge radius~\cite{Pohl10}, there have been four atomic hydrogen spectroscopic measurements of the proton charge radius~\cite{Beyer17,Fleurbaey18,Bezginov19,Grinin1061} with Bezginov {\it et al.}~\cite{Bezginov19} being a direct measurement of the hydrogen Lamb shift.

Beyer {\it et al.}~\cite{Beyer17} carried out a measurement of the $2S-4P$ transition of ordinary hydrogen atoms using a cryogenic beam of H atoms.  
 A major improvement over previous experiments in overcoming the limitation due to the electron-impact excitation used to produce atoms in the metastable $2S$ state is the use of the Garching $1S-2S$ apparatus~\cite{PhysRevLett.107.203001,PhysRevLett.110.230801} as a well-controlled cryogenic source of 5.8-K cold $2S$ atoms.  In this case, the $2S^{F=0}_{1/2}$ sublevel is almost exclusively populated via Doppler-free two-photon excitation without imparting additional momentum on the atoms. The line shifts due to quantum interference of neighboring atomic resonances, and the first-order Doppler shift are the two remaining major systematic issues of this experiment. In ~\cite{Beyer17}, in addition to the use of a cryogenic H source which reduces the thermal velocity of atoms by a factor of 10 compared with prior experiments, the employment of a specifically developed active fiber-based retroreflector~\cite{Beyer16} allows for a high level of compensation of the first-order Doppler shift -- 4 parts in 10$^6$ of the full collinear shift. To suppress the quantum interference effect in order to determine the absolute $2S-4P$ transition frequency, the experiment was designed to observe line shifts due to quantum interference effect and to simulate the line shifts fully using an atomic line shape model.  Finally the quantum interference effect is removed using the Fano-Voigt line shape to obtain the unperturbed transition frequency for both the $2S^{F=0}_{1/2} - 4P^{F=1}_{1/2}$ and the $2S^{F=0}_{1/2} - 4P^{F=1}_{3/2}$ transitions. Combining with previous precision measurements of the $1S-2S$ transition by the same group~\cite{PhysRevLett.107.203001,PhysRevLett.110.230801}, values for both the Rydberg constant and the proton charge radius were determined to be~\cite{Beyer17}: 
 \begin{eqnarray}
 R_{\infty} &=& 10 \, 973 \, 731.568 \, 076(96) \; {\rm m}^{-1}, 
 \nonumber \\
\langle r^2_{Ep} \rangle^{1/2} &=& 0.8335(95) \; {\rm fm}.  \nonumber 
 \end{eqnarray}
 The uncertainty on the proton charge radius from this single experiment is comparable to the prior aggregate atomic hydrogen world data. This result is consistent with the muonic hydrogen results on the proton charge radius, but 3.3 combined standard deviations smaller than the 2014 CODATA recommended value~\cite{codata14} based on previous world data from ordinary hydrogen.

Fleurbaey {\it et al.}~\cite{Fleurbaey18} in Paris reported a result on the proton charge radius and the Rydberg constant in 2018 by combining their measurement of the $1S - 3S$ transition from ordinary atomic hydrogen with the $1S - 2S$ transition measurement performed by the Garching group~\cite{PhysRevLett.107.203001}. 
The Paris experiment measured the $1S - 3S$ two-photon hydrogen transition frequency using a continuous-wave laser with a wavelength of 205 nm and through the Balmer-$\alpha$ $3S - 2P$ fluorescence detection. A room temperature atomic hydrogen beam was used in the experiment and the main systematic effect of the experiment is the second-order Doppler effect due to the room-temperature atomic velocity distribution. The results presented included data taken during two different periods (2013 and 2016-2017) with improvements taking place between the two periods.   
The reported results are~\cite{Fleurbaey18}:
\begin{eqnarray}
R_{\infty} &=& 10 \, 973 \, 731.568 \, 53(14) \; {\rm m}^{-1}, 
\nonumber \\
\langle r^2_{Ep} \rangle^{1/2} &=& 0.877(13) \; {\rm fm}. 
\nonumber 
\end{eqnarray}
While the extracted $r_{Ep}$ value is consistent with the CODATA2014~\cite{codata14} recommended value, it disagrees with the muonic hydrogen Lamb shift result~\cite{Antognini13} by 2.6 standard deviations.
This experiment and the aforementioned experiment~\cite{Beyer17} used a similar measurement technique in which two transition frequencies  are involved. Each transition is between two ordinary hydrogen energy levels, corresponding to two different principal quantum numbers $n_1$ and $n_2$ with at least one of them being a $S$ state. We note that both the Rydberg constant and the proton charge radius determined from the Paris experiment~\cite{Fleurbaey18} disagree with those from the Garching experiment~\cite{Beyer17} at a level of about 2 standard deviations. It will be important to resolve such a discrepancy especially by repeating the same transition, either the $1S-3S$ or the $2S-4P$ transition.   

To determine the proton charge radius from ordinary hydrogen spectroscopic measurements, one can also measure the Lamb shift (the $2S_{1/2}-2P_{1/2}$ transition) directly, in which case, the principal quantum numbers for the two states between the transition are the same, and as such the precision of the Rydberg constant from other experiments is sufficient and the Lamb shift measurement itself together with the state-of-the-art QED calculation is used to extract  $\langle r^2_{Ep} \rangle^{1/2}$.   The most recent $r_{Ep}$ determination~\cite{Bezginov19} from ordinary atomic hydrogen spectroscopy is such a measurement.
In the experiment by Bezginov {\it et al.}~\cite{Bezginov19}, a fast beam of hydrogen atoms was created by passing protons -- which were accelerated to 55 keV -- through a molecular hydrogen target chamber, and about half of the protons were neutralized into hydrogen atoms from collisions with the molecules, and about 4\% were created in the metastable $2S$ state. The experiment used two different radio frequency cavities to drive  the $2S$ state away from the $F=1$ substates so that only the $F=0$ substate survives. The transition between the $2S_{1/2}(F=0) \rightarrow 2P_{1/2} (F=1)$ is the Lamb shift measured in this experiment using the experimental technique of frequency-offset separated oscillatory field~\cite{PhysRevA.92.052504,PhysRevLett.121.143002}, which is a modified Ramsey technique of separated oscillatory fields~\cite{PhysRev.76.996}.  The measured transition frequency of $2S_{1/2}(F=0) \rightarrow 2P_{1/2} (F=1)$ from this experiment is 909.8717(32) MHz. The Lamb shift determined is 1057.8298(32) MHz after including the contribution from hyperfine structure, which is 147.9581 MHz~\cite{PhysRevA.93.022513}. The proton charge radius value deduced from this experiment is~\cite{Bezginov19}:
\begin{equation}
\langle r^2_{Ep} \rangle^{1/2} = 0.833(10) \; {\rm fm},
\end{equation}
which is consistent with the muonic hydrogen Lamb shift measurements~\cite{Pohl10,Antognini13}, the 2017 ordinary hydrogen measurement~\cite{Beyer17}, and the PRad result from electron scattering~\cite{Xiong19}.  It disagrees however with the Paris measurement~\cite{Fleurbaey18} at a level of about two standard deviations.  

\onecolumngrid
\begin{center}
\begin{figure}[h]
\vspace{-2.5cm}
\includegraphics[angle=-90,width=0.85\textwidth]{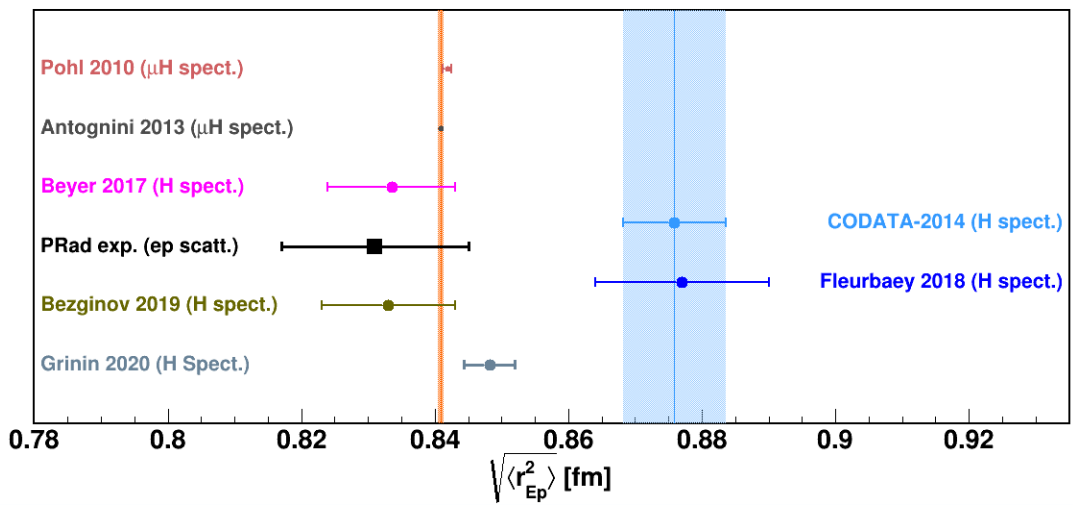}
\vspace{-2.cm}
  \caption{(Color online) The latest proton charge radius results from ordinary hydrogen spectroscopic measurements together with muonic hydrogen results and the CODATA-2014 recommended value based on ordinary hydrogen spectroscopy and the most recent result from electron scattering (figure credit: Jingyi Zhou). ~\label{fig:Hspect}}
\end{figure}
\end{center}
\twocolumngrid

Most recently, a new result on $\langle r^2_{Ep} \rangle^{1/2}$ from ordinary hydrogen spectroscopy has been published~\cite{Grinin1061}. This experiment measured the same $1S-3S$ transition as that of ~\cite{Fleurbaey18} but with significantly improved precision. Major improvements in reducing systematic uncertainties have been achieved by using a cold atomic beam and a two-photon direct frequency comb technique. The experiment also achieved an almost shot noise limited statistical uncertainty of 110 Hz. The unperturbed frequency for the $1S(F=1)-3S(F=1)$ transition determined from this experiment is 2,922,742,936,716.72(72) kHz, and $f_{1S-3S} ({\rm centroid}) = 2,922,743,278,665.79(72)$ KHz after subtracting the hyperfine shifts. Combing this new result on the $1S-3S$ transition with the $1S-2S$ transition frequency measured by the same group~\cite{PhysRevLett.110.230801} before, Grinin~{\it et al.} obtained~\cite{Grinin1061}: 
\begin{eqnarray}
R_{\infty} &=& 10 \, 973 \, 731.568 \, 226(38) \;{\rm m}^{-1}, 
\nonumber \\
\langle r^2_{Ep} \rangle^{1/2} &=& 0.8482(38) \; {\rm fm}.  \nonumber 
\end{eqnarray}
This extracted Rydberg constant is in agreement with the latest CODATA-18 recommended value. The new proton charge radius  result from~\cite{Grinin1061} is more than a factor of two more precise but also 2.9 standard deviations smaller compared with the CODATA-2014 recommended value from ordinary hydrogen spectroscopic measurements. It is more than a factor of three more precise, but 2.1 standard deviations smaller than the Paris result~\cite{Fleurbaey18}. Compared with muonic hydrogen results on $\langle r^2_{Ep} \rangle^{1/2}$, this new result from the $1S-3S$ transition is about two standard deviations larger. 
Fig.~\ref{fig:Hspect} shows the results on $\langle r^2_{Ep} \rangle^{1/2}$ from these four latest spectroscopic measurements using ordinary hydrogen atoms~\cite{Beyer17,Fleurbaey18,Bezginov19,Grinin1061} together with the muonic hydrogen results~\cite{Pohl10,Antognini13}. Also shown is the CODATA-2014~\cite{codata14} recommended value based on ordinary hydrogen spectroscopy as well as the most recent result from electron scattering~\cite{Xiong19}. 
While major progress has been made in recent years, and most of these recent measurements of the proton charge radius support a smaller value, the comparison of $\langle r^2_{Ep} \rangle^{1/2}$ extractions between electronic versus muonic systems is not fully settled. This situation highlights the importance of future high-precision scattering experiments, to improve on the result obtained by PRad. It is also highly desirable to have future spectroscopic measurements from ordinary hydrogen to achieve a comparable precision, i.e., a relative precision of 0.5\% or better. The PRad-II and other ongoing and upcoming scattering experiments will be discussed in the following section.

\section{Ongoing and upcoming experiments}

In this section we aim to briefly describe the current and planned experiments aimed at extracting the proton charge radius. Some of these plans have also 
been discussed in a recent review by Karr, Marchand, and Voutier~\cite{NatureReview20}.

\subsection{The MUSE experiment at PSI}

\begin{figure}[h]
\vspace{-1cm}
\includegraphics[width=0.48\textwidth]{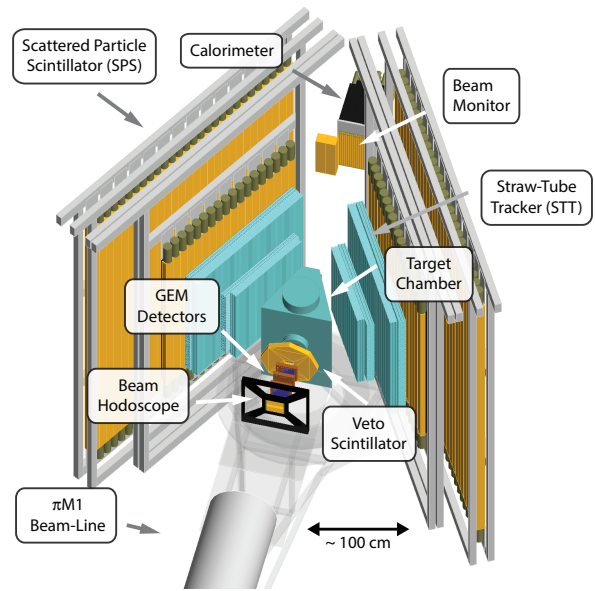}
\vspace{-2cm}
\caption[fig]{\label{fig:muse} (Color online) The schematics of the MUSE experiment at PSI (figure credit: Steffen Strauch).} 
\end{figure}

 The muonic hydrogen spectroscopic results on the proton charge radius ~\cite{Pohl10,Antognini13}  also motivated lepton-proton scattering measurements with muon beams. The MUon proton Scattering Experiment (MUSE)~\cite{muse,gilman2017technical} at PSI is currently ongoing in which measurements of lepton-proton elastic scattering cross sections utilizing both the $\mu^+$ and $\mu^-$ (muon) beams will be compared to those performed with electron and positron beams. The MUSE experiment uses the PSI $\pi M1$ beam line with $e^{\pm}$, and $\mu^{\pm}$ beams at incident momentum values of  115, 153 and 210 MeV/c to allow for simultaneous measurements of the $\mu^{\pm}p$ and $e^{\pm}p$ elastic scattering cross sections. The coverage of the scattering angle for the MUSE experiment is 20-100$^\circ$, corresponding to a $Q^2$ range of 0.0016 (with 115 MeV/c beam momentum) to 0.08 (GeV/c)$^2$ (210 MeV/c incident beam momentum). Due to the mass difference of $e^{\pm}$, and $\mu^{\pm}$, there is a small difference in the $Q^2$ coverage between the two.  The lowest $Q^2$ value reached by MUSE is comparable to that of the Mainz experiment~\cite{Bernauer10}, but much higher than that of the PRad experiment~\cite{Xiong19}, 0.0002 (GeV/c)$^2$.  In addition to the $\mu$ and $e$ beam particles, there are also pions in the $\pi M1$ mixed beam. Therefore, beam-line detectors for identifying various beam particles, determining the beam particle momentum and trajectories into the target, and counting the beam particles are important for the MUSE experiment. The beam-line detectors include beam hodoscope (fast scintillator array) measuring times relative to the accelerator RF to identify beam particle type, GEM detectors, veto scintillator, beam monitor and calorimeter. 
  A liquid hydrogen target is the main target for the production data taking with two symmetric spectrometers each equipped with detectors consisting of two scattered particle scintillator (SPS) paddles and two straw-tube trackers (STT). A schematics of the MUSE experiment is shown in Fig.~\ref{fig:muse}. 
The uncertainties from the MUSE experiment in the proton charge radius separately determined with $\mu^+ p$, $\mu^- p$, $e^+ p$, and $e^- p$ are expected to be nearly the same, which will be around 0.01 fm.  In addition to the determination of the proton charge radius, the MUSE experiment will allow for tests of the two-photon-exchange effect in lepton scattering by comparing the $\mu^{\pm} p$ and $e^{\pm} p$ cross section,  and a direct test of lepton universality.  More details about the MUSE experiment can be found in~\cite{gilman2021ReviewPSI}.

\subsection{The COMPASS++/AMBER Experiment at CERN}

The COMPASS collaboration proposed a precision measurement of elastic $\mu p$ scattering at high energy and low $Q^2$ with the M2 beam-line at CERN with COMPASS++/AMBER~\cite{Dreisbach:2019pkc}.  By carrying out a muon-proton scattering at high energies -- compared with low-energy lepton-proton scattering -- the proposed experiment has different, and in some cases favorable systematics. The COMPASS++/AMBER measurement of the proton radius will use 100 GeV muons of the CERN M2 beam-line. The hydrogen target will be an active target -- a high-pressure time projection chamber (TPC) -- in which the recoil protons will be measured for proton energies of 0.5 to 20 MeV.  For small-angle scattered muon detection, silicon detectors will be used for precision tracking. The triggers will be formed by scattered muons using the 200 $\mu$m SciFi stations, and the inner tracking and the ECAL of the COMPASS spectrometer will be used for measuring the scattered muons. The proposed experiment with 200 days of beam time will extract the proton electric form factor in a $Q^2$ range of 0.001 to 0.04 (GeV/c)$^2$ with relative point-to-point precision better than 0.001. The projected precision in the determination of the proton charge radius is expected to be better than 0.01 fm.

\subsection{The PRad-II experiment at Jefferson Lab}

The PRad experiment~\cite{Xiong19} has demonstrated the advantages of the
calorimetric method in $e-p$ scattering experiments to measure the
proton charge radius with high accuracy. However, it was the first time such a new method was employed and therefore it is important to push its limit to the ultimate precision.  
Recently the PRad collaboration proposed a new and upgraded experiment, PRad-II~\cite{PRad2,gasparian2020pradii} to the Jefferson Lab program advisory committee (PAC) to carry out the next generation of the proton charge radius measurement using an electromagnetic calorimeter together with two planes of tracking detectors with several major upgrades and improvements over the PRad experiment. The experiment has been approved by the PAC with the highest scientific rating.

\onecolumngrid
\begin{center}
\begin{figure}[h]
    \includegraphics[width=0.8\textwidth]{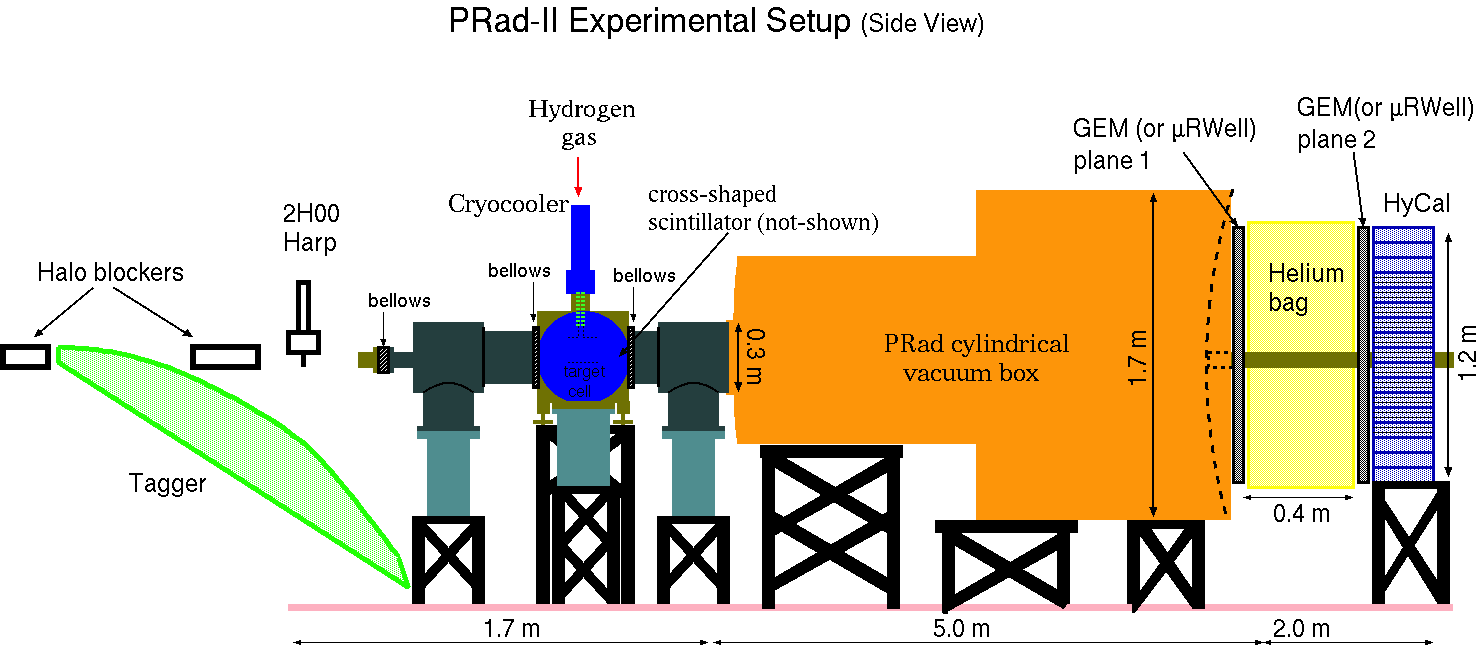}
\caption{(Color online) Schematic of the setup for the proposed PRad-II experiment. The incident electron beam is from left to right (figure credit: Dipangkar Dutta).
  \label{PRad-II_schematic}}
\end{figure}
\end{center}
\twocolumngrid

One important aspect of PRad-II compared with PRad is to reduce the statistical uncertainty of the electron-proton elastic scattering cross section measurement by a factor of 4.  Further,
a number of upgrades are proposed to improve the precision in determining the proton electric form factor and the charge radius significantly by reducing systematic uncertainties.  The upgrades include (i) adding a second plane of tracking detector for improving the tracking capability and further suppressing the beam-line related background; (ii) upgrading the HyCal by replacing its outer-region lead glass modules by PbWO$_4$ crystals to improve the detector resolutions and uniformity and suppress the inelastic contamination; (iii) adding a set of cross-shaped scintillator detectors in order to detect scattered electrons from $ep$ at scattering angles as forward as 0.5$^\circ$ still being cleanly separated from $ee$ scattering; (v) upgrading the HyCal readout to flash ADC to enhance the data taking rate; (v) adding a second beam halo blocker and with improved beam-line vacuum to further suppress the background; (vi) and future improved radiative correction calculations at the next-to-next-to-leading order (NNLO) for both $ep$ and $ee$ scattering. 
These upgrades and improvements will lead to the reduction of the overall
experimental uncertainty in the radius determination by a factor of 3.8 compared to PRad.  
As the muonic hydrogen result with its unprecedented precision ($\sim$0.05\%)
dominates the CODATA value of the proton charge radius, it is
critically important to help evaluate possible systematic uncertainties associated with muonic experiments using different experimental methods with high precision and different systematics. The PRad-II experiment with its projected
total uncertainty smaller than 0.5\% could potentially inform whether there is
any systematic difference in the radius results between $e-p$ scattering and muonic hydrogen measurements. 
The PRad-II will also be the first lepton scattering experiment to
reach a $Q^2$ range below 10$^{-4}$ GeV$^2$ with three proposed incident beam energies: 0.7, 1.4 and 2.1 GeV.

Fig.~\ref{PRad-II_schematic} shows the schematics of the proposed PRad-II setup. For the two tracking detectors, it was proposed to use the new $\mu$RWELL technology~\cite{Bencivenni_2015}, but GEM as was used in the PRad experiment will work also. Not shown in the figure are the cross-shaped scintillator detectors, which will be mounted inside the target chamber.

\begin{center}
\begin{figure}[h]
\vspace{-1cm}
\includegraphics[angle=-90,width=0.87\textwidth]{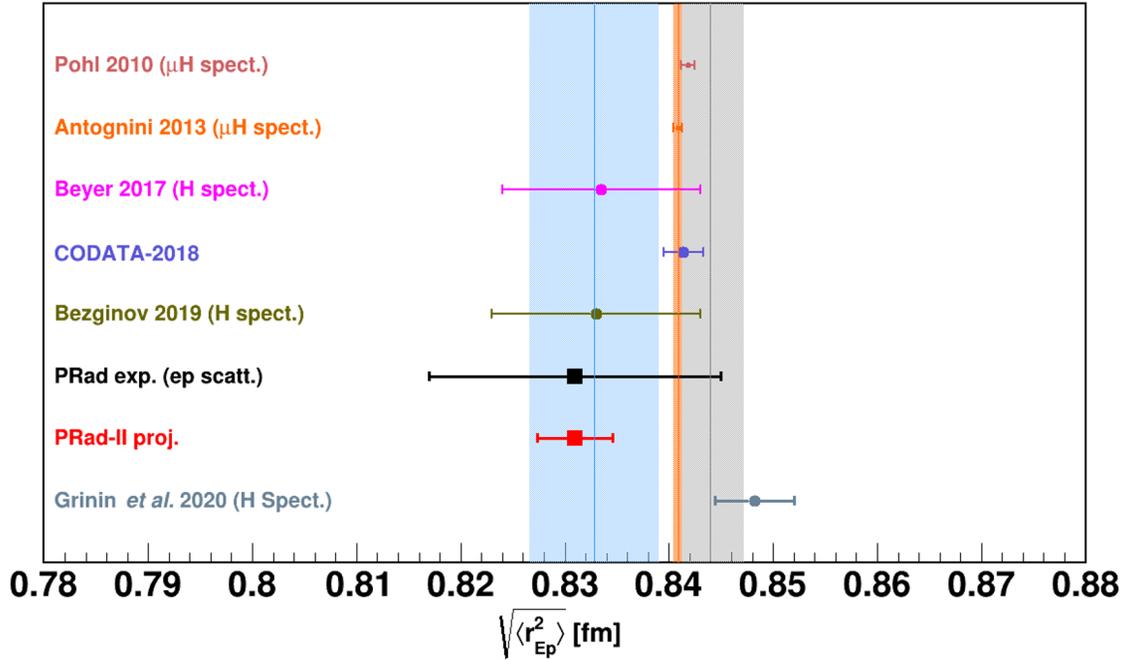}
\vspace{-1.1cm}
  \caption{(Color online) The PRad-II projection for  $\langle r^2_{Ep} \rangle^{1/2}$ with all proposed upgrades and improvements shown with a few selected results from other experiments and CODATA-2018 recommendations (see text) (figure credit: Jingyi Zhou).~\label{PRad-II_projection}}
\end{figure}
\end{center}

Fig.~\ref{PRad-II_projection} shows the projected radius measurement from PRad-II together with some of the most recent results on the proton radius including the $e-p$ scattering results~\cite{Xiong19}, the two muonic hydrogen results~\cite{Pohl10,Antognini13}, and the three recent atomic hydrogen spectroscopic results~\cite{Beyer17,Bezginov19,Grinin1061}. Also shown is the CODATA 2018~\cite{codata18} recommended value. The blue line and the band represent the weighted average of the $\langle r^2_{Ep} \rangle^{1/2}$ value and its uncertainty of the three proton radius values~\cite{Beyer17,Bezginov19,Xiong19} from ordinary hydrogen spectroscopy and electron-proton scattering.
The grey line and band are the results from weighted average of all four including the result from~\cite{Grinin1061}. This figure illustrates two points: (i) the importance of improving the precision of $\langle r^2_{Ep} \rangle^{1/2}$ measurement from electronic systems whether it be ordinary hydrogen spectroscopy or electron-proton scattering; (ii) additional new measurements from ordinary hydrogen in addition to the result from~\cite{Grinin1061} and the upcoming PRad-II will be essential to determine whether there is a difference between $\langle r^2_{Ep} \rangle^{1/2}$ determined between the electronic versus the muonic systems.

\onecolumngrid
\twocolumngrid

The fact that the CODATA 2018 value is dominated by the highly precise muonic hydrogen measurements~\cite{Pohl10,Antognini13} makes it crucial to further improve precisions in measuring the proton charge radius from lepton scattering and ordinary hydrogen spectroscopic experiments. For the PRad-II projection, it is shown with all proposed upgrades and improvements including an upgraded HyCal with all PbWO$_4$ crystals, two planes of tracking detectors based on the new $\mu$RWELL technology, and also the addition of a cross-shaped scintillator detector setup. The overall uncertainty in the proton radius will be reduced by a factor of 3.8 compared with the PRad result, corresponding to an absolute uncertainty of 0.0036 fm. This projected precision from PRad-II is slightly better than the 0.0038 fm precision from the latest hydrogen spectroscopy result of~\cite{Grinin1061}, which is the most precise measurement from ordinary hydrogen atomic spectroscopy.

If the PRad $\langle r^2_{Ep} \rangle^{1/2}$ value would prevail, the PRad-II result could signal a more than 2.7 standard deviations smaller than the muonic hydrogen result. 
While it does not seem possible in the foreseeable future for lepton-scattering experiments to reach the precision of muonic hydrogen spectroscopic measurements, the improvement of PRad-II is significant and will have the great potential to inform whether there is any systematic difference between muonic hydrogen results and results from electron scattering. The PRad-II measurement together with future improvements in ordinary hydrogen spectroscopic measurements will shed light on whether there is any systematic difference between the proton charge radius determined from electronic versus muonic systems. Therefore, they may uncover interesting new physics such as the violation of lepton universality.

\subsection{Future electron scattering experiments at Mainz}

There are two major new programs at Mainz University aimed at measuring the electron-proton elastic scattering at low $Q^2$, which will provide new results on the proton charge radius in the coming years.

The first is the PRES experiment~\cite{PRES,2019PPNL...16..524V,belostotski2019proton} at the Microtron MAMI in the A2 experimental hall. In this experiment a hydrogen time projection chamber will be used to measure recoil protons from $e-p$ elastic scattering in a $Q^2$ region from 0.001 to 0.04 (GeV/c)$^2$. Compared with other $e-p$ scattering experiments in which scattered electrons are commonly measured, the Mainz PRES experiment will have different systematics. The proposed experiment will aim at an absolute precision of 0.2~\% and a relative 0.1~\% in measuring the $e-p$ elastic scattering cross section. The PRES experiment is projected to  reach 0.5~\% statistical precision on $\langle r^2_{Ep} \rangle^{1/2}$, with systematic errors $\leq 0.3 \%$.  

\begin{figure}[h]
\centering
\vspace{.5cm}
\includegraphics[width=0.45\textwidth]{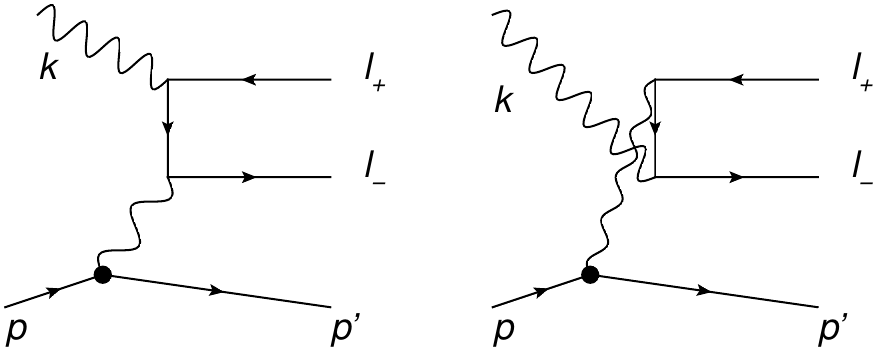}
\caption{Bethe-Heitler direct (left) and crossed (right) diagrams to the $\gamma p \to l^-l^+ p$ process, where the four-momenta of the external particles are: 
$k$ for the photon, $p (p^\prime)$ for initial (final) protons, and $l_-$, $l_+$ for the lepton pair.} 
\label{fig:bh}
\end{figure}

A further test of the lepton universality in the proton charge radius extraction was proposed in \cite{Pauk:2015oaa} through the photoproduction of a lepton pair on a proton target in the limit of small momentum transfer, in which this reaction is dominated by the Bethe-Heitler process shown in Fig.~\ref{fig:bh}. By detecting the recoiling proton in the $\gamma p \to l^- l^+ p$ reaction, it was shown that a measurement of a cross section ratio of $e^- e^+ + \mu^- \mu^+$ vs $e^- e^+$, above vs below dimuon threshold respectively, accesses the same information as muon vs electron scattering experiments. Furthermore such measurement is free from hadronic background if one performs the measurement in the di-lepton mass window between di-muon threshold and below $\pi \pi$ threshold. It thus complements a comparison of elastic $l - p$ scattering data, as the  overall normalization uncertainty drops out of the di-lepton photoproduction cross section ratio. The feasibility of such experiment using the TPC set-up at MAMI is currently under study~\cite{Sokhoyan20}. 

\begin{center}
\begin{figure}[h]
\vspace{.75cm}
\includegraphics[width=0.75\textwidth]{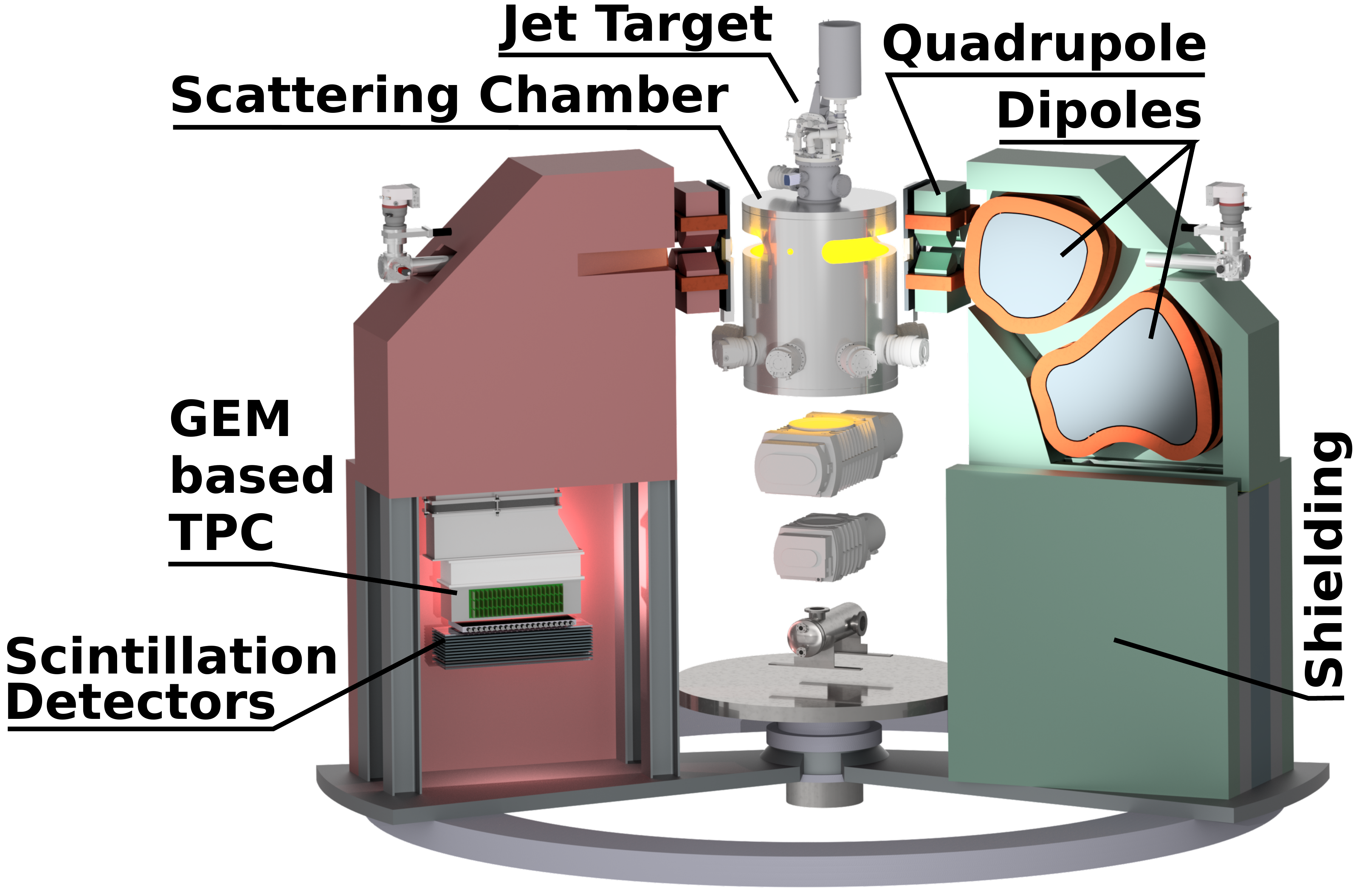}
\caption{
(Color online) The MAGIX high resolution dual-spectrometer setup at the MESA accelerator. The gas-jet target in the centre is also visible (figure credit: MAGIX Collaboration, \cite{Schlimme:2021gjx}).}
\label{fig:MAGIX}
\end{figure}
\end{center}

The second program in Mainz is centered around the Mainz Superconducting Energy Recovery Linac (MESA), which is a new accelerator presently under construction at the University of Mainz~\cite{Hug:2020miu}. MESA is designed as a recirculating superconducting linear accelerator which provides an external beam with high current and high degree of polarization. 
In the energy recovery mode, MESA will deliver an electron beam with 20 - 105 MeV and a current of 1 mA, which is ideal for precision experiments. 
The MAGIX experiment (Mainz Gas-Internal Target Experiment) at MESA will consist of a quadrupole in front of two medium sized dipole magnets, see Fig.~\ref{fig:MAGIX}. The compact design of the spectrometers will allow for a relative momentum resolution of order $10^{-4}$. For the focal-plane detector, a time projection chamber with open field cage and GEM readout is being developed~\cite{Caiazza:2020sda,Gulker:2019hht}. Finally, a windowless internal gas-jet target~\cite{Grieser:2018qyq}, which has already been commissioned at MAMI~\cite{Schlimme:2021gjx}, will be used.

With the MAGIX experiment at MESA, for the first time in hadron physics, an experiment will be developed, which combines the advantages of an ultra-light windowless gas target with the high intensity of an Energy Recovery Linac accelerator. This combination of a high beam intensity and a target, in which multiple scattering of the outgoing particles will be minimized, will lead to competitive luminosities in the range of $10^{35}$ cm$^{-2}$ s$^{-1}$, while providing at the same time a very clean experimental environment.
With the low beam energies of MESA, it will be possible to reach $Q^2$ values in $e-p$ scattering down to $10^{-4}$ (GeV/c)$^2$, and a relative precision on the proton electron form factor $G_{Ep}$ down to 0.05 \%. It will also significantly improve the determination of the proton magnetic radius~\cite{Bernauer20}.

\onecolumngrid
\twocolumngrid

\subsection{The UL${\rm Q}^2$ experiment at Tohoku University} 

The Ultra-Low ${\rm Q}^2$ (UL${\rm Q}^2$)~\cite{Suda18} collaboration is carrying out an electron-scattering experiment at Tohoku University using its 60 MeV electron linac.  This experiment will use the electron beam at energies from 20-60 MeV with a scattering angular range of 30 to 150$^\circ$, corresponding to a $Q^2$ range of 0.0003 to 0.008 (GeV/c)$^2$ for $e - p$ elastic scattering aiming at an absolute cross section measurement with a precision of 0.1~\%.
The UL${\rm Q}^2$ experiment will use a CH$_2$ target with elastic $e-^{12}$C as a reference reaction for normalization purpose. The root-mean-square charge radius of the $^{12}$C nucleus is known to a relative precision of $\sim 3 \times 10^{-3}$.  The proton electric form factor $G_{Ep}$ will be extracted using the Rosenbluth separation technique. To carry out this experiment, a new beam line and a new spectrometer have been built with single-sided silicon detectors -- developed for the J-PARC muon g-2 and the neutron electric dipole moment experiments~\cite{Sato:2017/P} -- as the focal plane detector.

\section{The deuteron charge radius}
\label{deuteronradius}

A less well known charge radius puzzle is concerning the deuteron, the simplest nucleus in nature which is loosely bound with a binding energy of 2.2 MeV. Like the proton, the deuteron charge radius can be determined by the extraction of the deuteron charge form factor, $G_{Cd}(Q^2)$ at low values of $Q^2$ from electron-deuteron elastic scattering first, and the subsequent extrapolation of the measured $G_{Cd}(Q^2)$ to the unmeasured region in order to determine its slope at $Q^2=0$.

The unpolarized elastic $e-d$ scattering cross section is described in the one-photon exchange picture as 

\begin{equation}
\frac{\mathrm{d}\sigma}{\mathrm{d}\Omega}\Lb E, \theta \Rb  = \sigma_{_{\!NS}} \left\{ A_{d}(Q^{2}) + B_{d}(Q^{2})\,
\tan^{2} \frac{\theta}{2}  \right\} ,
\label{eq:eqn_sigma}
\end{equation}
where $\sigma_{_{\!NS}}$ is the differential cross section for the elastic scattering from a point-like and spinless 
particle at a scattering angle $\theta$ and an incident energy $E$. 
For a spin-1 object such as the deuteron, its electromagnetic structure can be described by three form factors: the charge $G_{Cd}$, the magnetic dipole  $G_{Md}$, and the electric quadrupole $G_{Qd}$. The structure functions $A_{d}(Q^{2})$, $B_{d}(Q^{2})$ are related to these form factors via ~\cite{PhysRev.102.1586, Gourdin:1963}:
\begin{eqnarray}
A_{d}(Q^{2}) & = & G^2_{Cd}(Q^{2}) + \frac{2}{3}\,\tau_d  G^2_{Md}(Q^{2}) + \frac{8}{9}\,\tau_d^{2} G^2_{Qd}(Q^{2}),
\nonumber \\
B_{d}(Q^{2}) & = & \frac{4}{3}\,\tau_d (1 + \tau_d)  G^2_{Md}(Q^{2}),
\label{eq:eqn_deuteronstrucfunc}
\end{eqnarray}
with $\tau_d \equiv Q^{2}/(4M_{d}^{2})$, where $M_{d}$ is the deuteron mass. 
Also, there are the following additional relations:
\begin{displaymath}
G_{Cd}(0) = 1, 
\quad
G_{Md}(0) = \mu_{d},
\quad
G_{Qd}(0) = Q_d,  
\label{eq:eqn_deuteronstrucfunc2}
\end{displaymath}
with $\mu_d$ being the deuteron magnetic dipole moment (in units $e/(2 M_d)$), and $Q_d$, 
the electric quadrupole moment (in units $e/M_d^2$).
With three form factors, one needs to carry out three measurements with independent combinations of the three form factors in order to separate them for each $Q^2$ value. It was shown in \cite{Carlson:2008zc} how these three form factors allow to map out the transverse charge densities in a  deuteron, in a state of helicity $0$ or $\pm 1$, as viewed from a light front moving towards the deuteron. Furthermore, the charge densities for a transversely polarized deuteron are characterized by monopole, dipole and quadrupole patterns.

At low values of $Q^2$ most relevant for the charge radius determination, in the range $10^{-2}$ to $10^{-4}$ (GeV/c)$^2$, and small scattering angles, the unpolarized $e-d$ elastic scattering cross section is dominated by the deuteron charge form factor. One can therefore extract $G_{Cd}$ with negligible systematic uncertainties using data driven parameterizations for  $G_{Md}$, and $G_{Qd}$~\cite{DRad-fitter-PRC} from measured scattering cross section. The deuteron rms charge radius radius, $r_d$ can then be determined by
fitting the experimental $G_{Cd}$ data as a function of \(Q^2\), and calculating 
the slope of this function at \(Q^2=0\), according to
\begin{equation}
r_{d}  \equiv \sqrt{\langle r_d^{2} \rangle} = \left( -6  \left. \frac{\mathrm{d} G_{C}^{d}(Q^2)}
{\mathrm{d}Q^2} \right|_{Q^{2}=0} \right)^{1/2} ,
\label{eq:eqn_rd}
\end{equation}
in analogy to how $r_{Ep}$ is obtained. Zhou {\it et al.}~\cite{DRad-fitter-PRC} demonstrated how one can extract $r_d$ reliably using robust fitters.

Like the proton charge radius, the deuteron $r_d$ can also be determined from atomic spectroscopic measurements using ordinary deuterium or muonic deuterium atoms. The CREMA collaboration has reported a deuteron charge radius value  from a muonic spectroscopy-based measurement of three $2P \rightarrow 2S$ transitions in muonic deuterium atoms as~\cite{Pohl:2016} (labeled as $\mu$D 2016 in Fig.~27.)
\begin{equation}
r_{d} = 2.12562 \pm 0.00078 \; {\rm fm},
\label{eq:mudLS}
\end{equation}
which 
is 2.7 times more accurate but 7.5 standard deviations smaller than the CODATA-2010 recommended value \cite{Mohr10}. 
Newer values of $r_d$ based on the muonic deuterium spectroscopic measurement~\cite{Pohl:2016} with improved theoretical calculations are~\cite{HERNANDEZ2018377} 
\begin{equation*}
r_d = 2.12616 \pm 0.00090 \; {\rm fm},
\end{equation*}
and~\cite{PhysRevA.99.030501,PhysRevA.97.062511} 
\begin{equation*}
r_d = 2.12717 \pm 0.00082 \; {\rm fm}. 
\end{equation*}

From the spectroscopic measurement of $1S \rightarrow 2S$ transitions from ordinary deuterium atoms~\cite{Parthey:2010aya}, Pohl {\it et al.} 
extracted a deuteron radius value~\cite{Pohl_2017}:
\begin{equation*}
r_d = 2.1415 \pm 0.0045 \; {\rm fm}, 
\end{equation*}
which is 3.5 standard deviations larger than the extracted value of Eq.~(\ref{eq:mudLS}) from muonic deuterium atoms. 

Another spectroscopic method commonly used to extract the deuteron charge radius utilizes the isotope shift of the $1S \rightarrow 2S$ transition between atomic hydrogen and deuterium~\cite{PhysRevLett.80.468,Parthey:2010aya}, from which one can precisely determine the difference between the squares of the deuteron and proton charge radii~\cite{PhysRevA.83.042505}:
\begin{equation*}
r^{2}_d - r^{2}_p = 3.82007 (65) \; {\rm fm}^2.  
\end{equation*}
Combining the proton charge radius values with the isotope shift results, one can extract $r_d$. In fact, the CODATA-2010 recommended value for $r_d=2.1415 \,(21)$ fm used the isotope shift results on the radii and the proton charge radius values from electron scattering. 

\begin{center}
\begin{figure}[h]
\vspace{0.5cm}
\includegraphics[width=0.9\textwidth]{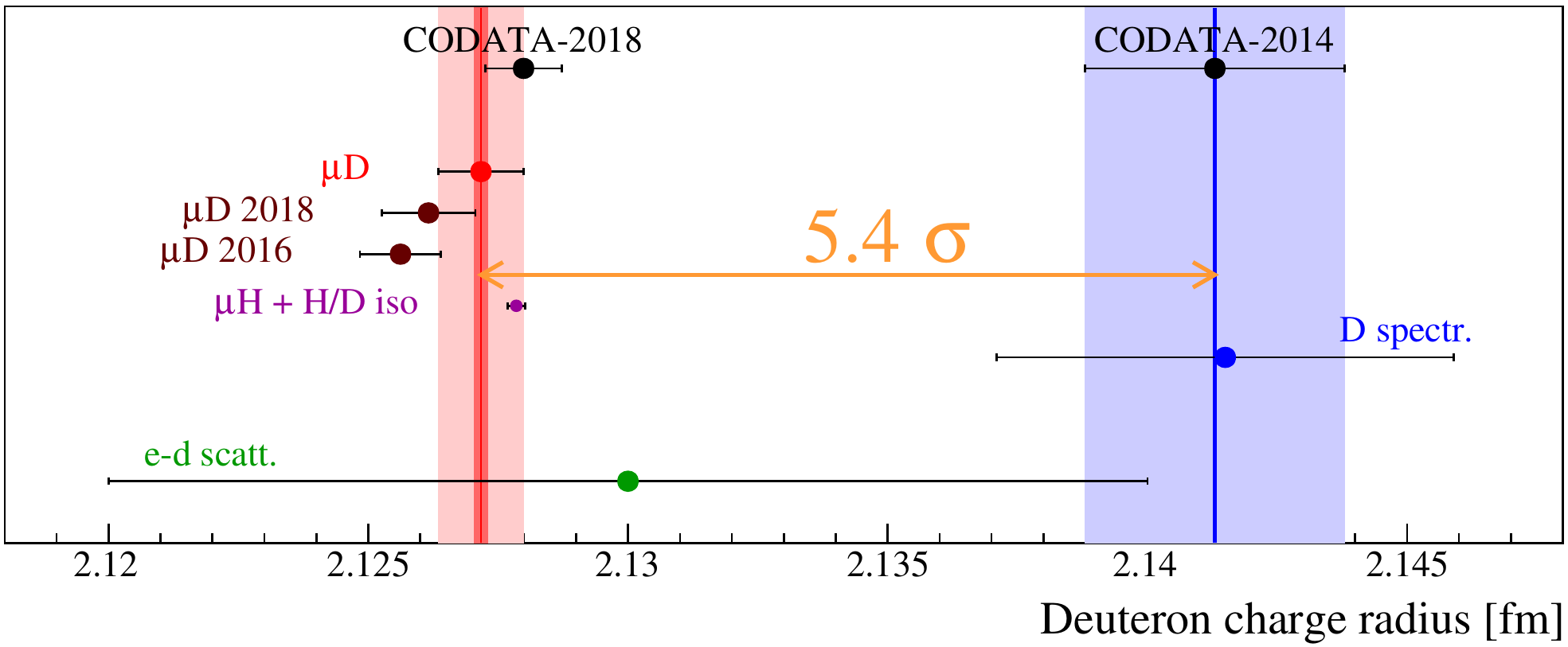}
\caption[fig]{\label{fig:DRad} (Color online) The existing results on the deuteron charge radius, see text for details (figure credit: Randolf Pohl).} 
\end{figure}
\end{center}

From the electron scattering side, all the elastic $e-d$ scattering measurements with rather large experimental uncertainties are not able to resolve the discrepancy between the $r_d$ values obtained from ordinary deuterium and muonic deuterium spectroscopic measurements. The re-analysis of world $e-d$ data gives~\cite{Sick:1998cvq}:
\begin{equation*}
r_{d} = 2.130 \pm 0.003 \, ({\rm stat.}) \pm 0.009 \, ({\rm syst.}) \; {\rm fm}. 
\end{equation*}
With rather large overall uncertainty, this $r_d$ value from the re-analysis is consistent with both the muonic deuterium result as well as that from ordinary deuterium spectroscopic measurements. Therefore, a significantly improved $r_d$ determination from a new electron-deuteron scattering experiment is needed to help resolve the current situation surrounding the deuteron charge radius. Fig.~\ref{fig:DRad} is a summary of results on the deuteron charge radius discussed above including the CODATA 2014 value~\cite{codata14} shown with the uncertainty as a band, and the CODATA 2018 recommended value~\cite{codata18}. Also included is an extraction of the $r_d$ using the isotope shift~\cite{PhysRevA.83.042505} and the muonic hydrogen result of the $r_{p}$~\cite{Antognini13}. The two latest extractions of the deuteron charge radius from the muonic deuterium measurement are labeled as $\mu$D 2018~\cite{HERNANDEZ2018377}, and $\mu$D~\cite{PhysRevA.99.030501,PhysRevA.97.062511}, respectively in Fig. 27. 

\onecolumngrid
\twocolumngrid

The PRad collaboration proposed a new electron-deuteron elastic scattering  experiment, called DRad~\cite{DRad},  using an apparatus modified from that for the proposed PRad-II experiment by installing a low-energy Silicon-based recoil detector in a cylindrical shape inside the windowless gas flowing target to detect the recoil deuterons in coincidence with the scattered electrons.  As demonstrated by the PRad experiment~\cite{Xiong19}, the proposed DRad experiment will also employ a well-known QED process, ~M{\o}ller~scattering to control the systematic uncertainties associated with measuring the absolute $e-d$ cross section. The DRad experiment will aim at an overall precision that is  0.22\% (relative) or better in the determination of the $r_d$, in an essentially model-independent way.

An elastic $e-d$ cross section measurement~\cite{Schlimme:2016wmj} was carried out at the Mainz Microtron several years ago in a momentum transfer squared range of $2.2 \times 10^{-3}$ to 0.28 (GeV/c)$^2$ with the goal of extracting the deuteron charge form factor and ultimately the deuteron charge radius. The data analysis is ongoing.  

\begin{center}
\begin{figure}[h]
\includegraphics[width=0.65\textwidth]{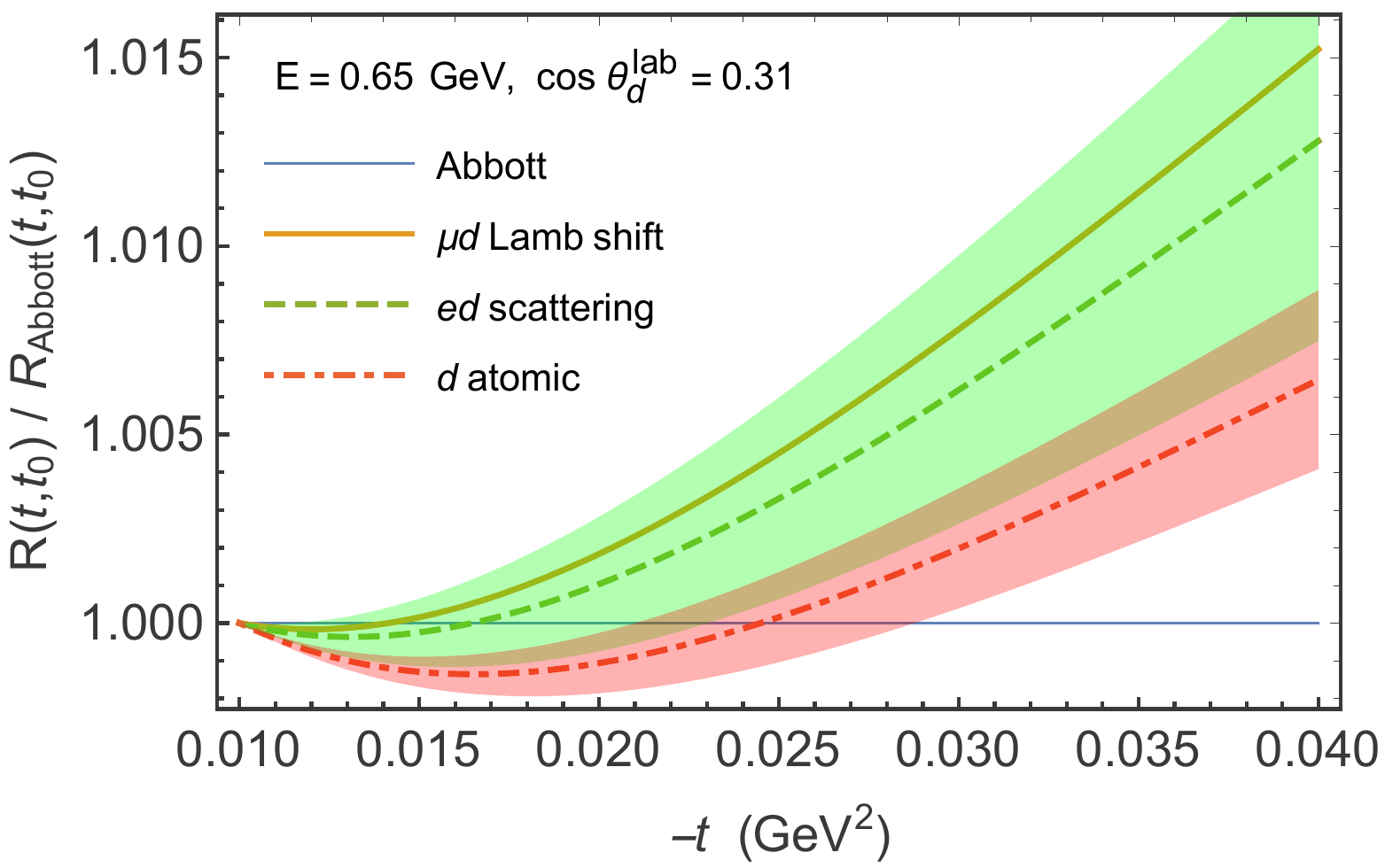}
\caption{(Color online) The momentum transfer 
$t$-dependence of the $\gamma d \to e^+e^- d$ cross section ratio $R(t,t_0)$, defined in Eq.~(\ref{eq:deutratioR}), for reference value $t_0 = -0.01$ GeV$^2$, at fixed deuteron lab angle, and for beam energy 0.65 GeV.  For convenience, the ratio is normalized to the result using Abbott {\it et al.}~form factors~\cite{Abbott:2000ak}. The curves and associated error bands are for different extractions of the deuteron charge radius (see text for details). Figure from \cite{Carlson:2018ksu}.
}
\label{fig:inelasticrelative}
\end{figure}
\end{center}

Furthermore, Carlson and Vanderhaeghen investigated the sensitivity of the cross section for lepton pair production off a deuteron target, $\gamma d \to e^+ e^- d$,  to the deuteron charge radius~\cite{Carlson:2018ksu}.  They demonstrated that for small momentum transfer this reaction is dominated by the Bethe-Heitler process, shown in Fig.~\ref{fig:bh}. 
They propose to measure the deuteron at a fixed angle, and scan the momentum transfer ($t$) dependence of the  $\gamma d \to e^+ e^- d$ cross section ratio defined as:
\begin{equation}
R(t,t_0) \equiv  \frac{d\sigma / dt \, dM_{ll}^2(t)}{  d\sigma / dt \, dM_{ll}^2(t_0)},
\label{eq:deutratioR}
\end{equation}
with $t = (p' - p)^2$ the momentum transfer, which is in one-to-one relation with the recoil deuteron lab momentum, $|\vec p^{\, \prime}|^{lab} = 2 M_d \sqrt{\tau_d (1 + \tau_d)}$, with $\tau_d \equiv -t/(4 M_d^2)$. 
Furthermore in Eq.~(\ref{eq:deutratioR}), $M^2_{ll}$ is the squared invariant mass of the dilepton pair, which at a fixed deuteron angle is a function of $t$, and the denominator in the ratio $R$ is the cross section for the same deuteron scattering angle and for a reference momentum transfer $t_0$.  
This ratio is shown in  Fig.~\ref{fig:inelasticrelative} for three extractions of the deuteron charge radius displayed in Fig.~\ref{fig:DRad}: 
the muonic deuterium Lamb shift value~\cite{Pohl:2016} (gold solid line, with uncertainty comparable to the width of the line);  $e$-$d$ elastic scattering value ~\cite{Sick:1998cvq} (green dashed line, with uncertainty limits indicated by the green band);   deuterium atomic spectroscopy value~\cite{Pohl_2017} (red dot-dashed line, with uncertainty limits indicated by the red band). One sees from Fig.~\ref{fig:inelasticrelative} 
that such cross section ratio measurement of about $0.1 \%$ relative accuracy could give a deuteron charge radius more accurate than the current $e-d$ scattering value~\cite{Sick:1998cvq} and sufficiently accurate to distinguish between the electronic and muonic atomic values.

\onecolumngrid
\twocolumngrid

\section{Conclusions}
\label{conclusions}

In this paper, we reviewed the experimental progress towards the resolution of the proton charge radius puzzle over the past decade as well as the related theoretical background and developments. In light of the latest precise determinations of the proton charge radius from both ordinary atomic hydrogen spectroscopic measurements, and the PRad electron scattering experiment, some might be tempted to conclude that the puzzle has been resolved.  
We point out however, while the recent experimental results prefer the CREMA value at about 0.84 fm, they are still within 3 standard deviations from the previously compiled value of about 0.88 fm. 
Furthermore, the most precisely determined value of $r_{Ep}$~\cite{Grinin1061} from ordinary hydrogen spectroscopy -- also the most recent measurement -- is about two standard deviations larger than the muonic hydrogen results. 
We believe more experiments, especially those with improved precision from electron scattering, and new results from muon scattering will be essential to fully resolve this puzzle.  To answer a more tantalizing question -- whether there is a difference in the proton charge radius determined from experiments involving electronic (e-p and ordinary hydrogen) versus muonic systems --  significantly improved precision from lepton scattering and also measurements from ordinary hydrogen spectroscopy with precision comparable to that of ~\cite{Grinin1061} will be critical. Pushing the precision frontier has more than once proven to be the harbinger of new discoveries.

\section{Acknowledgments}

H.G. thanks Xiangdong Ji for bringing the proton charge radius issue to her attention for the first time in the late 1990s. H.G. wishes to thank all members of the PRad Collaboration, especially Chao Peng, Weizhi Xiong, Xinzhan Bai, Chao Gu, Xuefei Yan, Dipangkar Dutta, Ashot Gasparian, Kondo Gnanvo, Mahbub Khandaker, Douglas Higinbotham, Nilanga Liyanage, Eugene Pasyuk, Jingyi Zhou, Yang Zhang and Vladimir Khachatryan.  H.G and M.V. also wish to thank Randolf Pohl for helpful discussions concerning the deuteron charge radius measurements, Ron Gilman for the MUSE experiment, and Jingyi Zhou for her assistance in making a number of figures used in this paper.

M.V. wishes to thank his collaborators on various works summarized here, especially C.E. Carlson, M. Gorchtein, M. Guidal, N. Kivel, C. Lorc\'e, V. Pascalutsa, B. Pasquini, V. Pauk, O. Tomalak.

The work of H.G. is supported in part by the U.S. Department of Energy under Contract No. DE-FG02-03ER41231.

The work of M.V. is supported by the Deutsche Forschungsgemeinschaft (DFG, German Research Foundation), in part through the Collaborative Research Center [The Low-Energy Frontier of the Standard Model, Projektnummer 204404729 - SFB 1044], and in part through the Cluster of Excellence [Precision Physics, Fundamental Interactions, and Structure of Matter] (PRISMA$^+$ EXC 2118/1) within the German Excellence Strategy (Project ID 39083149).

\bibliography{ProtonRadius-rmp-revtex5}

\end{document}